\documentclass[acmsmall]{acmart}

\usepackage{amsmath,amsfonts}
\usepackage{algorithmic}
\usepackage{graphicx}
\usepackage{textcomp}
\usepackage{xcolor,colortbl}
\usepackage{booktabs}
\usepackage{tikz,pgfplots}
\usepackage{pgfplots}
\usepackage{pgfplotstable}
\usepackage{standalone}
\usepackage{multirow}
\usepackage[tikz]{bclogo}
\usepackage[listings,skins,breakable]{tcolorbox}
\usepackage{csquotes}
\usepackage{amsmath}
\usepackage{url}
\usepackage[tikz]{bclogo}
\usetikzlibrary{arrows.meta}
\usepackage{fancyhdr}
\usepackage{threeparttable}
\usepackage{subcaption}
\usepackage{makecell}
 \usepackage{pbox}
\usepackage[ruled, linesnumbered, vlined, commentsnumbered]{algorithm2e}
\usepackage{tcolorbox}
\newtcolorbox{quotebox}{colback=steel!10,boxrule=0.4pt,colframe=black,fonttitle=\bfseries,top=2pt,bottom=2pt}
\usepackage{xfp}

\definecolor{result}{rgb}{0.1, 0.3, 0.5} 
\definecolor{steel}{rgb}{0, 0.2, 0.9} 

\mathchardef\mhyphen="2D
\newcommand{\vect}[1]{\boldsymbol{#1}}
\DeclareMathAlphabet\mathbfcal{OMS}{cmsy}{b}{n}

\newcolumntype{P}[1]{>{\centering\arraybackslash}m{#1}}
\newcolumntype{Y}{>{\centering\arraybackslash}X}

\usetikzlibrary{pgfplots.statistics}

  \newcommand{\squart}[4]{\begin{adjustbox}{max width=.1\textwidth}\begin{picture}(100,5)
    {\color{black}\put(0,5){\line(1,0){100}}\color{black}\put(0,5){\line(0,1){10}}\put(50,5){\line(0,1){10}}\put(100,5){\line(0,1){10}}\put(25,5){\line(0,1){5}}\put(75,5){\line(0,1){5}}\put(-2,-8){\LARGE$0$}\put(42,-8){\LARGE$0.5$}\put(96,-8){\LARGE$1$}}\end{picture}\end{adjustbox}}

  \newcommand{\quart}[4]{\begin{adjustbox}{max width=.1\textwidth}\begin{picture}(100,5)
    {\color{black}\put(#1,5){\line(1,0){#2}}\color{black}\put(#1,0){\line(0,1){10}}\color{black}\put(\fpeval{#1+#2},0){\line(0,1){10}}\color{steel}\put(#3,5){\circle*{7}}\color{black}\put(#3,5){\circle{7}}}\end{picture}\end{adjustbox}}
    
      \newcommand{\bquart}[4]{\begin{adjustbox}{max width=.1\textwidth}\begin{picture}(100,5)
    {\color{black}\put(#1,5){\line(1,0){#2}}\color{black}\put(#1,0){\line(0,1){10}}\color{black}\put(\fpeval{#1+#2},0){\line(0,1){10}}\color{red}\put(#3,5){\circle*{7}}\color{black}\put(#3,5){\circle{7}}}\end{picture}\end{adjustbox}}

      \newcommand{\quartexp}[4]{\begin{adjustbox}{max width=.1\textwidth}\begin{picture}(20,5)
    {\color{black}\put(#1,3){\line(1,0){#2}}\color{black}\put(#1,0){\line(0,1){6}}\color{black}\put(\fpeval{#1+#2},0){\line(0,1){6}}\color{steel}\put(#3,3){\circle*{4}}\color{black}\put(#3,3){\circle{4}}}\end{picture}\end{adjustbox}}
    
          \newcommand{\bquartexp}[4]{\begin{adjustbox}{max width=.1\textwidth}\begin{picture}(20,5)
    {\color{black}\put(#1,3){\line(1,0){#2}}\color{black}\put(#1,0){\line(0,1){6}}\color{black}\put(\fpeval{#1+#2},0){\line(0,1){6}}\color{red}\put(#3,3){\circle*{4}}\color{black}\put(#3,3){\circle{4}}}\end{picture}\end{adjustbox}}

\DeclareMathOperator*{\argmax}{argmax}
\DeclareMathOperator*{\argmin}{argmin}

\pgfplotsset{compat=1.11,
        /pgfplots/ybar legend/.style={
        /pgfplots/legend image code/.code={%
        \draw[##1,/tikz/.cd,bar width=3pt,yshift=-0.2em,bar shift=0pt]
                plot coordinates {(0cm,0.8em)};},
},
}
\AtBeginDocument{%
  \providecommand\BibTeX{{%
    \normalfont B\kern-0.5em{\scshape i\kern-0.25em b}\kern-0.8em\TeX}}}
    
\definecolor{mycolor}{rgb}{0.122, 0.435, 0.698}

\newtcbox{\mytag}{nobeforeafter,colframe=mycolor,colback=mycolor!30!white,boxrule=0.7pt,arc=0pt,
  boxsep=-3pt,left=6pt,right=6pt,top=4pt,bottom=5pt,tcbox raise base}

\setcopyright{acmlicensed}
\acmJournal{TOSEM}
\acmYear{2022} \acmVolume{1} \acmNumber{1} \acmArticle{1} \acmMonth{1} \acmPrice{15.00}\acmDOI{10.1145/3514233}

\begin{document}

\title[Pareto Search versus Weighted Search in Multi-Objective SBSE]{The Weights can be Harmful: Pareto Search versus Weighted Search in Multi-Objective Search-Based Software Engineering}

\author{Tao Chen}
\thanks{Both authors made commensurate contributions to
this research. Corresponding author: Tao Chen, t.t.chen@lboro.ac.uk}
\email{t.t.chen@lboro.ac.uk}

\affiliation{%
    \institution{Loughborough University}
  \country{United Kingdom}
}

\author{Miqing Li}
\affiliation{%
  \institution{University of Birmingham}
  \country{United Kingdom}
 }
\email{m.li.8@bham.ac.uk}

\renewcommand{\shortauthors}{Chen and Li.}

\ccsdesc[500]{Software and its engineering~Search-based software engineering}
\ccsdesc[300]{Software and its engineering~Empirical software validation}
\ccsdesc[300]{Software and its engineering~Software performance}


\begin{abstract}


In presence of multiple objectives to be optimized in Search-Based Software Engineering (SBSE),
Pareto search has been commonly adopted. 
It searches for a good approximation of the problem's Pareto optimal solutions, 
from which the stakeholders choose the most preferred solution according to their
preferences.
However, 
when clear preferences of the stakeholders 
(e.g., a set of weights which reflect relative importance between objectives) 
are available prior to the search,
weighted search is believed to be the first choice since it simplifies the search 
via converting the original multi-objective problem into a single-objective one 
and enable the search to focus on what only the stakeholders are interested in.

This paper questions such a \textit{``weighted search first''} belief.
We show 
that the weights can, in fact, be harmful to the search process even in the presence of clear preferences. 
Specifically, 
we conduct a large scale empirical study 
which consists of 38 systems/projects from three representative SBSE problems, 
together with two types of search budget and nine sets of weights, 
leading to 604 cases of comparisons. 
Our key finding is that weighted search reaches a certain level of solution quality 
by consuming relatively less resources at the early stage of the search; 
however, Pareto search is at the majority of the time (up to 77\% of the cases) significantly better than its weighted counterpart, 
as long as we allow a sufficient, but not unrealistic search budget. 
This is a beneficial result,
as it discovers a potentially new ``rule-of-thumb'' for the SBSE community: 
even when clear preferences are available, 
it is recommended to always consider Pareto search by default for multi-objective SBSE problems 
provided that solution quality is more important. 
Weighted search, 
in contrast, 
should only be preferred when the resource/search budget is limited, 
especially for expensive SBSE problems. 
This, 
together with other findings and actionable suggestions in the paper, 
allows us to codify pragmatic and comprehensive guidance on choosing weighted and Pareto search 
for SBSE under the circumstance that clear preferences are available. All code and data can be accessed at: \textcolor{blue}{\url{https://github.com/ideas-labo/pareto-vs-weight-for-sbse}}. 

\end{abstract}


\keywords{Search-based software engineering, multi-objective optimization, Pareto optimization, quality evaluation, quality indicator, user preference, configurable systems, adaptive systems, self-adaptive systems.}
\maketitle


\section{Introduction}
\label{sec:introduction}
	
Search-Based Software Engineering (SBSE) specializes the heuristic optimizers to automatically discover solutions for minimizing/maximizing objective(s) or for satisfying certain constraint(s) in various software engineering problems~\cite{Harman2012}. Over the past decades, SBSE has enjoyed a significant growth, as researches related to SBSE have spanned across different phases of software engineering, including requirements analysis~\cite{DBLP:journals/tosem/ZhangHORB18}, design~\cite{DBLP:conf/scam/AlizadehFK19}, testing~\cite{DBLP:journals/tec/WangAYL18}, deployment~\cite{Chen2018FEMOSAA}, and runtime self-adaptation~\cite{DBLP:journals/pieee/ChenBY20}.  


Many SBSE problems involve two or more objectives, which are more or less conflicting. 
For example, software testing needs to make a trade-off between coverage and cost; software configuration tuning involves conflicting objectives of latency and memory consumption. It is, therefore, an important engineering decision for one to choose how the relationship between the objectives can be formulated for the search algorithm to deal with. 
The SBSE community takes two alternative strategies in this engineering decision-making: Pareto search or weighted search (a.k.a. utility search). 
The former searches for a good approximation of the Pareto front,
from which the stakeholders make their choice~\cite{li2018critical,Li2020,Ramirez2019}. 
The latter directly searches for a single solution that maximizes the aggregated scalar fitness of the objectives (e.g., by weighted sum~\cite{Harman2012,10.1145/3392031,DBLP:conf/ssbse/BowersFC18}), 
on the basis of a set of weights (also called a weight vector) that reflects relative importance between the objectives.

In general, researchers in multi-objective SBSE choose one of the above two strategies 
according to availability and assumptions on the preferences of stakeholders: 
when no preferences (weights) between the objectives are available, 
Pareto search is undoubtedly chosen as it can reveal the entire Pareto front of solutions with rich diversity for one to examine without any prior information about the preferences~\cite{DBLP:journals/jss/CalinescuCGKP18,DBLP:journals/infsof/ChenLY19,DBLP:journals/tosem/ZhangHORB18,DBLP:conf/iccS/ChenBT13}. 
However, if clear preferences can be articulated, elicited, or even assumed, 
the weighted search would be used instead. This makes sense
since naturally the weights can simplify the problem and focus the search on the direction that is only of interest to the stakeholders 
(no waste of the resources on searching for solutions which are not of interest to the stakeholders)~\cite{DBLP:conf/scam/AlizadehFK19,DBLP:conf/icac/RamirezKCM09,DBLP:conf/icsm/AntoniolPH05,DBLP:journals/tec/WangAYL18,DBLP:conf/gecco/CanforaPEV05,DBLP:conf/icse/ChenB14,DBLP:journals/jss/SobhyMBCK20}. To confirm the prevalence of weighted search under such case, we conducted a pilot search over Google Scholar with the search string \texttt{``weights'' AND ``search based software engineering''} and randomly sampled 29 papers\footnote{Why 29? We obtained this number based on the equation of sample size by Kadam and Bhalerao~\cite{kadam2010sample} under the total number of papers returned by the search string (which is 1610) with 90\% confidence interval.} that assume weights are available between the objectives (surveys and tutorials are excluded). Among those, we found that 25 papers (i.e., 86\%) have chosen weighted search over its Pareto counterpart, which is clearly a large proportion. In addition, the weighted search has also been recommended in well-known SBSE roadmaps. For example, Harman~\cite{DBLP:conf/promise/Harman10} pinpoints that \textit{``where we know the relative weighting to be applied to the elements of $V$, we can simply use a single objective approach in which the overall fitness is a weighted sum of the predictive quality of each element in $V$.''}



Whilst it is clear that the Pareto search strategy is a good choice when no preferences are available, 
the strategy is also applicable when a set of weights is given. 
That is, 
we can run a Pareto optimizer that produces a set of well-distributed solutions (approximating the Pareto front), 
and then apply the given weights to cherry-pick a solution therein 
(i.e., the solution which is the most aligned with those weights).
This naturally raises a question: 
when there are (or assumed to have) clear preferences (i.e., weights), 
how does Pareto search perform in comparison with weighted search which has been believed to be well-suited to this situation in SBSE?


In SBSE, 
there exist some studies that have touched on the comparison between weighted search and Pareto search.
For weighted search, 
those studies use the given weight vector to simplify the problem and guide the search, 
but when it comes to comparing the results returned by weighted search with those by Pareto search,
they either considered generic quality indicators (e.g., hypervolume~\cite{Zitzler1998}) which are designed for Pareto search (such as \cite{DBLP:journals/tec/WangAYL18,DBLP:conf/gecco/PradhanWAY16,10.1145/3392031}), 
or the value on every objective of the SBSE problem, e.g., \cite{Zhang2007The}.
Such comparisons apparently disadvantage weighted search since the stakeholders' preferences (weights) are only used in the search but not in the evaluation. 
That is, 
to evaluate weighted search under a situation that the preferences are assumed to be unavailable. 
This certainly results in the conclusion that Pareto search is always better than weighted search~\cite{Zhang2007The,DBLP:journals/tec/WangAYL18,DBLP:conf/gecco/PradhanWAY16,10.1145/3392031}.
In this work, 
we aim to make a more fair and comprehensive comparison between weighted search and Pareto search 
under clear preferences in multi-objective SBSE.

\subsection{Hypothesis and Research Questions}

In this paper, we seek to understand whether Pareto search can serve as an equivalent alternative to weighted search for multi-objective SBSE problems 
under the circumstance that clear preferences are known. To this end, we conduct a confirmatory study, wherein our hypothesis is that:

	\begin{displayquote} 
		\textit{\textbf{Hypothesis:} Pareto search may be competitive with weighted search under clear preferences,
		i.e., a set of weights that reflect relative importance between the objectives, if the
		budget is sufficient.}
	\end{displayquote}

\noindent The rationale for this is that the Pareto search strategy searches for the whole Pareto front 
while the weighted search strategy searches for a single point on that Pareto front~\cite{10.1145/3392031,Emmerich2018} --- 
the result of the former with posterior cherry-picking can be similar to that of the latter provided that a sufficient budget
is allowed. 
This motivates us to reconsider the validity of the \textit{``weighted search first''} belief in multi-objective SBSE, 
given that it appears to be a general standard when clear preferences can be articulated, elicited or even assumed \cite{DBLP:conf/scam/AlizadehFK19,DBLP:conf/icac/RamirezKCM09,DBLP:conf/icsm/AntoniolPH05,DBLP:journals/tec/WangAYL18,DBLP:conf/gecco/CanforaPEV05,ding2006discussions,nan2019study,Harman2012,10.1145/3392031,DBLP:conf/ssbse/BowersFC18,DBLP:conf/sigsoft/ShahbazianKBM20,DBLP:conf/icse/ChenB14}. 
	






To verify the hypothesis, 
we systematically compare weighted search with Pareto search in SBSE in terms of solution quality (with respect to the given weight vector) and resources required to reach its certain level. 
This is achieved through a comprehensive empirical study consisting of 38 instances from three representative multi-objective SBSE problems with two or three objectives, covering a wide spectrum of characteristics, representations, search space, and objectives. This, together with two types of search budget (evaluations and time) and nine (four for three objective case) weight vectors, leads to 604 cases of investigation.


The comparison is nevertheless not straightforward, 
since the objectives in a multi-objective SBSE often come with rather different scales. 
Unlike Pareto search which can be scale-free, 
weighted search is significantly affected by the scale of different objectives.
Therefore, for using weighted search an additional decision is needed to make on 
how to normalize the objectives such that they become commensurable. 
The issue is a necessity applied to any optimizer for weighted search. 
While it has been shown that the best optimizer of weighted search depends on the SBSE problems and cases~\cite{Harman2012}, 
it is not previously known whether this is also the case for commonly used normalization methods or there is indeed a best one in general. Therefore, the first research question (RQ) we wish to answer is:

\begin{quotebox}
   \noindent
   \textit{\textbf{RQ1:} Is there a normalization method for weighted search that leads to the best solution in general across all multi-objective SBSE cases studied?}
\end{quotebox}

To that end, we compare four normalization methods (see Section~\ref{sec:rq1}) under four optimizers (i.e., Random Search~\cite{DBLP:conf/icse/ArcuriB11}, Hill Climbing with restart~\cite{Harman2012}, Simulated Annealing~\cite{DBLP:journals/pami/GranvilleKR94}, and Single-Objective Genetic Algorithm~\cite{goldberg2006genetic}). 
These optimizers have been widely used for weighted search in multi-objective SBSE according to the well-known SBSE surveys~\cite{Harman2012,DBLP:journals/infsof/ColanziAVFG20}. Investigating \textbf{RQ1} directly serves as the foundation to our next RQ:

\begin{quotebox}
   \noindent
   \textit{\textbf{RQ2:} Given a sufficient search budget, can Pareto search produce a generally competitive solution compared with its best weighted counterpart over the multi-objective SBSE cases considered?}
\end{quotebox}

Understanding \textbf{RQ2} requires us to choose a representative algorithm for Pareto and weighted search, respectively. 
In this work, for each case, we use the best optimizer and normalization method pair amongst the considered ones as the representative for weighted search, 
drawn from the results obtained in \textbf{RQ1}. 
Pareto search is represented by NSGA-II~\cite{Deb2002} --- arguably the most commonly used Pareto optimizer in multi-objective SBSE~\cite{Harman2012,Sayyad2013b,DBLP:journals/infsof/ColanziAVFG20}; and MOEA/D~\cite{DBLP:journals/tec/ZhangL07}, which is an optimizer that possesses many similarities with the weighted search.

Since the preferences between the objectives are described by a weight vector, 
an extended question of \textbf{RQ2} for us to examine is that:

\begin{quotebox}
   \noindent
   \textit{\textbf{RQ3:} Across multi-objective SBSE cases, 
   	do different weight vectors affect the comparative results between Pareto search and weighted search?}
\end{quotebox}

Apart from the quality of solutions, 
the resources required (i.e., search budget) also plays an integral role for software engineers to decide on whether Pareto or weighted search is a preferred strategy to handle multiple objectives in SBSE. 
Our final RQ thus is:

\begin{quotebox}
   \noindent
   \textit{\textbf{RQ4:} Overall, of Pareto search and weighted search, 
   	which is more resource efficient over different multi-objective SBSE cases?}
\end{quotebox}

As mentioned, we study this on two types of search budgets that reflect the resources, i.e., the number of evaluations and time. 
In particular, we seek to understand which consumes less resources for reaching a certain level of solution quality.


\subsection{Contributions}

The findings of our empirical study are encouraging yet surprising. 
The most unexpected result is that \textit{the weights can be considerably harmful to the search in multi-objective SBSE even under clear preferences}: given sufficient search budget, Pareto search is not only competitive with weighted search, 
but most of the time produce a significantly better solution than its weighted counterpart. 
Notably, a sufficient search budget does not have to be unrealistically high; rather, it is often reasonable in practice, 
e.g., it can be in the magnitude of seconds or less for some SBSE problems. 
Yet, this does not mean that weighted search can be completely abandoned: 
we confirm that it does consume less resources to reach a certain level of solution quality, 
hence it may still be preferred when the search budget is rather limited. Therefore, a key message we found from this work is that:

	\begin{displayquote} 
		\textit{\textbf{Key message:} When clear preferences (weights) are available in a multi-objective SBSE problem, the choice between Pareto search and its weighted counterpart can be a trade-off between the quality of solution and the provision of search budget.}
	\end{displayquote}


Specifically, our contributions are:

\begin{enumerate} 

\item An empirical study to understand the in-depth strengths/weaknesses of both Pareto and weighted search for multi-objective SBSE under clear preference. 
We find that:

\begin{itemize} 

\item[---] \textbf{RQ1:} The choice of the normalization methods can significantly affect the results of the weighted search 
and there does not exist a generally best one across all multi-objective SBSE cases. 
However, we do find that one often performs reasonably well 
(i.e., mostly the second best, if not the best) 
and one generally performs the worst (or the second-worst). This means that, when the objectives are of different scales, an additional process of finding the best normalization method is necessary for the weighted search to unlock its full potentials.

\item[---] \textbf{RQ2:} Pareto search can produce a significantly better solution than its best weighted counterpart for up to 77\% of the SBSE cases under sufficient search budget.

\item[---] \textbf{RQ3:} While the gain of Pareto search over its best weighted counterpart is mostly positive across the cases, 
the extent of which does vary depending on the weight vector. 
In particular, 
the maximum gain often appears under a certain range of weights in an SBSE problem. 
On the other hand, 
the lowest gain often occurs when the weights are closer to extreme values, e.g., $(0.1,0.9)$ and $(0.2,0.8)$.



\item[---] \textbf{RQ4:} The weighted search reaches a certain quality level by consuming less resources. 
However, the finding from \textbf{RQ2} suggests that it will not be able to reach the same quality level of Pareto search 
for most of the cases if the search is allowed to continue.

\end{itemize}

\item Actionable suggestions derived from the findings.

\item In-depth discussions on the reasons behind the above observations.

\item Drawing on the findings and suggestions from the RQs, 
we codify pragmatic and comprehensive guidance for the SBSE practitioners to decide on whether to use Pareto search or weighted search under an SBSE situation there are clear preferences available.

\end{enumerate} 

To promote open science practices, all source code and data of this work can be publicly accessed at our repository: \textcolor{blue}{\url{https://github.com/ideas-labo/pareto-vs-weight-for-sbse}}.

In what follows, this paper is organized as: Section~\ref{sec:bg} provides necessary background and ideational support of our hypothesis. Section~\ref{sec:cases} discusses the SBSE problems/instances studied and the rationale of these choices. Section~\ref{sec:empirical} justifies our designs of the empirical study. Section~\ref{sec:results} elaborates the findings, suggestions and reasons of observations, following by pragmatic guidance in Section~\ref{sec:guide}. Sections~\ref{sec:disccusion},~\ref{sec:related} and~\ref{sec:conclusion} present discussions, related work and conclusion, respectively.




\section{Theory}
\label{sec:bg}








Multi-objective optimization refers to mathematical optimization involving more than one objective to be tackled simultaneously.  
Without loss of generality, 
it can be generically expressed as:
\begin{align}
	\min \vect{f}(x) = (f_1(x), f_2(x),\cdots,f_m(x))
	\label{Eq:MOP}
	\end{align}
where $m$ is the number of objectives, 
and $x$ denotes a solution in the feasible solution space $X$, 
i.e., $x\in X\subset\mathbb{R}^n$ 
($n$ is the number of decision variables of the problem). 
As stated by Harman et al.~\cite{Harman2012}, 
in SBSE there are two fundamental components of an optimizer that one has to specialize: 
(i) the representation, 
i.e., how $x$ can be structured and changed; 
(ii) the objective function, 
i.e., how each single $f_i$ can be formulated to distinguish between the good and bad solutions. 
In the presence of multiple objectives,
a solution $x^1$ is said being better than $x^2$, 
called $x^1$ (Pareto) dominates $x^2$, 
	if and only if $x^1$ is not worse than $x^2$ for all the objectives and better for at least one objective.
	For a solution $x \in X$,
	if there is no solution in $X$ dominating $x$,
	then $x$ is Pareto optimal. 
	The set of all the Pareto optimal solutions is called the Pareto optimal set, 
	which can be prohibitively large or even infinite. 
	The image of the Pareto optimal solution set in the objective space is called the Pareto front.



\subsection{Pareto Search}

Since the optimum of a multi-objective problem is a Pareto front which can be infinite,
a straightforward strategy to tackle the problem is to search for an approximation set
that can well represent the front.
Afterwards,
from the approximation set, the stakeholders will choose their preferred one. 
This strategy is called Pareto search.
In many multi-objective SBSE problems~\cite{DBLP:journals/jss/CalinescuCGKP18,DBLP:journals/infsof/ChenLY19,DBLP:journals/tosem/ZhangHORB18}, 
Pareto search, working with a population-based optimizer (e.g., an evolutionary algorithm), 
is widely adopted,
where one individual in the population is used to represent a trade-off between objectives. 
Note that, by Pareto search, we refer to any optimizer that treats the objectives ``incomparably'' and searches for the entire Pareto front. As such, it includes not only optimizers that are based on Pareto-dominance relation (e.g., NSGA-II), 
but also those where multiple weight vectors are used (e.g., MOEA/D~\cite{DBLP:journals/tec/ZhangL07}), 
weight vectors are changed during the search (e.g., AdaW~\cite{Li2020weights}), 
or a quality indicator is used to guide the search (e.g., IBEA~\cite{DBLP:conf/ppsn/ZitzlerK04}), 
as long as they assume no clear preferences exist and seek to approximate the Pareto front.




\subsection{Weighted Search}

Another common strategy to deal with a multi-objective SBSE problem is to 
convert it into a single-objective problem through aggregating the objective functions (by a set of weights)~\cite{DBLP:conf/scam/AlizadehFK19,DBLP:conf/icac/RamirezKCM09,DBLP:conf/icsm/AntoniolPH05,DBLP:journals/tec/WangAYL18,DBLP:conf/gecco/CanforaPEV05}.
For example,
given a set of weights $(w_1, w_2,..., w_m)$ which satisfies $w_1+w_2+...+w_m=1$, 
the multi-objective problem in Eq.~(\ref{Eq:MOP}) can be converted into minimizing the weighted sum of the objectives: 
\begin{align}
	\min f_{ws}(x) = w_1f_1(x) + w_2f_2(x) +...+ w_mf_m(x)
	\label{Eq:SOP}
\end{align}
Although how to decide such a set of weights may be arguable for SBSE problems,
this strategy, 
as long as the weight vector can be confidently specified by the stakeholders,
is commonly believed to be the best practice that can lead to the most desired result, 
since it incorporates the stakeholders' preferences into the search and targets one single Pareto optimal solution, 
thus significantly simplifying the search problem~\cite{10.1145/3392031,Emmerich2018}. 
Note that in a specific SBSE scenario, 
it is possible that the stakeholders express a clear idea about relative importance between the objectives like 
\texttt{``the system latency is three times more important than the memory consumption''}; 
then the weights for the latency and memory consumption objectives can thus be 0.75 and 0.25, 
respectively~\cite{DBLP:conf/icac/RamirezCMB10,DBLP:conf/sigsoft/ShahbazianKBM20,DBLP:conf/ssbse/BowersFC18}.

\begin{figure}[t]
	\begin{center}
		\centering
		\begin{tabular}{@{}cc}
			\includegraphics[height=0.2\textheight]{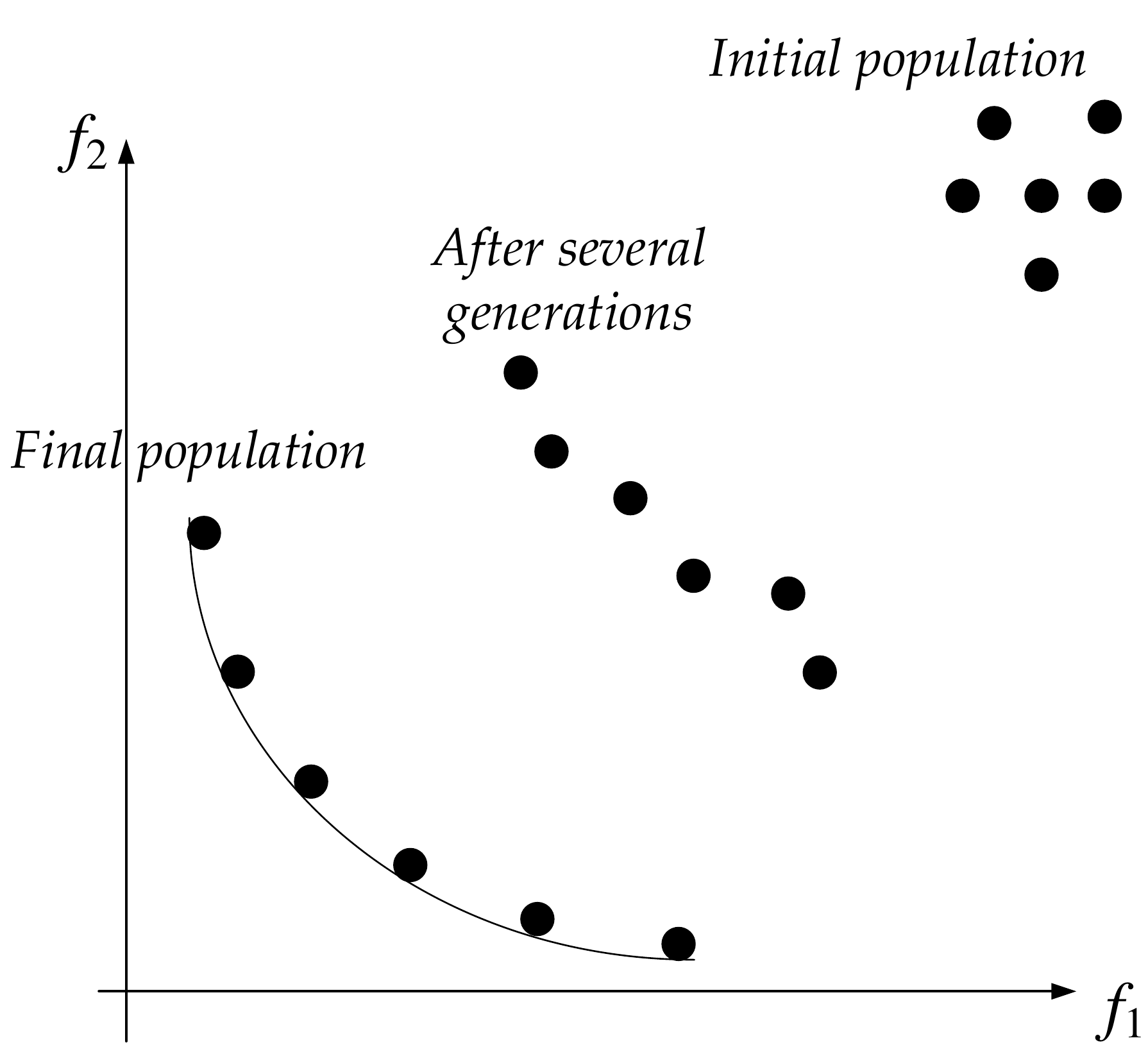}&
			\hspace{1cm}
			\includegraphics[height=0.2\textheight]{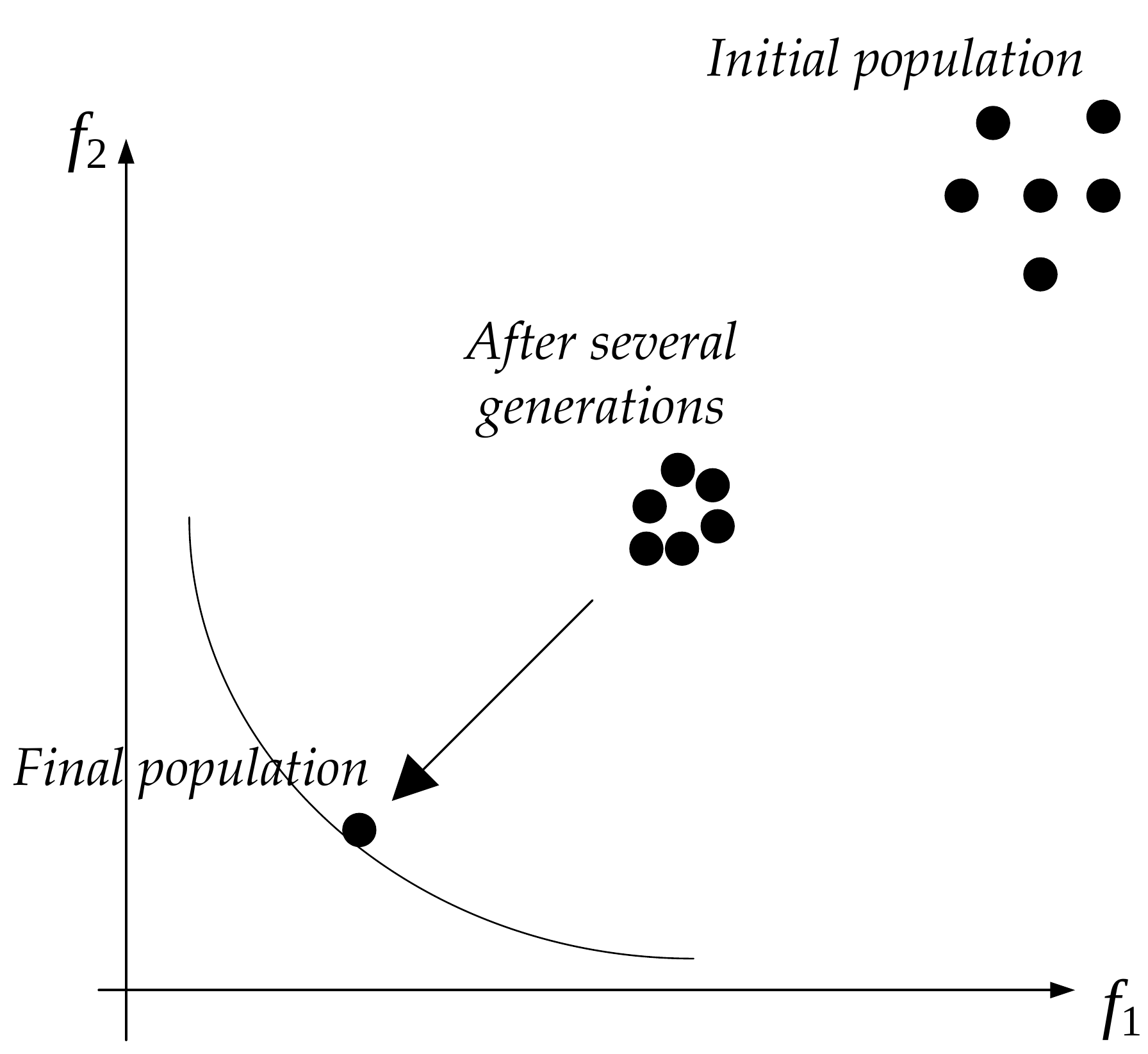}\\
			(a) Pareto search & (b) Weighted search \\
		\end{tabular}
	\end{center}
	\caption{Illustration of Pareto search and weighted search in a bi-objective minimization scenario,
		where the curve represents the Pareto front. 
		For weighted search, the weight vector is $(0.5,0.5)$ and the population seeks to converge into one point on the Pareto front.}
	\label{Fig:ParetoVSweighted}
\end{figure}

Figure~\ref{Fig:ParetoVSweighted} illustrates how the two strategies differ with respect to their search process 
(under a population-based optimizer, e.g., genetic algorithm). 



\subsection{Why Pareto Search can be Competitive to Weighted Search?}
\label{sec:theory-reason}

It is commonly accepted that Pareto search can be a ``go-to'' solution 
when the preferences of the stakeholders cannot be provided. 
Yet,
we argue that even when the preferences can be confidently specified,
there is still a theoretical possibility that Pareto search may outperform weighted search as well.
There are two reasons supporting this. 
First, 
compared to weighted search, 
Pareto search may not easily get stuck in local optima,
particularly when the objectives are conflicting.
This is because the solutions during the search process are often incomparable 
(i.e., Pareto nondominated to each other) 
and the population-based search can maintain a set of diverse solutions.
In contrast, 
fine-grained comparability of the scalar fitness in weighted search may not be able 
to maintain some solutions which can help jump out of local optima.

\begin{figure*}[t!]
	\centering
	\includegraphics[scale=0.35]{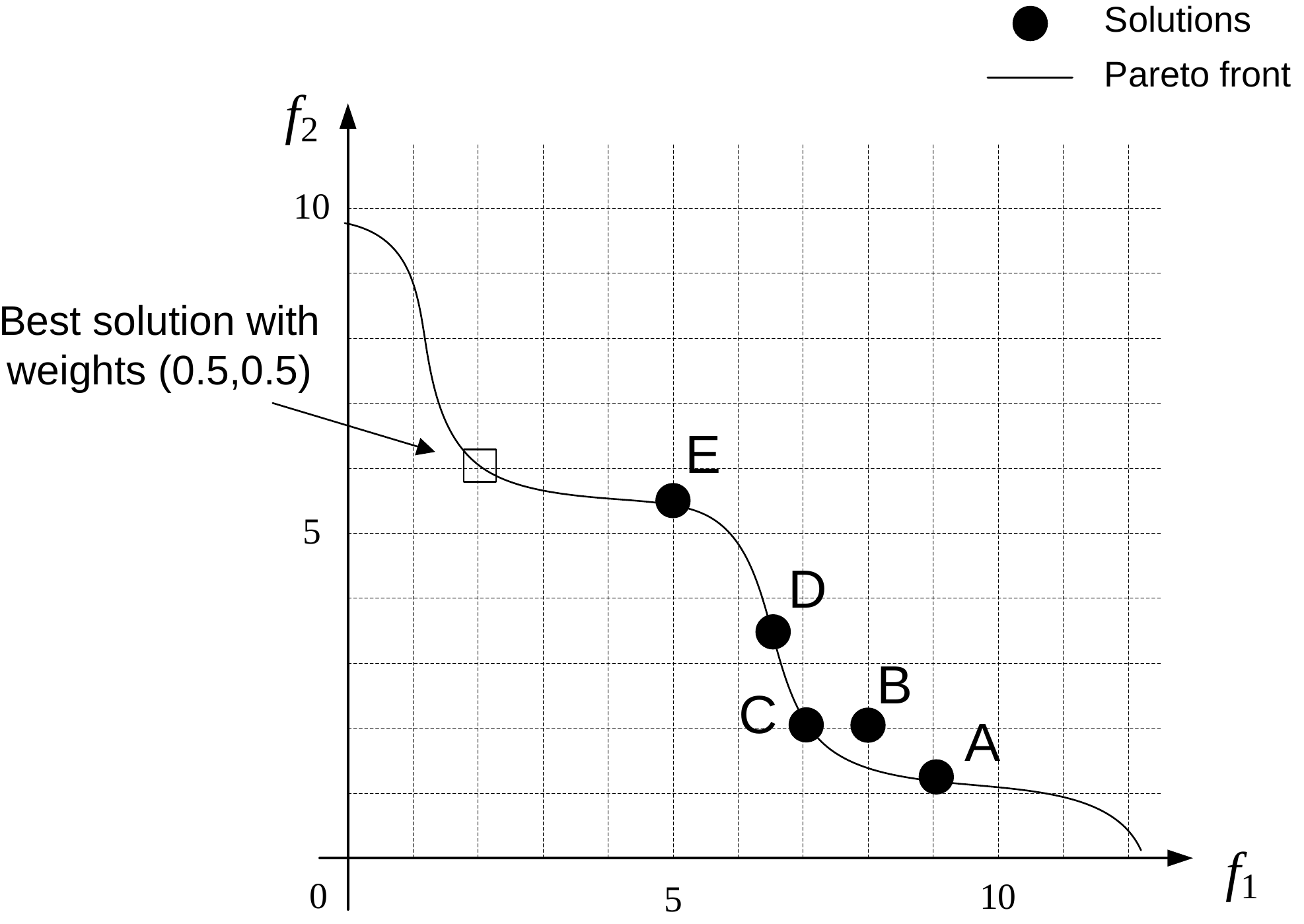}
	\caption{An illustration of why Pareto search may be easier to find the global optimum than weighted search.
		Of five solutions $\mathbf{A} (9,1.25), \mathbf{B} (8,2), \mathbf{C} (7,2), \mathbf{D} (6.5,3.5)$ and $\mathbf{E} (5,5.5)$,
		$\mathbf{A}, \mathbf{C}, \mathbf{D}, \mathbf{E}$ are nondominated with each other.
		The solution $\mathbf{B}$ is dominated by $\mathbf{C}$ and thus is regarded as the worst solution for Pareto search.
		For weighted search, 
		under the weight vector $(0.5, 0.5)$, 
		the score of $\mathbf{A}, \mathbf{B}, \mathbf{C}, \mathbf{D}$ and $\mathbf{E}$ is 
		$5.125, 5, 4.5, 5$ and $5.25$, respectively.
		Apparently, $\mathbf{E}$ has the worst score and is likely to be eliminated for weighted search.
		However,
		$\mathbf{E}$ is actually closer than the other four solutions 
		to the possible best solution obtained for the problem under the weight vector $(0.5,0.5)$
		(i.e., the square in the figure and its score is $4$).
		Preserving $\mathbf{E}$ is beneficial to help find the global optimum of the problem.}
	\label{fig:Pareto}
\end{figure*}

Consider a bi-objective minimization example shown in Figure~\ref{fig:Pareto} 
where there are five solutions $\mathbf{A} (9,1.25), \mathbf{B} (8,2), \mathbf{C} (7,2), \mathbf{D} (6.5,3.5)$ and $\mathbf{E} (5,5.5)$.
Among them, 
$\mathbf{A}, \mathbf{C}, \mathbf{D}, \mathbf{E}$ are nondominated with each other,
while $\mathbf{B}$ is dominated by $\mathbf{C}$.
If one would like to identify the worst solution of them and eliminate it 
(e.g. because the population capacity is four),
then $\mathbf{B}$ will be that solution for Pareto search.
Now, let us say that the weight vector specified is $(0.5, 0.5)$.
Then, 
the score of the five solutions $\mathbf{A}, \mathbf{B}, \mathbf{C}, \mathbf{D}$ and $\mathbf{E}$ for weighted search is 
$5.125, 5, 4.5, 5$ and $5.25$, respectively, according to $f_{ws}(x)=w_{1}f_1(x)+w_{2}f_2(x)$.
Apparently, $\mathbf{E}$ has the worst score and is likely to be eliminated in weighted search.
However,
$\mathbf{E}$ is actually closer than the other four solutions 
to the optimal point on the Pareto front under the considered weight vector 
(i.e., the square in the figure and its score is $4$).
Apparently, 
removing $\mathbf{E}$ makes it harder to approach the optimal point later on during the search.
In fact,
the region where the solutions $\mathbf{A}, \mathbf{B}, \mathbf{C}, \mathbf{D}$ are located can be seen as a local optimal region,
and the solution $\mathbf{E}$ is outside but $\mathbf{E}$ has a worse score,
thus likely to be eliminated during weighted search.

The second reason is that in weighted search the weights, 
which are specified to reflect the stakeholders' preferences between the objectives, 
may not be able to do so throughout the search. 
Since the objectives from most SBSE problems come with radically different scales and there are often some unknown bounds~\cite{Li2020}, the bounds of the objectives found during the search, 
which are used to perform the normalization of the objectives to make them comparable in the objective aggregation,
can be very different from the real bounds for the problem. 
Working with such bounds, 
the weight vector, which is determined on the basis of the real bounds,
may easily drive the search into some areas that are not in line with the stakeholders' preferences.

In the rest of this paper, 
we will check if this theoretical possibility actually occurs in real-world SBSE problems.

\section{Subject SBSE Problems}
\label{sec:cases}


\begin{table}[t!]
\caption{Top 5 most common SBSE problems summarized from~\cite{Harman2012,Sayyad2013b,Li2020,DBLP:journals/infsof/ColanziAVFG20}.}
\label{tb:prob}
\setlength{\tabcolsep}{1mm}
\centering
\footnotesize
\begin{tabular}{cc}\toprule

\textbf{SBSE Problem}&\textbf{Paper Count}\\

\midrule

Test Case Generation (TCG)&59\\
\rowcolor{steel!10}Next Release Problem (NRP)&28\\
Software Product Line Engineering (SPLE)&17\\
\rowcolor{steel!10}Software Configuration Tuning (SCT)&16\\
Web Service Composition (WSC)&16\\

\bottomrule
\end{tabular}
\centering
\end{table}


To ensure the external validity of our empirical study, we need to choose a set of diverse SBSE problems from different domains, covering a wide spectrum of characteristics, objectives, search spaces, the dimensionality of variables, and the concrete software systems/projects. To this end, in Jan 2021, we searched over Google Scholar for well-known SBSE surveys published since 2010 with a search string \texttt{``survey'' AND ``search based software engineering"} and considered surveys that meet the following criteria:

\begin{itemize}
    \item It covers all SBSE problems in general rather than focusing on SBSE for a particular domain. This is important to avoid bias since we are interested in finding the most common SBSE problems studied.
    \item It does not focus on circumstances where the weighted search is impractical. For example, many objective SBSE problem is one such case where the weights are too difficult to be specified with the increasing dimension of the objectives.
    \item When multiple surveys come from the same research group, only the most cited one is considered.
\end{itemize}

From the above, we identified several well-known SBSE surveys~\cite{Harman2012,Sayyad2013b,Li2020,DBLP:journals/infsof/ColanziAVFG20}, covering the SBSE papers in the past two decades. Since those surveys contain readily available classification of the SBSE papers (e.g., appendix in~\cite{Harman2012} and Table A1 in~\cite{Li2020}), we summarize the top 5 most popular problems, as shown in Table~\ref{tb:prob}. Then, we read through those papers with the following selection criteria in mind:

\begin{itemize}
    \item[---] \textbf{Criterion 1:} To make sure our findings are meaningful, both Pareto search and weighted search should have been used by more than one paper for the SBSE problem.
    \item[---]  \textbf{Criterion 2:} The objectives to be optimized can be used directly to assess the quality of an optimizer for the SBSE problem, i.e., the objective is monotonically correlated with the common indicator used in the evaluation of the SBSE problem, hence the evaluation of the weighted score in weighted search is meaningful.
    \item[---]  \textbf{Criterion 3:} The SBSE problem should have readily available real-world software or data.
    
\end{itemize}

The investigation has led us to rule out the TCG problem as it violates criterion 2. This makes sense, since a higher testing coverage (a key search objective in TCG) may not necessarily detect more faults (the common metric and ultimate purpose of software testing in the evaluation). Indeed, studies~\cite{DBLP:conf/icse/GaySWH14,DBLP:conf/fase/StaatsGWH12} have shown that there is a nonmonotonic and unclear correlation between the level of coverage and the number of faults detected. We also do not consider SPLE since we found no paper that aims for a weighted (or single-objective) search when addressing the problem, thereby it does not meet criterion 1. We eventually chose SCT, WSC, and NRP as the subject SBSE problems in this work, as they satisfy all the criteria above\footnote{Without loss of generality, we convert all maximizing objectives into minimizing one by multiplying $-1$.}. 

Once the SBSE problems have been identified, we read the related papers from the surveys and investigate the concrete subject systems/projects used by the most recent work. In particular, we eliminated those subjects that contain missing data or do not work as specified in the original paper. In summary, our empirical study was conducted based upon:

\begin{itemize}
    \item 10 software systems/environments for configuration tuning used by~\cite{DBLP:conf/mascots/JamshidiC16,nair2018finding};
    \item 13 system workflows for service composition used by~\cite{DBLP:journals/infsof/ChenLY19,PubSub10849_Chen,Wada2012E};
    \item 15 software projects/versions for planning requirements in the next release used by~\cite{DBLP:journals/tosem/ZhangHORB18,DBLP:journals/access/GengYJZLGX18}.
    \end{itemize}

We would like to stress that these selected problems and subjects are by no means to be comprehensive; rather, they are representative and convenient samples of SBSE problems. This is because our aim is not to exhaustively cover all SBSE problems, but as a first step to validate our hypothesis on a set of representative ones. We hope to spark a dialogue about new research opportunities regardless of whether our hypothesis can be confirmed: a positive outcome would be surprising and exciting, which encourages the SBSE community to reconsider the key criteria to choose between Pareto and weighted search in future work when weights are available (at least for the SBSE problems studied); otherwise, negative results could imply that an extended study may be required to fully confirm the invalidity of our hypothesis. In what follows, we specify the three SBSE problems in detail.

\subsection{Software Configuration Tuning (SCT)}

\begin{table}
\caption{Software systems for {SCT}.}
\label{tb:spo}
\setlength{\tabcolsep}{1.2mm}
\centering
\footnotesize
\begin{tabular}{clccl}\toprule

\textbf{System}&$\lvert \mathbfcal{O} \rvert$&$\lvert \mathbfcal{C} \rvert$&$\lvert \mathbfcal{S} \rvert$&\textbf{Description}\\

\midrule

\textsc{wc-c1-3d}&Latency and throughput&3&1,343&{Apache Storm with Word Count on OpenNebula (1 CPU)}\\

\rowcolor{steel!10}\textsc{wc-c3-3d}&Latency and throughput&3&1,512&{Apache Storm with Word Count on OpenNebula (3 CPU)}\\

\textsc{wc-c4-3d}&Latency and throughput&3&756&{Apache Storm with Word Count on Amazon (2 CPU)}\\

\rowcolor{steel!10}\textsc{wc-c5-5d}&Latency and throughput&5&1,080&{Apache Storm with Word Count on Azure (1 CPU)}\\

\textsc{rs-c3-6d}&Latency and throughput&6&3,839&{Apache Storm with Rolling Sort on OpenNebula (3 CPU)}\\

\rowcolor{steel!10}\textsc{wc-c1-6d}&Latency and throughput&6&2,880&{Apache Storm with Word  Count on OpenNebula (1 CPU)}\\

\textsc{llvm}&{Latency and memory}&11&1,023&{A compiler profiled by the standard benchmark program}\\

\rowcolor{steel!10}\textsc{trimesh}&{Latency and \# Iteration \:}&13&239,260&{A library to
manipulate random triangle meshes}\\

\textsc{vp8}&{Latency, energy and CPU load}&11&2,736&{A video encoder for video processing}\\

\rowcolor{steel!10}\textsc{hsqldb}&{Latency, energy and CPU load}&15&864&{A SQL database for large volume of data}\\

\bottomrule
\end{tabular}
\centering
 \begin{tablenotes}
    \footnotesize
    \item $\lvert \mathbfcal{O} \rvert$, $\lvert \mathbfcal{C} \rvert$ and $\lvert \mathbfcal{S} \rvert$ denote objectives, \# configuration options and search space, respectively. 
    \end{tablenotes}
\end{table}

\subsubsection{Problem}

Many software systems are highly configurable and adaptable (at design time or runtime)~\cite{nair2018finding,DBLP:conf/icac/RamirezKCM09,Chen2018FEMOSAA,DBLP:conf/sigsoft/0001L21,ChenLiDOS22,DBLP:journals/csur/ChenBY18,DBLP:journals/tsc/ChenB17,DBLP:journals/computer/ChenB15,DBLP:conf/icse/Chen19b}, which raises a search problem and opportunity for one to tune their configuration options for multiple performance concerns, such as latency, throughput, and memory consumption. According to the literature, SCT has been widely studied in SBSE, e.g.,~\cite{DBLP:journals/jss/CalinescuCGKP18,DBLP:conf/mascots/JamshidiC16,nair2018finding,DBLP:conf/icac/RamirezKCM09,Chen2018FEMOSAA,DBLP:conf/sigsoft/0001L21,ChenLiDOS22,DBLP:journals/corr/abs-2112-07303,DBLP:conf/icse/LiX0WT20,DBLP:conf/kbse/LiXCT20}. As mentioned, we select 10 commonly used real-world software systems and their environments from the literature~\cite{DBLP:conf/mascots/JamshidiC16,nair2018finding}, which are specified in Table~\ref{tb:spo}. As can be seen that some software systems do not involve an intractable search space; 
however, the solution evaluations in all of them are expensive. For example, \textsc{wc-c4-3d} requires several hours to explore only a small proportion of the search space~\cite{DBLP:conf/mascots/JamshidiC16}. This renders the exhaustive or linear search unrealistic. 



\subsubsection{Representation}

The configuration (solution) of a software system in {SCT} can be represented by a vector $\mathbf{\overline{c}}=(x_1,x_2,...,x_n)$, whereby $x_n$ denotes the $n$th configuration option that can be tuned~\cite{DBLP:conf/mascots/JamshidiC16,Chen2018FEMOSAA,nair2018finding}. Since both the categorical and numeric options in configurable software can be discretized~\cite{Chen2018FEMOSAA}, each $x_n$ is associated with a pre-defined list of possible values. 
Taking \textsc{wc-c1-3d} as an example, its configuration can be represented as $\mathbf{\overline{c}}=($\texttt{max\_spout}$,$ \texttt{spliters}$,$ \texttt{counters}$)$ where \texttt{max\_spout}$=(1,2,...,1000,10000)$, \texttt{spliters}$=(1,2,...,6)$ and \texttt{counters}$=(1,2,...,18)$. A particular configuration could be $(1000,5,15)$.

\subsubsection{Objective}

As we see from Table~\ref{tb:spo}, all software systems have two or three objectives to be optimized, which can be written as. 
\begin{equation}
\label{eq:spo}
 \argmax/\argmin \, \langle f_1(\mathbf{\overline{c}}), f_2(\mathbf{\overline{c}}) \rangle \text{ or } \langle f_1(\mathbf{\overline{c}}), f_2(\mathbf{\overline{c}}), f_3(\mathbf{\overline{c}})\rangle 
\end{equation}
Albeit work exists on performance modeling for configurable and adaptable systems~\cite{DBLP:journals/tse/ChenB17,DBLP:conf/icse/ChenB14,DBLP:conf/ucc/ChenBY14}, there are no well-defined objective functions for $f_1$, $f_2$, and $f_3$ in {SCT}; 
thereby to guarantee accuracy, every evaluation needs to be done by profiling the software under a benchmark~\cite{DBLP:journals/jss/CalinescuCGKP18,DBLP:conf/mascots/JamshidiC16,nair2018finding,DBLP:conf/sigsoft/0001L21,ChenLiDOS22,DBLP:journals/corr/abs-2112-07303}. For instance, optimizing \textsc{llvm} involves configuring the software, running it to compile a standard benchmark program, and recording the results as the objective values thereafter. This is also the reason behind the expensiveness for {SCT}.

\begin{table}
\caption{System workflows for {WSC}.}
\label{tb:wsc}
\centering
\footnotesize
\begin{tabular}{clccl}\toprule

\textbf{Workflow}&$\lvert \mathbfcal{O} \rvert$&$\lvert \mathbfcal{A} \rvert$&$\lvert \mathbfcal{S} \rvert$&\textbf{Description}\\

\midrule

\textsc{5as-1}&Cycle time and cost&5&1.08$\times10^{10}$&\makecell{1 parallel and 3 sequential connectors}\\

\rowcolor{steel!10}\textsc{5as-2}&Cycle time and cost&5&1.25$\times10^{10}$&\makecell{1 parallel and 2 sequential connectors}\\

\textsc{5as-3}&Cycle time and cost&5&1.73$\times10^{10}$&\makecell{4 sequential connectors}\\

\rowcolor{steel!10}\textsc{10as-1}&Cycle time and cost&10&1.23$\times10^{20}$&\makecell{3 parallel and 4 sequential connectors}\\

\textsc{10as-2}&Cycle time and cost&10&2.45$\times10^{20}$&\makecell{2 parallel and 3 sequential connectors}\\

\rowcolor{steel!10}\textsc{10as-3}&Cycle time and cost&10&2.23$\times10^{20}$&\makecell{1 parallel and 2 sequential connectors}\\

\textsc{15as-1}&Cycle time and cost&15&2.12$\times10^{30}$&\makecell{1 parallel and 2 sequential connectors}\\

\rowcolor{steel!10}\textsc{15as-2}&Cycle time and cost&15&3.17$\times10^{30}$&\makecell{4 parallel and 6 sequential connectors}\\

\textsc{15as-3}&Cycle time and cost&15&2.60$\times10^{30}$&\makecell{6 parallel and 7 sequential connectors}\\

\rowcolor{steel!10}\textsc{50as}&Cycle time and cost&50&1.86$\times10^{202}$&\makecell{10 parallel and 29 sequential connectors}\\

\textsc{5as-3o}&Cycle time, cost and latency&5&1.73$\times10^{10}$&\makecell{4 sequential connectors}\\

\rowcolor{steel!10}\textsc{10as-3o}&Cycle time, cost and latency&10&2.23$\times10^{20}$&\makecell{1 parallel and 2 sequential connectors}\\

\textsc{15as-3o}&Cycle time, cost and latency&15&2.60$\times10^{30}$&\makecell{6 parallel and 7 sequential connectors}\\

\bottomrule
\end{tabular}
\centering
 \begin{tablenotes}
    \footnotesize
    \item $\lvert \mathbfcal{O} \rvert$, $\lvert \mathbfcal{A} \rvert$ and $\lvert \mathbfcal{S} \rvert$ denote the \# objectives, \# abstract services and search space, respectively. The objectives for all workflows are cycle time and cost. There is also a different number of abstract services in the group of each parallel connector. 
    \end{tablenotes}
\end{table}

\subsection{Web Service Composition (WSC)}

\subsubsection{Problem}


Service-oriented software system is a workflow of inter-connected abstract services (e.g., via parallel or sequential connectors), each of which can be realized by a readily concrete service. The search problem is to select a set of concrete services from an explosively large pool of candidates with different performance guarantees and costs, such that the overall performance and cost of the workflow are optimized~\cite{DBLP:journals/infsof/ChenLY19,Wada2012E,DBLP:conf/icws/KumarB0B19}. Such a nature of {WSC} is again well-fit with the purpose of SBSE, as what has been widely studied from the literature~\cite{DBLP:journals/infsof/ChenLY19,PubSub10849_Chen,Wada2012E,DBLP:conf/icpads/KumarBCLB18,DBLP:conf/icse/Kumar0BB20}. Here, we choose 13 commonly used system workflows from the existing work~\cite{DBLP:journals/infsof/ChenLY19,PubSub10849_Chen} as shown in Table~\ref{tb:wsc}, in which the performance and cost of the candidate concrete services are sampled from the real-world dataset named WS-DREAM~\cite{zheng2012investigating}. Their diverse numbers of abstract services and connectors result in different candidate concrete services, hence different scales of the search space.


\subsubsection{Representation}

In {WSC}, the composition (solution) of a system workflow is represented as $\mathbf{\overline{s}}=(s_1,s_2,...,s_n)$, whereby $s_n$ denotes the $n$th abstract service that needs to be realized by a concrete service~\cite{DBLP:journals/infsof/ChenLY19}. For each $s_n$, the service broker would discover a list of $m$ candidate concrete services, which provide the same functionality but differ in terms of performance and cost. For example, \textsc{5as-1} has five abstract services and thus its solution is represented as  $\mathbf{\overline{s}}=(s_1,s_2,...,s_5)$. Each abstract service may have a different number of candidate concrete services from which one needs to be selected; hence we have $s_1=(\mathbf{\overline{s}}_{1,1},\mathbf{\overline{s}}_{1,2},...,\mathbf{\overline{s}}_{1,109})$, $s_2=(\mathbf{\overline{s}}_{2,1},\mathbf{\overline{s}}_{2,2},...,\mathbf{\overline{s}}_{2,94})$, and so forth. Each concrete service is also represented as a vector of its performance and cost, e.g., if cycle time and cost of the composition are of concern, then we can have $\mathbf{\overline{s}}_{1,1}=(34s,\$14.33)$. In such case, a particular solution can be $((34,14.33),(13,5.42),...,(74,49.07))$.

\subsubsection{Objective}

From Table~\ref{tb:wsc} we see that all system workflows seek to optimize cycle time and cost; or cycle time, cost, and latency, whose objective functions have been well-defined in the literature~\cite{DBLP:journals/infsof/ChenLY19,PubSub10849_Chen,Wada2012E,DBLP:journals/jss/RamakrishnanK20}:
\begin{equation}
\label{eq:wsc}
\argmin \, \langle f_{time}(\mathbf{\overline{s}}), f_{cost}(\mathbf{\overline{s}})\rangle \text{ or } \langle f_{time}(\mathbf{\overline{s}}), f_{cost}(\mathbf{\overline{s}}), f_{latency}(\mathbf{\overline{s}})\rangle
\end{equation}
\begin{equation}
\label{eq:wsc1}
f_{time}(\mathbf{\overline{s}})=\mathbf{Max} \, T_{s_i}; \; f_{cost}(\mathbf{\overline{s}})=\sum^n_{i=1} C_{s_i}; \; f_{latency}(\mathbf{\overline{s}})=\sum^n_{i=1} L_{s_i}
\end{equation}
whereby $T_{s_i}$, $C_{s_i}$, and $L_{s_i}$ are the cycle time, cost, and latency of the concrete service selected for $s_i$, respectively; $n$ is the number of abstract services. In essence, the overall cycle time of a workflow represents the maximum time for which a service needs to hold each request under processing. It is, therefore, often considered as the reciprocal of throughput and hence equals to the worst cycle time achieved by an abstract service (hence indicating the bottleneck). The overall cost (latency), in contrast, is the sum of cost (average delay) on all selected concrete services for the abstract services. All of them are to be minimized.



\subsection{Next Release Planning (NRP)}

\subsubsection{Problem}


As software evolves, there is often a large number of stakeholders involved and their preferences on each requirement, together with the cost of requirement fulfillment, can differ significantly~\cite{Zhang2007The}. Here, the search problem is what requirements should be fulfilled for the next release such that some goals, e.g., importance and cost, are optimal. The NRP problem has been widely studied in SBSE~\cite{Zhang2007The,Li2014Robust,DBLP:journals/tosem/ZhangHORB18,DBLP:journals/access/GengYJZLGX18}, from which we choose 15 releases dataset that was mined from real-world software projects and their versions. As shown in Table~\ref{tb:nrp}, each software project/version involves a set of randomly sampled requirements to fulfill. 

\begin{table}
\caption{Software projects/versions for {NRP}.}
\label{tb:nrp}
\centering
\footnotesize
\begin{tabular}{clccl}\toprule

\textbf{Project/Version}&$\lvert \mathbfcal{O} \rvert$&$\lvert \mathbfcal{R} \rvert$&$\lvert \mathbfcal{S} \rvert$&\textbf{Description}\\

\midrule

\textsc{nrp-e1}&Rank score and cost&143&1.10$\times10^{43}$&Eclipse with 536 stakeholders\\

\rowcolor{steel!10}\textsc{nrp-e2}&Rank score and cost&123&1.06$\times10^{37}$&Eclipse with 491 stakeholders\\

\textsc{nrp-e3}&Rank score and cost&47&1.41$\times10^{14}$&Eclipse with 456 stakeholders\\

\rowcolor{steel!10}\textsc{nrp-e4}&Rank score and cost&139&6.97$\times10^{41}$&Eclipse with 399 stakeholders\\

\textsc{nrp-g1}&Rank score and cost&46&2.20$\times10^{12}$&Gnome with 445 stakeholders\\

\rowcolor{steel!10}\textsc{nrp-g2}&Rank score and cost&91&2.48$\times10^{27}$&Gnome with 315 stakeholders\\

\textsc{nrp-g3}&Rank score and cost&102&5.07$\times10^{30}$&Gnome with 423 stakeholders\\

\rowcolor{steel!10}\textsc{nrp-g4}&Rank score and cost&138&3.48$\times10^{41}$&Gnome with 294 stakeholders\\

\textsc{nrp-m1}&Rank score and cost&117&1.66$\times10^{35}$&Mozilla with 768 stakeholders\\

\rowcolor{steel!10}\textsc{nrp-m2}&Rank score and cost&78&3.02$\times10^{23}$&Mozilla with 617 stakeholders\\

\textsc{nrp-m3}&Rank score and cost&56&7.20$\times10^{16}$&Mozilla with 765 stakeholders\\

\rowcolor{steel!10}\textsc{nrp-m4}&Rank score and cost&140&1.39$\times10^{42}$&Mozilla with 568 stakeholders\\

\textsc{nrp-e-3o}&Rank score, cost and coverage&47&1.41$\times10^{14}$&Eclipse with 456 stakeholders\\

\rowcolor{steel!10}\textsc{nrp-g-3o}&Rank score, cost and coverage&46&2.20$\times10^{12}$&Gnome with 445 stakeholders\\

\textsc{nrp-m-3o}&Rank score, cost and coverage&56&7.20$\times10^{16}$&Mozilla with 765 stakeholders\\

\bottomrule
\end{tabular}
\centering
 \begin{tablenotes}
    \footnotesize
     \item $\lvert \mathbfcal{O} \rvert$, $\lvert \mathbfcal{R} \rvert$ and $\lvert \mathbfcal{S} \rvert$ denote the \# objectives, \# requirements and search space, respectively. The objectives for all projects/versions are importance score and cost.
    \end{tablenotes}
\end{table}

\subsubsection{Representation}

The representation of the release plan (solution) for NRP is a vector $\mathbf{\overline{r}}=(r_1,r_2,...,r_n)$ where $r_n$ is the $n$th requirement that can be selected to fulfill in the next software release~\cite{Zhang2007The,DBLP:journals/tosem/ZhangHORB18}. $\mathbf{\overline{r}}$ comes in a binary form and thereby $r_n$ can only be set as either 0 or 1, meaning that $r_n$ is not selected or selected, respectively. Considering \textsc{nrp-e3}, the release plan can be represented as $\mathbf{\overline{r}}=(r_1,r_2,...,r_{47})$ and a particular one could be $(0,1,...,1)$.

\subsubsection{Objective}

As shown in Table~\ref{tb:nrp}, we use two or three common objectives for all software projects/versions, namely penalty score, cost, and coverage, which have the objective functions as below~\cite{Zhang2007The,Li2014Robust,DBLP:journals/tosem/ZhangHORB18,DBLP:journals/access/GengYJZLGX18,DBLP:conf/ssbse/AraujoP14}: 
\begin{equation}
\label{eq:nrp}
\argmin \, \langle f_{score}(\mathbf{\overline{r}}), f_{cost}(\mathbf{\overline{r}})\rangle \text{ or } \langle f_{score}(\mathbf{\overline{r}}), f_{cost}(\mathbf{\overline{r}}), f_{coverage}(\mathbf{\overline{r}})\rangle
\end{equation}
\begin{equation}
\label{eq:nrp1}
f_{score}(\mathbf{\overline{r}})=({{\sum^n_{i=1}\sum^m_{j=1} r_i \times I_{j}}})^{-1}; \; f_{cost}(\mathbf{\overline{r}})=\sum^n_{i=1} r_i \times C_{i}; \; f_{coverage}(\mathbf{\overline{r}})=\sigma(r_i,R_{i})
\end{equation}
where there are $n$ requirements and $m$ stakeholders. $I_{j}$, $C_{i}$, and $R_{i}$ are respectively the satisfaction level of the $i$th requirement from the $j$th stakeholder, the related cost for fulfilling the $i$th requirement $r_i$, and the ratio of fulfilled requirement for $r_i$. $\sigma(\cdot)$ denotes the standard deviation across all stakeholders. As mentioned, the variable $r_i$ can be either 0 or 1 only. All objectives are to be minimized.

\section{Empirical Study Design}
\label{sec:empirical}


We empirically investigate Pareto and weighted search on all the SBSE problems and their systems/projects from Section~\ref{sec:cases}. In particular, each case of the SBSE problems is repeated 100 runs. To ensure realism, we use a cluster of machines with Intel i7 2.8GHz CPU and 8GB RAM. All experiment code was implemented in Java based on jMetal~\cite{DBLP:journals/aes/DurilloN11} and Opt4J~\cite{DBLP:conf/gecco/LukasiewyczGRT11}. In what follows, we will discuss the settings in greater detail.



\subsection{Optimizers}
\label{sec:alg}

Although the existing belief is to use weighted search when the weights can be explicitly given, the underlying optimizer can vary. Indeed, there is a vast set of optimizers being used for weighted search in the SBSE problems, as summarized by several surveys~\cite{Harman2012,Sayyad2013b,DBLP:journals/infsof/ColanziAVFG20,DBLP:journals/corr/abs-2001-08236}. To mitigate the threat to construct validity in our empirical study, we investigate four optimizers for weighted search:

\begin{itemize}

\item Random Search (RS);
\item Hill Climbing with restart (HC);
\item Simulated Annealing (SA)~\cite{DBLP:journals/pami/GranvilleKR94}; 
\item Single-Objective Genetic Algorithm (SOGA)~\cite{goldberg2006genetic}. 

\end{itemize}

At this point, it is natural to ask why we chose those optimizers for weighted search. Indeed, since we are challenging the \textit{``weighted search first''} belief under clear preferences, it is desired to examine all possible optimizers that have ever been used for weighted search (or single-objective search) in SBSE. However, it is fundamentally unrealistic to do so given the resource constraint~\cite{Harman2012}. Instead, we seek to examine the most widely used ones in SBSE. 


\begin{table}
\caption{Commonality of optimizers used from~\cite{DBLP:journals/infsof/ColanziAVFG20}.}
\label{tb:alg-count}
\centering
\footnotesize
\begin{tabular}{lll}\toprule

\textbf{Acronym}&\textbf{Optimizer}&\textbf{Paper Count}\\

\midrule

GA&(Single-Objective) Genetic Algorithm&26\\
\rowcolor{steel!10}NSGA-II&Non-dominated Sorting Genetic Algorithm-II&15\\
SA&Simulated Annealing&10\\
\rowcolor{steel!10}HC&Hill Climbing&7\\
GP&Genetic Programming&5\\
\rowcolor{steel!10}MOSA&Many-Objective Sorting Algorithm&4\\
ACO&Ant Colony Optimization&4\\
\rowcolor{steel!10}CP&Constraint Programming&2\\
IGA&Interactive Genetic Algorithm&2\\
\rowcolor{steel!10}LIPBS&Linearly Independent Path based Search&2\\
MIO&Many Independent Objective algorithm&2\\
\rowcolor{steel!10}SPEA2&Strength Pareto Evolutionary Algorithm&2\\

\bottomrule
\end{tabular}

\end{table}

To understand what are the most prevalent optimizers in SBSE, we refer to well-known SBSE surveys by \citeauthor{Harman2012}~\cite{Harman2012} and \citeauthor{DBLP:journals/infsof/ColanziAVFG20}~\cite{DBLP:journals/infsof/ColanziAVFG20}. As shown in Table~\ref{tb:alg-count}, \citeauthor{DBLP:journals/infsof/ColanziAVFG20} summarized the optimizers used in the papers published at SSBSE over the past ten years, and concluded that SOGA is the most widely used one while SA and HC are ranked as the 3rd and 4th most popular ones, respectively. Likewise,  \citeauthor{Harman2012} has also confirmed that \textit{``the most widely used are local search, Simulated Annealing (SA), Genetic Algorithms
(GAs), Genetic Programming (GP), and Hill Climbing (HC)''}. In particular, they showed that SOGA, HC, and SA are significantly more promising than the other optimizers in SBSE. The commonality of those optimizers has also been confirmed by studies of the three SBSE problems we examine~\cite{Harman2012,DBLP:journals/tsc/JatothGB17,DBLP:conf/mascots/JamshidiC16}. Note that we ruled out the basic local search as HC and SA are both parts of it; 
we do not consider GP 
since it is designed to search for an optimal program rather than a solution vector that minimizes/maximizes objectives/criteria. MOSA and ACO are also ruled out as the former aims for the case with more than three objectives while the latter works better on path-finding problems. The remaining optimizers are much rarely used minorities. Although not as part of the above, we additionally take RS into account, as recommended by \citet{DBLP:conf/icse/ArcuriB11}, it should serve as a baseline for any SBSE problem.

For Pareto search, 
we use two optimizers:

\begin{itemize}
    \item Non-dominated Sorting Genetic Algorithm-II (NSGA-II)~\cite{Deb2002}
    \item Multi-objective Evolutionary Algorithm Based on Decomposition (MOEA/D)~\cite{DBLP:journals/tec/ZhangL07}
\end{itemize}

NSGA-II is chosen because of its predominant appearance in SBSE. As shown in Table~\ref{tb:alg-count}, \citeauthor{DBLP:journals/infsof/ColanziAVFG20}~\cite{DBLP:journals/infsof/ColanziAVFG20} rank NSGA-II as the second most popular optimizers in ten year's SSBSE papers (among those for weighted/single-objective search). Similarly, \citeauthor{Sayyad2013b}~\cite{Sayyad2013b} confirm that NSGA-II has been used by 53\% of the papers studied --- over $4\times$ more frequent than the 2nd most popular one. The same trend has also been observed for the three SBSE problems studied in this work, as discussed in their corresponding reviews~\cite{DBLP:journals/tosem/ZhangHORB18,DBLP:journals/corr/abs-2001-08236,DBLP:journals/tsc/JatothGB17}. In contrast, despite rarely being used for SBSE, we examine MOEA/D (with a weighted sum scalar function and its dynamic bounds for normalization) because it possesses many similarities with the weighted search, as it seeks to approximate the Pareto front via internal weight vectors. 
Indeed, it could be fruitful if more advanced ones for Pareto search are examined. 
However, 
NSGA-II and MOEA/D, despite being developed for quite a while,
have still shown their competitiveness on a lot of instances recently~\cite{Li2021}.
Moreover, 
if our hypothesis can be confirmed under a very basic optimizer for Pareto search, 
then examining more advanced ones would not change our conclusion.

Indeed, the optimizers for weighted search and those for Pareto search may have similar or rather different designs. Yet, a critical aspect that distinguishes between them is related to how the solutions are preserved into the next iterations. To give a concrete example of comparison, Algorithm~\ref{alg:nsgaii-sga} compares the key steps of NSGA-II and SOGA. As can be seen, 
although the optimizers can be set to use the identical mating selector, crossover, and mutation operators, there is a key difference in the criterion used in the selection processes (i.e., mating selection and surviving selection), 
in which the NSGA-II applies non-dominated sorting and crowding distance while SOGA sorts the solutions based on their weighted aggregation\footnote{We use the weighted sum in this work due to its prevalence~\cite{Harman2012,10.1145/3392031,DBLP:conf/ssbse/BowersFC18}.}. 

\begin{algorithm}[t]
    \DontPrintSemicolon
    \footnotesize
    \caption{Unified code for NSGA-II and SOGA.}
    \label{alg:nsgaii-sga}
    \KwIn{Search space $\mathcal{M}$; objective functions $\mathbf{\overline{F}}$; weight vector $\mathbf{\overline{w}}$}
    \KwOut{Solution set $\mathcal{S'}$ or the best solution $\mathcal{S}$}
    Randomly initialize a population of $n$ solutions $\mathcal{P}$\\
    \While{The search budget is not exhausted}
    {  
      $\mathcal{P'}=\emptyset$\\
      \While{$\mathcal{P'}<n$}
      { 
        \lIf{NSGA-II}{
        $\{s_x,s_y\}\leftarrow$\textsc{mating($\mathcal{P}$)}
        } \lElse {
        $\{s_x,s_y\}\leftarrow$\textsc{mating($\mathbf{\overline{w}},\mathcal{P}$)}
        }
        $\{o_x,o_y\}\leftarrow$\textsc{doCrossoverAndMutation($\mathcal{M}, s_x,s_y$)}\\
        \textsc{evaluate($o_x, o_y, \mathbf{\overline{F}}$)}\\
        $\mathcal{P'}\leftarrow\mathcal{P'}\bigcup\{o_x,o_y\}$\\
      }
      \lIf{NSGA-II}{
        $\mathcal{U}\leftarrow$\textsc{nondominatedSort($\mathcal{P}\bigcup\mathcal{P'}$)}
        } \lElse {
        $\mathcal{U}\leftarrow$\textsc{weightedAggregationSort($\mathbf{\overline{w}},\mathcal{P}\bigcup\mathcal{P'}$)}
      }
      $\mathcal{P}\leftarrow$top $n$ solutions from $\mathcal{U}$\\  
    }
       \lIf{NSGA-II}{
        \Return $\mathcal{S'}\leftarrow$\textsc{nondominatedSolutions($\mathcal{P}$)}
        } \lElse {
        \Return $\mathcal{S}\leftarrow$\textsc{bestWeightedSolution($\mathbf{\overline{w}},\mathcal{P}$)}
      }
\end{algorithm}

\subsection{Settings}
\label{sec:settings}

\subsubsection{Components}

We define the neighborhood radius in HC and SA as the solutions that differ on exactly one decision variable (e.g., a configuration option in SCT or a requirement in NRP). Such a definition of the neighborhood has been recommended by Harman~\cite{DBLP:conf/icse/Harman07}, who states that for most SBSE problems, the neighbour in HC and SA is often a ``small mutation away''. Indeed, this has been widely applied in the three SBSE problems studied with promising results, e.g.,~\cite{DBLP:conf/cloudcom/WangLT12,DBLP:conf/sigecom/MenasceBD01} for SCT;~\cite{DBLP:conf/icws/KleinIH11} for WSC;~\cite{DBLP:conf/eurocon/MausaGBP13,DBLP:conf/icsm/BakerHSS06} for NRP. As for SOGA, NSGA-II, and MOEA/D, the most common binary tournament is used for mating selection on all the SBSE problems~\cite{Sayyad2013On}. For SCT and {WSC}, we apply the boundary mutation and uniformed crossover in all systems, as used in prior work~\cite{Chen2018FEMOSAA,DBLP:journals/infsof/ChenLY19}. {NRP} differently uses bitflip mutation and one-point crossover, which are recommended in the literature~\cite{DBLP:journals/tosem/ZhangHORB18}.

\subsubsection{Parameters and Search Budget}
\label{sec:budget-setting}

Under all SBSE problems and their systems/projects, we use the same parameter values (e.g., population size, mutation rate, and crossover rate) for both SOGA and NSGA-II, as shown in Table~\ref{tb:settings}. These setting are identical to what have been commonly used from the literature~\cite{Chen2018FEMOSAA,DBLP:journals/infsof/ChenLY19,DBLP:journals/tosem/ZhangHORB18}. This fits our purpose well as we intend to compare Pareto and weighted search under the most common practices.

Ideally, we would like the comparisons to be conducted on the true convergence, i.e., the best-weighted result that can be possibly achieved given an unlimited search budget. This is nevertheless not practical. Therefore, comparing them under a fixed identical search budget is more plausible. However, to avoid the outcomes of premature convergence (which is always possible) from dominating the comparisons, in this work we at least seek to allow all optimizers to reach a reasonable convergence: a degree of convergence where the best-found solution (albeit still possible to be a local optimum) 
does not change for some number of iterations under a search budget. In what follows, we set two metrics to represent an identical search budget under each of which Pareto and weighted search can be compared with reasonable convergence. 

\begin{table}
\caption{Parameter settings and budget.}
\label{tb:settings}
\setlength{\tabcolsep}{1.8mm}
\footnotesize
\centering
\begin{tabular}{ccccccc}\toprule

\textbf{System/Project}&$\lvert \mathbfcal{M} \rvert$&$\lvert \mathbfcal{C} \rvert$&$\lvert \mathbfcal{P} \rvert$&$\lvert \mathbfcal{G} \rvert$&$\lvert \mathbfcal{E} \rvert$&$\lvert \mathbfcal{T} \rvert$\\
\midrule

\textsc{wc-c1-3d}&0.1&0.9&20&30&600&49.31mins\\
\rowcolor{steel!10}\textsc{wc-c3-3d}&0.1&0.9&20&30&600&38.17mins\\
\textsc{wc-c5-5d}&0.1&0.9&20&30&600&164mins\\
\rowcolor{steel!10}{\textsc{llvm}}&0.1&0.9&20&25&500&26.31mins\\
\textsc{wc-c4-3d}&0.1&0.9&20&15&300&170mins\\
\rowcolor{steel!10}{\textsc{rs-c3-6d}}&0.1&0.9&50&30&1.5$\times 10^3$&60.54mins\\
{\textsc{wc-c1-6d}}&0.1&0.9&50&30&1.5$\times 10^3$&102.98mins\\
\rowcolor{steel!10}\textsc{trimesh}&0.1&0.9&100&100&$10^4$&84.35mins\\
{\textsc{vp8}}&0.1&0.9&30&20&600&233.33mins\\
\rowcolor{steel!10}\textsc{hsqldb}&0.1&0.9&50&30&1.5$\times 10^3$&81.25mins\\
\hline

\textsc{5as-1}&0.1&0.9&100&300&3.0$\times 10^4$&2.051s\\
\rowcolor{steel!10}\textsc{5as-2}&0.1&0.9&100&300&3.0$\times 10^4$&5.563s\\
\textsc{5as-3}&0.1&0.9&100&300&3.0$\times 10^4$&3.59s\\
\rowcolor{steel!10}\textsc{10as-1}&0.1&0.9&100&300&3.0$\times 10^4$&4.943s\\
\textsc{10as-2}&0.1&0.9&100&300&3.0$\times 10^4$&3.936s\\
\rowcolor{steel!10}\textsc{10as-3}&0.1&0.9&100&300&3.0$\times 10^4$&4.61s\\
\textsc{15as-1}&0.1&0.9&100&300&3.0$\times 10^4$&4.027s\\
\rowcolor{steel!10}\textsc{15as-2}&0.1&0.9&100&300&3.0$\times 10^4$&4.986s\\
\textsc{15as-3}&0.1&0.9&100&300&3.0$\times 10^4$&4.079s\\
\rowcolor{steel!10}\textsc{50as}&0.02&0.8&100&500&5.0$\times 10^4$&45.802s\\
\textsc{5as-3o}&0.1&0.9&100&300&3.0$\times 10^4$&4.731s\\
\rowcolor{steel!10}\textsc{10as-3o}&0.1&0.9&100&300&3.0$\times 10^4$&4.801s\\
\textsc{15as-3o}&0.1&0.9&100&300&3.0$\times 10^4$&4.89s\\

\hline

\textsc{nrp-e1}&0.01&0.8&100&200&2.0$\times 10^4$&0.624s\\
\rowcolor{steel!10}\textsc{nrp-e2}&0.01&0.8&100&200&2.0$\times 10^4$&1.66s\\
\textsc{nrp-e3}&0.01&0.8&100&200&2.0$\times 10^4$&2.306s\\
\rowcolor{steel!10}\textsc{nrp-e4}&0.01&0.8&100&200&2.0$\times 10^4$&2.943s\\
\textsc{nrp-g1}&0.01&0.8&100&200&2.0$\times 10^4$&1.695s\\
\rowcolor{steel!10}\textsc{nrp-g2}&0.01&0.8&100&200&2.0$\times 10^4$&6.82s\\
\textsc{nrp-g3}&0.01&0.8&100&200&2.0$\times 10^4$&2.061s\\
\rowcolor{steel!10}\textsc{nrp-g4}&0.01&0.8&100&200&2.0$\times 10^4$&2.256s\\
\textsc{nrp-m1}&0.01&0.8&100&200&2.0$\times 10^4$&1.868s\\
\rowcolor{steel!10}\textsc{nrp-m2}&0.01&0.8&100&200&2.0$\times 10^4$&4.711s\\
\textsc{nrp-m3}&0.01&0.8&100&200&2.0$\times 10^4$&2.762s\\
\rowcolor{steel!10}\textsc{nrp-m4}&0.01&0.8&100&200&2.0$\times 10^4$&3.175s\\
\textsc{nrp-e-3o}&0.01&0.8&100&200&2.0$\times 10^4$&2.568s\\
\rowcolor{steel!10}\textsc{nrp-g-3o}&0.01&0.8&100&200&2.0$\times 10^4$&2.254s\\
\textsc{nrp-m-3o}&0.01&0.8&100&200&2.0$\times 10^4$&2.578s\\

\bottomrule
\end{tabular}
\centering
 \begin{tablenotes}
    \footnotesize
    \item $\lvert \mathbfcal{M} \rvert$, $\lvert \mathbfcal{C} \rvert$, $\lvert \mathbfcal{P} \rvert$, $\lvert \mathbfcal{G} \rvert$, $\lvert \mathbfcal{E} \rvert$, $\lvert \mathbfcal{T} \rvert$ denote the mutation rate, crossover rate, population size, \# generations, \# evaluations budget, and time budget, respectively.
    \end{tablenotes}
\end{table}

\begin{itemize}
    \item[---] \textbf{Evaluation budget:} In SBSE, using an identical number of evaluations in the comparisons is a common practice~\cite{Praditwong2011Software,DBLP:journals/tec/WangAYL18}. 
   To find such a fixed number of evaluations for each system/project, we firstly conduct preliminary runs on all optimizers and use the smallest evaluation number that satisfies all the following criteria:
    \begin{itemize}
    \item[\footnotesize$\bullet$] To ensure reasonable convergence under RS, HC, SA, and SOGA, the solution (or population) should converge to one solution point with no improvement in the last 5\% of the evaluations for at least 90\% of the repeated runs.
    \item[\footnotesize$\bullet$] To achieve reasonable convergence while respecting the diversity in NSGA-II, the population's solutions should change by less than 5\% in the last 5\% of the evaluations for at least 90\% of the repeated runs. A similar setting has been used in SBSE~\cite{DBLP:journals/ase/GerasimouCT18}. 
    \item[\footnotesize$\bullet$] Each run can be completed within three hours. This is to ensure a reasonable effort and resources required for concluding the empirical study.
    \end{itemize}
    
The results are shown in Table~\ref{tb:settings} (column $\lvert \mathbfcal{E} \rvert$). 


    \item[---] \textbf{Time budget:} Since the clock time is also an important factor for the practical scenarios of SBSE and an identical number of evaluations may not imply the same time consumption, in this work, we additionally compare Pareto and weighted search under an identical time budget. In particular, for each system/project, we record the longest time, $t_{max}$, taken by Pareto or weighted search (all optimizers) to complete one run using the fixed number of evaluations in Table~\ref{tb:settings}. We then allow whichever optimizer that uses less time, if any, to run up to $t_{max}$. In this way, we give the ones that are originally less time-consuming a fair chance to improve (e.g., escape from premature convergence and local optima). 
    
    As can be seen from Table~\ref{tb:settings}  (column $\lvert \mathbfcal{T} \rvert$), the time required to reach reasonable convergence may differ radically across the SBSE problems, due primarily to the time required to evaluate a solution --- it could be up to hours-long for expensive problems like SCT, but can be as low as a few seconds (or less) for others such as WSC and NRP.
    
    
\end{itemize}

\subsubsection{Possible Weight Vectors}


To avoid bias to a particular assumption of stakeholders' preferences (weights), we compare Pareto and weighted search under 9 uniformly distributed sets of weights, $(0.1,0.9), (0.2,0.8),..., (0.9,0.1)$, for the two objective case. Such a setting covers a wide spectrum of the weights in SBSE, as what has been used from the literature~\cite{Zhang2007The}, while being realistic enough for us to complete the empirical study. For SBSE problems with three objectives, we use three edge weight vector, $(0.1,0.1,0.8)$, $(0.1,0.8,0.1)$, and $(0.8,0.1,0.1)$, together with a middle one, i.e., $(0.33,0.33,0.33)$, as the representatives.

\subsubsection{Normalization Methods}

It is not uncommon to have conflicting objectives that are of radically different scales in SBSE (e.g., the latency and throughput in {SCT}); normalization is, therefore, crucial for weighted search to make different objectives commensurable. In this work, we consider four normalization methods that are widely used in SBSE for weighted search:

\begin{itemize}

\item \texttt{Dynamic:} In this method, the SBSE objectives are normalized by using their upper and lower bounds: ${v-v_{min}}\over{v_{max}-v_{min}}$, whereby $v$ is the raw objective value; $v_{max}$ and $v_{min}$ are the upper and lower bounds for that objective, respectively. However, since one may not normally know $v_{max}$ and $v_{min}$ from the beginning, 
it adopts a dynamic method wherein the objective values are normalized using the maximal and minimal values found so far as the search proceeds. This is a common method to normalize objectives when weighted search is used~\cite{ding2006discussions,nan2019study,Harman2012,10.1145/3392031,DBLP:conf/ssbse/BowersFC18,DBLP:conf/sigsoft/ShahbazianKBM20}. 

\item \texttt{Fixed:} This method is similar to \texttt{Dynamic}, but differs in the sense that the bounds are known a priori. Thus, the weights are static and no dynamic updates occur during the search. Clearly, this is only applicable to certain SBSE scenarios, as in existing work~\cite{DBLP:conf/icac/RamirezKCM09,DBLP:conf/icse/ChengGS06}. When the bounds of a SBSE problem is not known naturally, in this work, we use the bounds obtained via a preliminary single-objective search that explores the extreme values of each objective.

\item \texttt{Ratio:} Here, the objectives are converted by using ${v} \over {v+1}$, whereby $v$ is the raw objective value. This method has been used in~\cite{DBLP:journals/tec/WangAYL18,DBLP:conf/gecco/PradhanWAY16} .

\item \texttt{None:} This is a baseline method that no normalization is applied at all, 
despite the fact that objectives may be of completely different scales in SBSE problems.

\end{itemize}


\subsection{Analysis and Comparison}
\label{sec:comp}

\subsubsection{Metric of Solution Quality}

Since there are clear preferences (weights) between the objectives, 
we know which solution the stakeholders favor the most, i.e., by Eq.~(\ref{Eq:SOP}). 
To compare the final quality of both strategies, we directly compare the weighted score, i.e., the scalar value produced by the weighted aggregation function from Eq.~(\ref{Eq:SOP}). For those that maintain a population (i.e., SOGA and NSGA-II), we use the best scalar value obtained by their population of solutions over the weight vector. 
To ensure an accurate comparison,
we use the range of the estimated Pareto front\footnote{The estimated Pareto front refers to the non-dominated solutions of all the solutions generated by all optimizers over all the runs (and weight vectors).} 
as the posterior bounds in the normalization~\cite{li2019quality}. Since we convert all objectives to be minimized, the smaller the weighted score, the better.




\subsubsection{Statistical Validation}

We test the statistical significance and effect size of the comparisons using:

\textbf{Wilcoxon rank-sum test~\cite{Wilcoxon1945IndividualCB} (U-test):} This was chosen because of its statistical power on pairwise comparisons~\cite{Wilcoxon1945IndividualCB}, which fits precisely our needs. It is also a non-parametric and non-paired test that makes little assumption about the underlying distribution of the data and has been recommended in SBSE~\cite{DBLP:journals/infsof/KampenesDHS07,DBLP:conf/icse/ArcuriB11}. In this work, we follow the common significance level as $\alpha=0.05$.
    
    
\textbf{$\mathbf{\hat{A}_{12}}$ effect size~\cite{Vargha2000ACA}:} We measure the pairwise effect size to evaluate the probability that one is better than the other. According to Vargha and Delaney~\cite{Vargha2000ACA}, when comparing Pareto and weighted search in our experiments, $\hat{A}_{12}=0.5$ means they are equivalent. $\hat{A}_{12}>0.5$ and $A_{12}<0.5$ denote that Pareto search and weighted search is better for more than 50\% of the runs, respectively; they have also suggested that $\hat{A}_{12} \geq 0.56$ or $\hat{A}_{12} \leq 0.44$ are considered as non-trivial effect sizes. In particular, $0.56 \leq \hat{A}_{12} < 0.64$ (or 0.36 $ < \hat{A}_{12} \leq 0.44$), $0.64 \leq \hat{A}_{12} < 0.72$ (or 0.28 $< \hat{A}_{12} \leq 0.36$), and $\hat{A}_{12} \geq 0.72$ (or $\hat{A}_{12} \leq 0.28$) indicate small, medium, and large effect, respectively.
    
    
\textbf{Scott-Knott test~\cite{MittasA13}:} Wilcoxon rank-sum test and $\hat{A}_{12}$ only work for pairwise comparison. Therefore, when comparing multiple subjects (e.g., the four normalization methods for \textbf{RQ1} and selecting the best optimizer/normalization pair of weighted search for \textbf{RQ2}-\textbf{RQ4}), we apply the Scott-Knott test --- a recursive clustering based on pairwise comparisons --- to rank their weighted score over 100 runs, as recommended by \citeauthor{MittasA13}~\cite{MittasA13}. In a nutshell, Scott-Knott sorts the list of treatments (the optimizers and/or normalization method) by their median weighted scores. Next, it splits the list into two sub-lists with the largest expected difference~\cite{xia2018hyperparameter}. For example, suppose that we compare $A$, $B$ and $C$, a possible split could be: $\{A, B\}$ and $\{C\}$, with the rank of 1 and 2, respectively. This means that, in the statistical sense, $A$ and $B$ perform similarly, but they are significantly better than $C$. Formally, Scott-Knott test aims to find the best split by maximizing the difference $\Delta$ in the expected mean before and after each split:
\begin{equation}
    \Delta = \frac{|l_1|}{|l|}(\overline{l_1} - \overline{l})^2 + \frac{|l_2|}{|l|}(\overline{l_2} - \overline{l})^2
\end{equation}
whereby $|l_1|$ and $|l_2|$ are the sizes of two sub-lists ($l_1$ and $l_2$) from list $l$ with a size $|l|$. $\overline{l_1}$, $\overline{l_2}$, and $\overline{l}$ denote their mean weighted score.

During the splitting, we apply a statistical hypothesis test $H$ to check if $l_1$ and $l_2$ are significantly different. This is done by using bootstrapping and $\hat{A}_{12}$~\cite{Vargha2000ACA} (a non-parametric effect size metric). If that is the case, Scott-Knott recurses on the splits. In other words, we divide the classifiers into different sub-lists if both bootstrap sampling and effect size test suggests that a split is statistically significant (with a confidence level of 99\%) and not a small effect $\hat{A}_{12} \geq 0.56$. The sub-lists are then ranked based on their mean weighted score.

In contrast to other non-parametric statistical tests that require correction on multiple comparisons (e.g., Kruskal-Wallis test), Scott-Knott test offers the following advantages:

 \begin{itemize}
     \item It does not require posterior correction, as the comparisons are essential conducted in a pairwise manner.
    \item It does not only show whether some treatments are statistically different or not, but also indicates which one is better than another, i.e., by means of ranking.
 \end{itemize}


\section{Results}
\label{sec:results}

In this section, we present and discuss the results from our empirical study with the aim to address the questions posed in Section~\ref{sec:introduction}. The complete data of all cases can be found in our supplementary file: \textcolor{blue}{\url{https://github.com/ideas-labo/pareto-vs-weight-for-sbse/blob/main/supplementary.pdf}}.


\begin{figure}[t!]
  \centering
  \begin{tikzpicture}
\begin{axis}[
ymajorgrids=true,
height=5cm, width=12cm,
grid style = {dashed},
legend pos=outer north east,
ymin=0,  ymax=3,
enlarge x limits=0.3,
ybar=0pt,
bar width=21pt,
xtick={1,2,3,4},
xticklabels={\textsc{Dynamic},\textsc{Fixed},\textsc{Ratio},\textsc{None}},
ylabel=Average Scott-Knott rank,
xticklabel style={rotate=0},
legend columns=1,
legend cell align={left},
nodes near coords,
  legend style={nodes={scale=1, transform shape},draw=none,fill=none,at={(0.75,0.98)}},
]


\addplot[fill=blue!55,draw=black]
coordinates { 
(1,1.62407407407)
(2,1.47361111111)
(3,2.23564814815)
(4,1.65277777778)
};

\addplot[fill=yellow!55,draw=black]
coordinates { 
(1,1.59375)
(2,1.57421875)
(3,1.93359375)
(4,1.7109375)
};

\addlegendentry{2 Objectives}
\addlegendentry{3 Objectives}

\end{axis}
\end{tikzpicture}
  \caption{The average Scott-Knott ranks over 2160 cases with 2 objectives (30 systems/projects $\times$ 4 optimizers $\times$ 9 weight vectors $\times$ 2 types of budget) and 256 cases with 3 objectives (8 systems/projects $\times$ 4 optimizers $\times$ 4 weight vectors $\times$ 2 types of budget).}
  \label{fig:rq1-summary}
\end{figure}
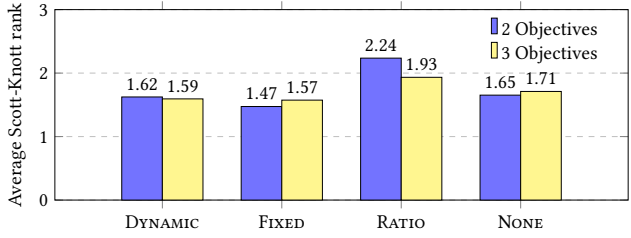

\subsection{RQ1: Normalization Methods for Weighted Search}
\label{sec:rq1}

\subsubsection{Method}

To study \textbf{RQ1}, we compare the four normalization methods under each of the optimizers (i.e., RS, HC, SA, and SOGA) for weighted search. We do that based on all 38 systems/projects of the SBSE problems, weight vectors, and types of search budget. In each case, we use Scott-Knott test to rank the normalization methods and the weighted scores are also reported.

\begin{table*}[t!]
\caption{Scott-Knott ranks and the weighted scores for the normalization methods on two objectives SBSE problems. A cell summarizes 72 cases (4 algorithms $\times$ 9 weight vectors $\times$ 2 types of budget) on 100 runs each. $\mathbfcal{R}$ denotes the total rank over all cases, hence a 72 means that a method has been ranked the first under each case. \quartexp{0}{20}{10}{20} denotes the 25th, 50th, and 75th percentile of the weighted score. \bquartexp{0}{20}{10}{20} denotes the best median among others. The best rank for each system has been highlighted.}
    \label{tb:rq1}
    \footnotesize
    \setlength{\tabcolsep}{1mm}
  \begin{center}

    \begin{tabular}{cc@{}cc@{}cc@{}cc@{}cc}
        \begin{tabular}{llcc}
            \cellcolor[gray]{1}\textbf{} & &\cellcolor[gray]{1}$\mathbfcal{R}$  & \cellcolor[gray]{1}\textbf{Quantiles}  \\
            \hline

\texttt{Dynamic}&&114&  \bquart{0.0}{0.6540482810291642}{0.29964170256126055}{100}\\
\texttt{Fixed} &&110&  \bquart{0.0}{0.6043336979011558}{0.29964170256126055}{100}\\ 
\texttt{Ratio}&&114&  \quart{0.06259602338819183}{99.9374039766118}{0.4091847434905963}{100}\\ 
\texttt{None}&&\cellcolor{steel!10}\textbf{104}&  \bquart{0.0}{0.499402837602101}{0.29964170256126055}{100}\\  &&&  \squart{1}{1}{1}{100}\\ 
        \end{tabular} & 
        &
       \begin{tabular}{lcccl}
                     \cellcolor[gray]{1}\textbf{}  &\cellcolor[gray]{1}$\mathbfcal{R}$  & \cellcolor[gray]{1}\textbf{Quantiles}  \\
            \hline
&112&  \bquart{0.0}{0.04345578237268617}{0.02075206599819772}{100}\\ 
&114&  \bquart{0.0}{0.04407477822848394}{0.02090412518480909}{100}\\ 
&126&  \quart{0.007635337886804661}{99.99236466211318}{50.08797482938875}{100}\\ 
&\cellcolor{steel!10}\textbf{98}&   \bquart{0.0}{0.04068210133507966}{0.01976060828692093}{100}\\  &&  \squart{1}{1}{1}{100}\\ 
        \end{tabular} &
        &
       \begin{tabular}{lcccl}
                    \cellcolor[gray]{1}\textbf{}  &\cellcolor[gray]{1}$\mathbfcal{R}$  & \cellcolor[gray]{1}\textbf{Quantiles}  \\
            \hline
&\cellcolor{steel!10}\textbf{72}&  \quart{0.0}{11.211530612152494}{5.2456423624724415}{100}\\ 
&76&  \bquart{0.0}{11.299801577064816}{5.031443005287632}{100}\\ 
&148& \quart{35.713035135724894}{64.2869648642751}{67.87488699400126}{100}\\ 
&74&   \quart{0.0}{11.211530612152494}{5.2456423624724415}{100}\\  &&  \squart{1}{1}{1}{100}\\ 

        \end{tabular} &
        &
       \begin{tabular}{lcccl}
                    \cellcolor[gray]{1}\textbf{}  &\cellcolor[gray]{1}$\mathbfcal{R}$  & \cellcolor[gray]{1}\textbf{Quantiles}  \\
            \hline
&\cellcolor{steel!10}\textbf{72}&  \bquart{100}{0.0}{100}{100}\\ 
&\cellcolor{steel!10}\textbf{72}&  \bquart{100}{0.0}{100}{100}\\ 
&\cellcolor{steel!10}\textbf{72}&  \bquart{100}{0.0}{100}{100}\\ 
&\cellcolor{steel!10}\textbf{72}&  \bquart{100}{0.0}{100}{100}\\ &&  \squart{1}{1}{1}{100}\\ 
        \end{tabular}&
        &
       \begin{tabular}{lcccl}
                    \cellcolor[gray]{1}\textbf{}  &\cellcolor[gray]{1}$\mathbfcal{R}$  & \cellcolor[gray]{1}\textbf{Quantiles}  \\
            \hline
&\cellcolor{steel!10}\textbf{106}&  \bquart{0.0}{98.11019514718409}{43.49222322841968}{100}\\ 
&\cellcolor{steel!10}\textbf{106}&   \quart{0.0}{98.11019514718409}{44.35700509505781}{100}\\ 
&116&  \quart{0.0}{100.0}{49.899196665199106}{100}\\ 
&118&  \quart{0.0}{98.11019514718409}{48.27157336696332}{100}\\  &&  \squart{1}{1}{1}{100}\\ 
        \end{tabular}  \\
               (a). SCT: \textsc{wc-c1-3d}&& (b). SCT: \textsc{wc-c3-3d} && (c). SCT: \textsc{wc-c4-3d}&& (d). SCT: \textsc{wc-c5-5d}&& (e). SCT: \textsc{rs-cd-6d}

\\
        \\
      
\begin{tabular}{llcc}
             \cellcolor[gray]{1}\textbf{} & &\cellcolor[gray]{1}$\mathbfcal{R}$  & \cellcolor[gray]{1}\textbf{Quantiles}  \\
            \hline

\texttt{Dynamic}&&118&  \quart{16.28400535254356}{83.71599464745644}{61.45000816020318}{100}\\ 
\texttt{Fixed}&&120&  \quart{16.28400535254356}{80.57391019988849}{61.45000816020318}{100}\\ 
\texttt{Ratio}&&\cellcolor{steel!10}\textbf{98}&  \bquart{0.0}{64.62835147247665}{32.39878739252126}{100}\\ 
\texttt{None}&&110&  \quart{0.16922331256585627}{85.92064366929748}{48.51356943249895}{100}\\  &&&  \squart{1}{1}{1}{100}\\ 
        \end{tabular} & 
        &
       \begin{tabular}{lcccl}
                    \cellcolor[gray]{1}\textbf{}  &\cellcolor[gray]{1}$\mathbfcal{R}$  & \cellcolor[gray]{1}\textbf{Quantiles}  \\
            \hline
&106&  \quart{2.030050988430217}{3.646773838361586}{2.5839784527298413}{100}\\ 
&\cellcolor{steel!10}\textbf{78}&  \bquart{0.0}{2.9702031067369195}{2.4061118357528932}{100}\\ 
&172&  \quart{24.93317297105011}{75.0668270289499}{59.85194721596397}{100}\\
&110&  \quart{2.030050988430217}{5.065980045059535}{2.737451933664233}{100}\\ &&  \squart{1}{1}{1}{100}\\ 
        \end{tabular} &
        &
       \begin{tabular}{lcccl}
                    \cellcolor[gray]{1}\textbf{}  &\cellcolor[gray]{1}$\mathbfcal{R}$  & \cellcolor[gray]{1}\textbf{Quantiles}  \\
            \hline
&\cellcolor{steel!10}\textbf{94}&  \bquart{0.1332967786246219}{66.28845238013818}{33.24376371839139}{100}\\ 
&112&   \quart{14.671142025788317}{85.32885797421169}{53.996595847216376}{100}\\ 
&112&  \quart{17.276313392432076}{66.89105122805309}{50.84766092416191}{100}\\ 
&96&  \quart{0.0}{66.3676564796563}{33.16292946324728}{100}\\  &&  \squart{1}{1}{1}{100}\\ 
        \end{tabular} &
        &
       \begin{tabular}{lcccl}
                   \cellcolor[gray]{1}\textbf{}  &\cellcolor[gray]{1}$\mathbfcal{R}$  & \cellcolor[gray]{1}\textbf{Quantiles}  \\
            \hline
&150&  \quart{15.398046358704729}{39.223761507698526}{27.54159475973198}{100}\\ 
&149&  \quart{15.318863361588036}{40.73063697978033}{27.785332122144187}{100}\\ 
&175&  \quart{23.368424727072018}{76.63157527292798}{61.088721731832436}{100}\\ 
&\cellcolor{steel!10}\textbf{74}&  \bquart{0.0}{14.46556123586702}{4.8530255241488165}{100}\\  &&  \squart{1}{1}{1}{100}\\ 
        \end{tabular}&
        &
       \begin{tabular}{lcccl}
                     \cellcolor[gray]{1}\textbf{}  &\cellcolor[gray]{1}$\mathbfcal{R}$  & \cellcolor[gray]{1}\textbf{Quantiles}  \\
            \hline
&\cellcolor{steel!10}\textbf{85}&  \bquart{0.0}{38.80610351966513}{22.645413574128153}{100}\\ 
&88& \quart{0.0}{40.17858020871456}{22.627702859875857}{100}\\ 
&182&  \quart{26.223270761815225}{73.77672923818479}{60.568494149002575}{100}\\  
&137&  \quart{13.917541476969033}{38.148713834896654}{37.46681029885222}{100}\\  &&  \squart{1}{1}{1}{100}\\ 
        \end{tabular}  \\
            (f). SCT: \textsc{wc-c1-6d}&& (g). SCT: \textsc{llvm} && (h). SCT: \textsc{trimesh}&& (i). WSC: \textsc{5as-1} && (j). WSC: \textsc{5as-2}
            
            \\
            \\

\begin{tabular}{llcc}
             \cellcolor[gray]{1}\textbf{} & &\cellcolor[gray]{1}$\mathbfcal{R}$  & \cellcolor[gray]{1}\textbf{Quantiles}  \\
            \hline

\texttt{Dynamic}&&101&  \quart{1.2713041716524012}{17.919719070963676}{9.523168148949269}{100}\\ 
\texttt{Fixed}&&111&  \quart{3.8124586441420036}{16.0277031258265}{9.9751680903197}{100}\\
\texttt{Ratio}&&198&  \quart{27.692674110120542}{72.30732588987944}{65.55811419562794}{100}\\ 
\texttt{None}&&\cellcolor{steel!10}\textbf{94}&  \bquart{0.0}{14.932449140782465}{7.069438513846545}{100}\\  &&&  \squart{1}{1}{1}{100}\\ 

        \end{tabular} & 
        &
       \begin{tabular}{lcccl}
                  \cellcolor[gray]{1}\textbf{}  &\cellcolor[gray]{1}$\mathbfcal{R}$  & \cellcolor[gray]{1}\textbf{Quantiles}  \\  
            \hline
&132&  \quart{19.113777885035276}{34.896388674790856}{35.13656932020115}{100}\\  
&132&   \quart{19.064874335420217}{35.314897107167695}{35.3772759614371}{100}\\ 
&173&  \quart{22.548607607967924}{77.45139239203208}{48.97391974326306}{100}\\ 
&\cellcolor{steel!10}\textbf{73}&  \bquart{0.0}{35.15293677068417}{10.655828054665266}{100}\\  &&  \squart{1}{1}{1}{100}\\ 
        \end{tabular} &
        &
       \begin{tabular}{lcccl}
                    \cellcolor[gray]{1}\textbf{}  &\cellcolor[gray]{1}$\mathbfcal{R}$  & \cellcolor[gray]{1}\textbf{Quantiles}  \\
            \hline
&112&  \quart{0.06713982890116674}{45.89295775507703}{23.526083855085258}{100}\\ 
&134&  \quart{2.015535356763251}{44.3226085861825}{24.10030829528178}{100}\\ 
&210&   \quart{26.65852540547005}{73.34147459452996}{55.230349679500634}{100}\\ 
&\cellcolor{steel!10}\textbf{76}&   \bquart{0.0}{32.23503366641988}{11.304790359450394}{100}\\  &&  \squart{1}{1}{1}{100}\\ 
        \end{tabular} &
        &
       \begin{tabular}{lcccl}
                   \cellcolor[gray]{1}\textbf{}  &\cellcolor[gray]{1}$\mathbfcal{R}$  & \cellcolor[gray]{1}\textbf{Quantiles}  \\
            \hline
&109&  \quart{0.0}{52.38352939740156}{26.973049315274345}{100}\\ 
&122& \quart{1.4489308033043713}{51.083571810040866}{26.98852405312334}{100}\\ 
&196&  \quart{26.192619560746536}{73.80738043925348}{51.28739264773593}{100}\\  
&\cellcolor{steel!10}\textbf{79}&  \bquart{0.43267491863890395}{36.377675931955146}{8.438949483648582}{100}\\  &&  \squart{1}{1}{1}{100}\\ 
        \end{tabular}&
        &
       \begin{tabular}{lcccl}
                     \cellcolor[gray]{1}\textbf{}  &\cellcolor[gray]{1}$\mathbfcal{R}$  & \cellcolor[gray]{1}\textbf{Quantiles}  \\
            \hline
&121&  \quart{17.94600164116997}{46.218292929315446}{35.63325624139688}{100}\\ 
&123&  \quart{18.437887354164324}{45.5872988450986}{35.99709629789996}{100}\\
&167& \quart{18.712794506689793}{81.28720549331021}{49.331688425089936}{100}\\ 
&\cellcolor{steel!10}\textbf{75}&  \bquart{0.0}{42.5892557338341}{16.469280965088185}{100}\\ &&  \squart{1}{1}{1}{100}\\ 
        \end{tabular}  \\
         (k). WSC: \textsc{5as-3}&& (l). WSC: \textsc{10as-1} && (m). WSC: \textsc{10as-2}&& (n). WSC: \textsc{10as-3} && (o). WSC: \textsc{15as-1}
            
            \\
            \\

\begin{tabular}{llcc}
             \cellcolor[gray]{1}\textbf{} & &\cellcolor[gray]{1}$\mathbfcal{R}$  & \cellcolor[gray]{1}\textbf{Quantiles}  \\
            \hline

\texttt{Dynamic}&&103&  \quart{10.947980020360495}{60.9293793098774}{40.80138074254576}{100}\\ 
\texttt{Fixed}&&135&  \quart{12.608886196448584}{61.417961175539304}{43.52933045399906}{100}\\ 
\texttt{Ratio}&&194&   \quart{20.58851958636621}{79.4114804136338}{51.51289228369911}{100}\\
\texttt{None}&&\cellcolor{steel!10}\textbf{75}&   \bquart{0.0}{52.50836768508368}{19.477733677293}{100}\\  &&&  \squart{1}{1}{1}{100}\\ 

        \end{tabular} & 
        &
       \begin{tabular}{lcccl}
                    \cellcolor[gray]{1}\textbf{}  &\cellcolor[gray]{1}$\mathbfcal{R}$  & \cellcolor[gray]{1}\textbf{Quantiles}  \\
            \hline
&\cellcolor{steel!10}\textbf{96}&  \bquart{0.0}{69.45331943474352}{33.50241756021043}{100}\\
&117&  \quart{2.109459829239889}{69.28784759992837}{34.9032815772552}{100}\\
&169&  \quart{9.841211571493044}{90.15878842850695}{43.17580154446228}{100}\\ 
&105&  \quart{1.544776787691992}{47.152808719851286}{16.991623473851853}{100}\\  &&  \squart{1}{1}{1}{100}\\ 
     \end{tabular} &
        &
       \begin{tabular}{lcccl}
                    \cellcolor[gray]{1}\textbf{}  &\cellcolor[gray]{1}$\mathbfcal{R}$  & \cellcolor[gray]{1}\textbf{Quantiles}  \\
            \hline
&136&  \quart{3.1546481022771}{83.68589990739915}{25.269173020652616}{100}\\  
&122&  \quart{2.74571578183953}{84.81357218661651}{24.35821684868136}{100}\\ 
&170& \quart{3.448760634403433}{96.55123936559657}{29.07642529731338}{100}\\ 
&\cellcolor{steel!10}\textbf{72}&  \bquart{0.0}{32.22971145222564}{14.123789734150433}{100}\\  &&  \squart{1}{1}{1}{100}\\ 
        \end{tabular} &
        &
       \begin{tabular}{lcccl}
                   \cellcolor[gray]{1}\textbf{}  &\cellcolor[gray]{1}$\mathbfcal{R}$  & \cellcolor[gray]{1}\textbf{Quantiles}  \\
            \hline
&151& \quart{12.11435041134388}{22.9005651901968}{21.98274675857006}{100}\\  
&\cellcolor{steel!10}\textbf{86}&  \bquart{0.0}{26.161988459318913}{2.724612408799873}{100}\\ 
&175&  \quart{24.26990645419396}{32.0101348332616}{33.59971380801482}{100}\\
&144&  \quart{14.464345085924881}{85.53565491407512}{35.38647258104989}{100}\\  &&  \squart{1}{1}{1}{100}\\ 

        \end{tabular}&
        &
       \begin{tabular}{lcccl}
                     \cellcolor[gray]{1}\textbf{}  &\cellcolor[gray]{1}$\mathbfcal{R}$  & \cellcolor[gray]{1}\textbf{Quantiles}  \\
            \hline
&147&  \quart{7.252488693382301}{20.183659246174134}{16.366363600781693}{100}\\ 
&\cellcolor{steel!10}\textbf{85}&  \bquart{0.0}{20.095644090605987}{3.2660109105113393}{100}\\ 
&178&   \quart{19.333971925817927}{26.516794177935083}{27.796926911826716}{100}\\ 
&154& \quart{11.479568458480916}{88.52043154151909}{29.998444755598243}{100}\\ &&  \squart{1}{1}{1}{100}\\ 
        \end{tabular}  \\
        (p). WSC: \textsc{15as-2}&& (q). WSC: \textsc{15as-3} && (r). WSC: \textsc{50as}&& (s). NRP: \textsc{nrp-e1} && (t). NRP: \textsc{nrp-e2}

\\
\\

\begin{tabular}{llcc}
             \cellcolor[gray]{1}\textbf{} & &\cellcolor[gray]{1}$\mathbfcal{R}$  & \cellcolor[gray]{1}\textbf{Quantiles}  \\
            \hline

\texttt{Dynamic}&&\cellcolor{steel!10}\textbf{74}&  \bquart{0.0}{10.816195363298428}{6.010647025537896}{100}\\  
\texttt{Fixed}&&137&   \quart{7.544477762511443}{8.912442794628278}{13.694191319941408}{100}\\ 
\texttt{Ratio}&&138&  \quart{7.898096131757496}{40.67677038938396}{13.053037455225027}{100}\\ 
\texttt{None}&&197&   \quart{14.741129108835151}{85.25887089116483}{31.433155361521933}{100}\\ &&&  \squart{1}{1}{1}{100}\\ 
        \end{tabular} & 
        &
       \begin{tabular}{lcccl}
                    \cellcolor[gray]{1}\textbf{}  &\cellcolor[gray]{1}$\mathbfcal{R}$  & \cellcolor[gray]{1}\textbf{Quantiles}  \\
            \hline
&152&   \quart{9.18644829448514}{18.68152893190084}{18.293747431026713}{100}\\ 
&\cellcolor{steel!10}\textbf{86}&  \bquart{0.0}{19.44267859441527}{1.7517921192271688}{100}\\ 
&184&  \quart{18.940039928263847}{23.56888061156767}{26.133921245742574}{100}\\ 
&154&  \quart{12.46238720736558}{87.53761279263442}{28.620243828870255}{100}\\  &&  \squart{1}{1}{1}{100}\\ 

     \end{tabular} &
        &
       \begin{tabular}{lcccl}
                    \cellcolor[gray]{1}\textbf{}  &\cellcolor[gray]{1}$\mathbfcal{R}$  & \cellcolor[gray]{1}\textbf{Quantiles}  \\
            \hline
&141&  \quart{8.469329914575964}{13.358455790452899}{17.092906181030756}{100}\\ 
&\cellcolor{steel!10}\textbf{78}&  \bquart{0.0}{10.672242988644163}{4.070979241116215}{100}\\ 
&170&   \quart{14.56419138779277}{25.896500634344804}{27.23679073625388}{100}\\ 
&171&  \quart{8.91022242053533}{91.08977757946467}{22.662003378826583}{100}\\  &&  \squart{1}{1}{1}{100}\\ 

        \end{tabular} &
        &
       \begin{tabular}{lcccl}
                   \cellcolor[gray]{1}\textbf{}  &\cellcolor[gray]{1}$\mathbfcal{R}$  & \cellcolor[gray]{1}\textbf{Quantiles}  \\
            \hline
&146& \quart{11.092853571361205}{16.804330970659745}{19.796780332815278}{100}\\ 
&\cellcolor{steel!10}\textbf{85}&  \bquart{0.0}{16.699473125279887}{2.967747772807922}{100}\\ 
&192& \quart{24.196614295966945}{13.843199160040387}{30.70256902061613}{100}\\ 
&154& \quart{8.615258004700097}{91.38474199529989}{26.3661824833086}{100}\\  &&  \squart{1}{1}{1}{100}\\ 
        \end{tabular}&
        &
       \begin{tabular}{lcccl}
                     \cellcolor[gray]{1}\textbf{}  &\cellcolor[gray]{1}$\mathbfcal{R}$  & \cellcolor[gray]{1}\textbf{Quantiles}  \\
            \hline
&152&  \quart{9.287333005409826}{16.946301617084107}{16.903722424533782}{100}\\ 
&\cellcolor{steel!10}\textbf{86}&   \bquart{0.0}{17.916994784143192}{1.632239734589432}{100}\\ 
&176&  \quart{17.863476241685014}{27.044320505170028}{25.775198806873888}{100}\\ 
&149&  \quart{10.844356413653074}{89.15564358634693}{25.775198806873888}{100}\\ &&  \squart{1}{1}{1}{100}\\ 
        \end{tabular}  \\
      (u). NRP: \textsc{nep-e3}&& (v). NRP: \textsc{nrp-e4} && (w). NRP: \textsc{nrp-g1}&& (x). NRP: \textsc{nrp-g2} && (y). NRP: \textsc{nrp-g3}
     
     \\
     \\

     \begin{tabular}{llcc}
             \cellcolor[gray]{1}\textbf{} & &\cellcolor[gray]{1}$\mathbfcal{R}$  & \cellcolor[gray]{1}\textbf{Quantiles}  \\
            \hline

 \texttt{Dynamic}&&147&   \quart{12.374852774039548}{18.421264622911533}{20.59028433213088}{100}\\ 
\texttt{Fixed}&&\cellcolor{steel!10}\textbf{86}&  \bquart{0.0}{22.10417296405589}{2.407393649135468}{100}\\ 
\texttt{Ratio}&&184&  \quart{25.21837684411851}{20.465245822685432}{33.4776481118785}{100}\\ 
\texttt{None}&&152&  \quart{14.187282259863245}{85.81271774013676}{30.487736784125406}{100}\\  &&&  \squart{1}{1}{1}{100}\\ 

        \end{tabular} & 
        &
       \begin{tabular}{lcccl}
                    \cellcolor[gray]{1}\textbf{}  &\cellcolor[gray]{1}$\mathbfcal{R}$  & \cellcolor[gray]{1}\textbf{Quantiles}  \\
            \hline

&\cellcolor{steel!10}\textbf{110}&  \bquart{0.0}{16.71811777173573}{6.88805122760889}{100}\\ 
&115&  \quart{0.6523059869622279}{12.214667938146258}{8.204406088351057}{100}\\ 
&156&  \quart{9.034365770235251}{43.82733554307643}{20.554820986979454}{100}\\ 
&155& \quart{7.587051284274695}{92.41294871572532}{21.04928995779485}{100}\\ 
&&  \squart{1}{1}{1}{100}\\ 
     \end{tabular} &
        &
       \begin{tabular}{lcccl}
                    \cellcolor[gray]{1}\textbf{}  &\cellcolor[gray]{1}$\mathbfcal{R}$  & \cellcolor[gray]{1}\textbf{Quantiles}  \\
            \hline

&139&  \quart{5.031781997612107}{16.54277059590393}{12.49737305773704}{100}\\ 
&\cellcolor{steel!10}\textbf{85}&  \bquart{0.0}{14.474671444414533}{2.0482833256450093}{100}\\ 
&184& \quart{15.54049105064644}{22.56324430563477}{22.411676514750752}{100}\\ 
&159& \quart{8.890686930586964}{91.10931306941305}{24.204620553561114}{100}\\  
&&  \squart{1}{1}{1}{100}\\ 
        \end{tabular} &
        &
       \begin{tabular}{lcccl}
                   \cellcolor[gray]{1}\textbf{}  &\cellcolor[gray]{1}$\mathbfcal{R}$  & \cellcolor[gray]{1}\textbf{Quantiles}  \\
            \hline
&\cellcolor{steel!10}\textbf{80}&  \bquart{0.0}{12.508660567987134}{6.415470655649955}{100}\\ 
&140&  \quart{7.195684291015519}{12.369041595013433}{13.690342546350209}{100}\\ 
&138&  \quart{5.1779316724125115}{43.431923368317825}{13.87255507320672}{100}\\ 
&189&  \quart{14.144993454152598}{85.8550065458474}{25.25068193308455}{100}\\ 
&&  \squart{1}{1}{1}{100}\\ 
        \end{tabular}&
        &
       \begin{tabular}{lcccl}
                     \cellcolor[gray]{1}\textbf{}  &\cellcolor[gray]{1}$\mathbfcal{R}$  & \cellcolor[gray]{1}\textbf{Quantiles}  \\
            \hline
&128&  \quart{1.6564113638028488}{17.338579086830425}{9.438479185349268}{100}\\ 
&\cellcolor{steel!10}\textbf{93}&  \bquart{0.0}{13.739316107956041}{4.343942701788036}{100}\\
&162&  \quart{6.681122996645385}{43.312205084536345}{20.836034751291447}{100}\\ 
&150&  \quart{7.9513622569857505}{92.04863774301425}{22.656481651061895}{100}\\ 
&&  \squart{1}{1}{1}{100}\\ 
        \end{tabular}  \\
       (z). NRP: \textsc{nep-g4}&& (aa). NRP: \textsc{nrp-m1} && (bb). NRP: \textsc{nrp-m2}&& (cc). NRP: \textsc{nrp-m3} && (dd). NRP: \textsc{nrp-m4}
     
    \end{tabular}
   \end{center}
\end{table*}

\begin{table*}[t!]
\caption{Scott-Knott ranks and the weighted scores for the normalization methods on three objectives SBSE problems. A cell summarizes 32 cases (4 algorithms $\times$ 4 weight vectors $\times$ 2 types of budget) on 100 runs each. Formats are the same as Table~\ref{tb:rq1}.}
    \label{tb:rq1-3}
    \footnotesize
    \setlength{\tabcolsep}{1.5mm}
  \begin{center}

    \begin{tabular}{cc@{}cc@{}cc@{}cc}
        \begin{tabular}{llcc}
            \cellcolor[gray]{1}\textbf{} & &\cellcolor[gray]{1}$\mathbfcal{R}$  & \cellcolor[gray]{1}\textbf{Quantiles}  \\
            \hline

\texttt{Dynamic}&&64&  \quart{4.331158064183889}{41.19506812320192}{7.0011831884968645}{100}\\ 
\texttt{Fixed} &&\cellcolor{steel!10}\textbf{40}&  \bquart{2.5485335355613694}{2.787110402382816}{3.563546539091644}{100}\\ 
\texttt{Ratio}&&54&  \quart{0.0}{26.09611959646862}{2.616961512462431}{100}\\ 
\texttt{None}&&82&  \quart{6.029059340953847}{93.97094065904616}{40.79166356868651}{100}\\  &&& \squart{1}{1}{1}{100}\\ 
        \end{tabular} & 
        &
       \begin{tabular}{lcccl}
                     \cellcolor[gray]{1}\textbf{}  &\cellcolor[gray]{1}$\mathbfcal{R}$  & \cellcolor[gray]{1}\textbf{Quantiles}  \\
            \hline
&\cellcolor{steel!10}\textbf{32}&  \bquart{0.07882175141549173}{55.568588999638095}{15.673674308827811}{100}\\ 
&\cellcolor{steel!10}\textbf{32}&  \quart{0.0}{71.983818157347}{16.23179498783406}{100}\\ 
&34& \quart{2.46330510454759}{59.755941937503565}{17.3088382052765}{100}\\ 
&36&   \quart{2.7537742481724874}{97.24622575182752}{15.303980054530017}{100}\\ && \squart{1}{1}{1}{100}\\ 

        \end{tabular} &
        &
       \begin{tabular}{lcccl}
                    \cellcolor[gray]{1}\textbf{}  &\cellcolor[gray]{1}$\mathbfcal{R}$  & \cellcolor[gray]{1}\textbf{Quantiles}  \\
            \hline
&44& \quart{0.0}{17.46473702985858}{6.72585031065293}{100}\\
&82&  \quart{5.607004164016086}{47.80008897637398}{34.236504654300944}{100}\\
&91&  \quart{5.607004164016086}{47.80008897637398}{34.236504654300944}{100}\\
&\cellcolor{steel!10}\textbf{41}&   \bquart{0.03729043178173853}{22.719688658237445}{3.341874789025712}{100}\\  && \squart{1}{1}{1}{100}\\ 
        \end{tabular} &
        &
       \begin{tabular}{lcccl}
                    \cellcolor[gray]{1}\textbf{}  &\cellcolor[gray]{1}$\mathbfcal{R}$  & \cellcolor[gray]{1}\textbf{Quantiles}  \\
            \hline
&51&  \quart{7.627338498826552}{65.84403724781319}{16.749192257178752}{100}\\ 
&83& \quart{7.627338498826552}{84.36556438109065}{18.648721415418184}{100}\\ 
&102&  \quart{7.627338498826552}{92.37266150117344}{20.678796390498295}{100}\\ 
&\cellcolor{steel!10}\textbf{32}&  \bquart{0.0}{35.3321299946951}{11.14676404463341}{100}\\  && \squart{1}{1}{1}{100}\\ 

        \end{tabular} 
        \\

               (a). SCT: \textsc{vp8}&& (b). SCT: \textsc{hsqldb} && (c). WSC: \textsc{5as-3o}&& (d). WSC: \textsc{10as-3o}

\\
        \\
        
             \begin{tabular}{llcc}
              \cellcolor[gray]{1}\textbf{} & &\cellcolor[gray]{1}$\mathbfcal{R}$  & \cellcolor[gray]{1}\textbf{Quantiles}  \\
            \hline

\texttt{Dynamic}&&56&  \quart{7.627338498826552}{65.84403724781319}{16.749192257178752}{100}\\ 
\texttt{Fixed} &&67&  \quart{7.627338498826552}{84.36556438109065}{18.648721415418184}{100}\\ 
\texttt{Ratio}&&77&  \quart{7.627338498826552}{92.37266150117344}{20.678796390498295}{100}\\ 
\texttt{None}&&\cellcolor{steel!10}\textbf{42}&  \bquart{0.0}{35.3321299946951}{11.14676404463341}{100}\\ 
&&&  \squart{1}{1}{1}{100}\\ 
        \end{tabular} & 
        &
       \begin{tabular}{lcccl}
        \cellcolor[gray]{1}\textbf{}  &\cellcolor[gray]{1}$\mathbfcal{R}$  & \cellcolor[gray]{1}\textbf{Quantiles}  \\
            \hline
&50&  \quart{7.999691340415685}{48.07508287943913}{26.875412345816194}{100}\\ 
&\cellcolor{steel!10}\textbf{32}&  \bquart{0.0}{45.6775656939541}{8.355759771620368}{100}\\ 
&48&  \quart{9.795251127527736}{49.97461874146412}{34.99585183175869}{100}\\ 
&74& \quart{17.952726544147957}{82.04727345585205}{62.7780176746448}{100}\\ 
&& \squart{1}{1}{1}{100}\\ 
        \end{tabular} &
        &
       \begin{tabular}{lcccl}
        \cellcolor[gray]{1}\textbf{}  &\cellcolor[gray]{1}$\mathbfcal{R}$  & \cellcolor[gray]{1}\textbf{Quantiles}  \\
            \hline
&60&  \quart{23.992970396958334}{44.43724667032439}{48.232538287417555}{100}\\ 
&\cellcolor{steel!10}\textbf{34}&  \bquart{0.0}{55.72236149780192}{12.652760112014441}{100}\\ 
&48&  \quart{12.00141389792017}{51.916337383211186}{21.489123759764354}{100}\\ 
&66&  \quart{15.752374132231083}{84.24762586776892}{56.60712568172148}{100}\\ 
&& \squart{1}{1}{1}{100}\\ 
        \end{tabular} &
        &
       \begin{tabular}{lcccl}
        \cellcolor[gray]{1}\textbf{}  &\cellcolor[gray]{1}$\mathbfcal{R}$  & \cellcolor[gray]{1}\textbf{Quantiles}  \\
            \hline

&51&  \quart{10.407256275687004}{64.37002824185394}{36.10291304229987}{100}\\ 
&\cellcolor{steel!10}\textbf{33}&  \bquart{0.0}{64.28404396835008}{6.917227205970893}{100}\\ 
&41&   \quart{0.1702099485510396}{78.7213844710177}{17.711116840195196}{100}\\ 
&65&  \quart{14.200167650763808}{85.7998323492362}{58.79400302123143}{100}\\ 

&& \squart{1}{1}{1}{100}\\ 
        \end{tabular} 
        \\

               (e). WSC: \textsc{15as-3o}&& (f). NRP: \textsc{nrp-e-3o} && (g). NRP: \textsc{nrp-g-3o}&& (h). NRP: \textsc{nrp-m-3o}

        \\

    \end{tabular}
   \end{center}
\end{table*}

\subsubsection{Result}

As can be seen from Figure~\ref{fig:rq1-summary}, it appears to be that, regardless of the number of objectives, the \texttt{Fixed} has the best Scott-Knott rank across all the cases, which is also similar to that of \texttt{Dynamic}. The reaming two, particularly the \texttt{Ratio}, performs considerably worse than the others. However, obtaining the bounds to be used in the \texttt{Fixed} method can be time-consuming when the true bounds are not naturally known beforehand.

To take a closer look at each system/project, the two objective case has been illustrated in Table~\ref{tb:rq1}. Here, for SCT, we see that the best normalization method varies quite differently depending on the actual software system. Yet, the worst one generally has considerably bad results and larger variation. On the results under WSC and NRP, we see quite different observations: \texttt{None} tends to be the best in general for the former while \texttt{Fixed} is often better than the other three for the latter. Their advantages are both statistically significant and to a considerable extent. \texttt{Ratio}, in contrast, often perform the worst across the systems/projects. The same trends can be observed for the three objective case, as shown in Table~\ref{tb:rq1-3}.

The above shows a clear sign that the best normalization method 
for weighted search highly depends on the system/project of a SBSE problem in hand, meaning that for the best result of the weighted search, an extra step is required for deciding on the best normalization method. In contrast, this is not required for the Pareto search as it tends to be less sensitive to the objective scales.


\begin{quotebox}
   \noindent
   \textit{\textbf{Finding:} There does not exist a generally best normalization for weighted search across the multi-objective SBSE problems and the systems/projects. Therefore, when necessary, an extra process of identifying the best normalization method is needed in order to achieve the best result.}
\end{quotebox}

\subsubsection{Implication}

From the results for \textbf{RQ1}, we note that, though there is not a generally best normalization method, \texttt{Dynamic} tends to be a safe option as it often performs the second-best (if not the best) while certainly is never the worst. This matches with its overall 2nd ranking among the cases from Figure~\ref{fig:rq1-summary}.
However, while the \texttt{Fixed} and \texttt{None} was ranked as the 1st and 3rd on the overall ranking, the results tend to differ from that when looking at the detailed systems/projects: the former has considerably worse results than \texttt{Dynamic} for WSC, while the latter is much more inferior to \texttt{Dynamic} under NRP. In contrast, \texttt{Ratio} performs the worst in general across the systems/projects and SBSE problems, which is consistent with its overall ranking, and therefore it can be ruled out from the comparison when the resource is limited.

As such, we suggest the following for SBSE practitioners:

\begin{quotebox}
   \noindent
   \textit{\textbf{Suggestion:} Do the following when using weighted search for a multi-objective SBSE problem:
\begin{enumerate}
\item Experimentally comparing the normalization methods (at least the four in this work) whenever the conditions permitted. 
The \texttt{Ratio} can be omitted in the case that the resource is limited.
\item When (1) is not possible, using \texttt{Dynamic} by default as it is a generally safer option.
\end{enumerate}}
\end{quotebox}

\subsubsection{Reason}

Clearly, the performance of the normalization method \texttt{None} highly depends on the scale of objectives in the SBSE problem. 
For example, 
in NRP the general range of the objectives in the two objective case is $[\lvert0.0035\rvert,\lvert0.11\rvert]$ and $[1.0,504.0]$, respectively, 
and therefore \texttt{None} can easily lead to biased search against the first objective.


Consider the normalization methods \texttt{Ratio} and \texttt{Dynamic}.
It may not be difficult to understand why \texttt{Ratio} performs worse than \texttt{Dynamic}.
In contrast to \texttt{Dynamic} which always transforms the objective into a range of $[0.0, 1.0]$,
\texttt{Ratio} (defined as ${v} \over {v+1}$) is actually still affected by the scale of each individual objective --- 
a very small or very big objective value $v$ will squish the normalized value into a tiny part of the range $[0.0,1.0]$. 
For example, 
the cost objective under WSC has a range of $[14.64,122.46]$, 
and after transformed it will become $[0.94,0.99]$.

For the normalization method \texttt{Fixed},
it however may not be easy to understand why it does not always perform best, 
given the fact that it uses the known bounds of the SBSE problem's objectives 
that should have been the most accurate information. 
Here we use an example to explain why the \texttt{Fixed} method does not always work. 

Consider a bi-objective minimization problem where the bound of the first objective is $[0.0,1.0]$ 
and the bound of the second objective is $[0.0, 10.0]$, 
and they are known prior to the search. 
Let us say that the stakeholders equally like the two objectives
and the solution $(0.5, 5)$ be the most preferred one. 
For the method \texttt{Fixed}, 
the weight $(10/11, 1/11)$ is chosen since the range of the second objective is ten times larger than that of the first objective. 
But during the search, 
there may just be a small portion of the space accessible, 
especially at the beginning stage. 
For example, 
the initial population only covers the range $[9.0,10.0]$ on the second objective 
(but covers the full range $[0.0,1.0]$ on the first objective). 
In this case, 
the weights $(10/11, 1/11)$ will likely lead to solutions close to the point $(0.0+1/11, 9.0+10/11)$ to be preferred. 
This will drive the search away from the region of the desired solution $(0.5, 5)$. 
In multi-objective SBSE, 
it is not uncommon that the accessibility of the objectives is different during the search. 
Take WSC as an example, the value range of cycle time is less accessible compared with that of the cost (and latency).
The is due to the way of how the objectives are calculated determining that the cycle time is much less sensitive to different compositions than the cost (and latency), 
as only the concrete service with the maximum cycle time would be used while the cost is always the summation of all.





\subsection{RQ2: Quality of Solution}
\label{sec:rq2}

\subsubsection{Method}

To answer \textbf{RQ2}, 
we perform pairwise comparisons between Pareto and weighted search using all the 38 systems/projects of the SBSE problems, 
the nine sets of weights, and the two types of search budget (evaluation and time), leading to a total of 604 cases of investigation (540 for two objective and 64 for three objective case). In each of those cases, we extract the best optimizer (i.e., amongst RS, HC, SA, and SOGA) and its normalization method (i.e., amongst \texttt{Dynamic}, \texttt{Fixed}, \texttt{Ratio}, and \texttt{None}), denoted as $W_{best}$, to compare with the NSGA-II and MOEA/D independently. To identify the best, we leverage on the result from \textbf{RQ1} to find the one from the best Scott-Knott rank; if there are multiple optimizers (and their normalization methods) in the best rank, we use the pair with the best median (and smallest IQR, if needed) weighted score as the best. This makes sense since we are only interested in the result of the best optimizer/normalization method for the weighted search against that of the Pareto counterpart. For each case, both Wilcoxon rank-sum test and $\hat{A}_{12}$ are used (over 100 runs) to test the significance of the comparison on the resulted weighted score between $W_{best}$ and NSGA-II (or MOEA/D).

 \begin{figure}[t!]
\centering
  \includegraphics[width=\textwidth]{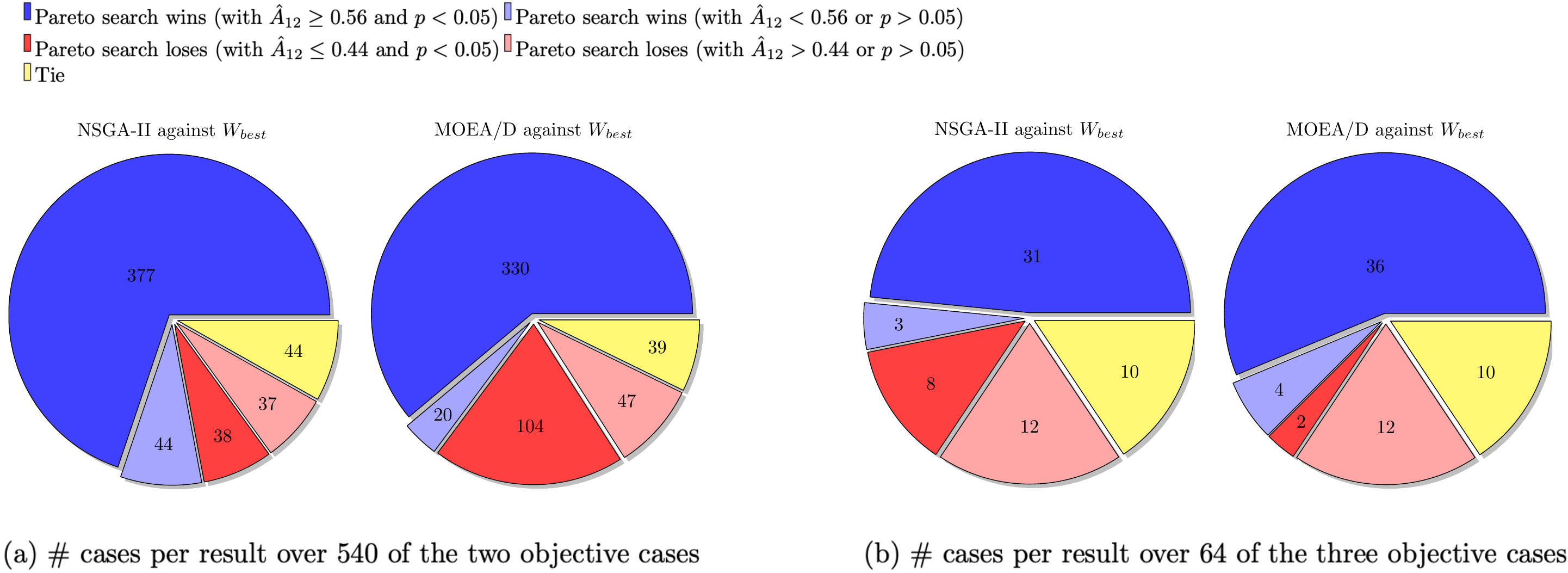}
  \caption{The count on the pairwise comparison results between Pareto search and the best weighted search ($W_{best}$ for each case, as identified from \textbf{RQ1}) over (a) 540 cases with 2 objectives (30 systems/projects $\times$ 9 weight vectors $\times$ 2 types of budget) and (b) 64 cases with 3 objectives (8 systems/projects $\times$ 4 weight vectors $\times$ 2 types of budget). The Pareto search with a win, lose, and tie refer to the case of $\hat{A}_{12} > 0.5$, $\hat{A}_{12} < 0.5$,  and $\hat{A}_{12} = 0.5$, respectively. When there is a $\hat{A}_{12} \geq 0.56$ (or $\hat{A}_{12} \leq 0.44$) and $p<0.05$, we say that the result is statistically significant.}
  
  \label{fig:rq2-summary}
 \end{figure}

\subsubsection{Result}

As we can see from Figure~\ref{fig:rq2-summary}, for two objective case, Pareto search wins on 77\% of the cases (421 out of 540) for NSGA-II and 65\% of the cases (350 out of 540) for MOEA/D; loses on 14\% cases (75 out of 540) for NSGA-II and 27\% cases (151 out of 540) for MOEA/D; there is a 9\% tie (44 cases) for NSGA-II and 7\% tie (39 cases) for MOEA/D. Particularly, in the cases where Pareto search wins, 377 (for NSGA-II) and 330 (for MOEA/D) of them come with statistical significance and a large effect size. In contrast, on the 75 (for NSGA-II) and 151 (for MOEA/D) cases where Pareto search loses, only 38 (for NSGA-II) and 104 (for MOEA/D) of them exhibit statistical significance and non-trivial effect size. The above is also consistent for the three objective case, where the Pareto search wins on 53\% cases (34 out of 64) for NSGA-II and 62\% (40 out of 64) for MOEA/D, on the majority of which are statistically significant. The improvement over the weighted search degrades slightly though.

To confirm whether the above observation applied to the systems/projects and SBSE problems, Tables~\ref{tb:rq2} and~\ref{tb:rq2-3} show the detailed results for two and three objective case, respectively. Here, we see that the overall conclusion remains unchanged for both numbers of the objectives: Pareto search often wins with a reasonably good degree at the 25th, 50th, and 75th percentiles for a majority of the systems/projects (up to $1.77\times$ median improvement). Another observation is that the results are consistent across the different SBSE problems and their systems/projects, which further confirms the generality of the conclusion.

This is a rather surprising outcome 
as weighted search is guided by the exact weight vector used in the comparison, 
and hence it was generally believed to be better by the SBSE community. 
The result confirms our hypothesis and even reveals the significant superiority of Pareto search under clear preferences for \textbf{RQ2}.


\begin{quotebox}
   \noindent
   \textit{\textbf{Finding:} Pareto search is significantly better for up to 77\% cases of the multi-objective SBSE cases under reasonable convergence.}
\end{quotebox}


\subsubsection{Implication}

Our findings for \textbf{RQ2} demonstrate that the \textit{weighted search first} belief is problematic, 
as Pareto search can perform considerably better in general as long as there is a sufficient search budget. Recall that from Table~\ref{tb:settings}, we note that such a budget differ significantly depending on the systems/projects and SBSE problems in hand, but essentially they do not have to be unrealistically high: it could well be a few hundreds of evaluations or in the magnitude of seconds.

\begin{table*}[t!]
\caption{Pairwise comparisons (by $\hat{A}_{12}$ and Wilcoxon rank-sum test) and the weighted scores for Pareto and weighted search on two objectives SBSE problems. \quartexp{0}{20}{10}{20} denotes the 25th, 50th, and 75th percentile of the weighted score. A pair summarizes 18 cases (9 weight vectors $\times$ 2 types of budget) on 100 runs each. $W_{best}$ represents the best single objective algorithm and the normalization for the weighted search. $\mathbfcal{W}$ and $\mathbfcal{L}$ show how many cases the Pareto search wins ($\hat{A}_{12} > 0.5$) and loses ($\hat{A}_{12} < 0.5$) to the weighted counterpart, respectively. $\mathbfcal{T}$ denotes tie ($\hat{A}_{12} = 0.5$). The comparisons where the Pareto search and its weighed counterpart wins more are shown in \setlength{\fboxsep}{1.5pt}\colorbox{steel!10}{blue} and \setlength{\fboxsep}{1.5pt}\colorbox{red!30}{red}, respectively. The number in the bracket shows how many win/lose is statistically significant, i.e., $\hat{A}_{12} \geq 0.56$ (or $\hat{A}_{12} \leq 0.44$) and $p<0.05$. The one that with more statistically significant wins is highlighted in bold.}

    \label{tb:rq2}
    \footnotesize
    \setlength{\tabcolsep}{1mm}
  \begin{center}
    \begin{adjustbox}{max width = \textwidth}

    \begin{tabular}{cc@{}cc@{}cc@{}cc@{}cc}
        \begin{tabular}{llP{0.73cm}P{0.73cm}P{0.5cm}c}
            \cellcolor[gray]{1}\textbf{} & &\cellcolor[gray]{1}$\mathbfcal{W}$ &\cellcolor[gray]{1}$\mathbfcal{L}$  
            &\cellcolor[gray]{1}$\mathbfcal{T}$ & \cellcolor[gray]{1}\textbf{Quantiles}  \\
            \hline
W$_{best}$&&-&-&-& \quart{0.0}{75.0}{36.75078864353313}{100}\\ 
NSGA-II&&\cellcolor{steel!10}\textbf{18(8)}& \cellcolor{steel!10}0(0)& \cellcolor{steel!10}0& \quart{0.0}{50.00000000000001}{24.999999999999996}{100}\\ 
MOEA/D&&\cellcolor{red!10}8(6)&\cellcolor{red!10}\textbf{10(10)}&\cellcolor{red!10}0&  \quart{7.833859095688747}{92.16614090431125}{50.00000000000001}{100}\\ &&&&&  \squart{1}{1}{1}{100}\\ 
        \end{tabular} & 
        &
       \begin{tabular}{lcP{0.73cm}P{0.73cm}P{0.3cm}cl}
                        \cellcolor[gray]{1}\textbf{} & &\cellcolor[gray]{1}$\mathbfcal{W}$ &\cellcolor[gray]{1}$\mathbfcal{L}$   &\cellcolor[gray]{1}$\mathbfcal{T}$& \cellcolor[gray]{1}\textbf{Quantiles}  \\
            \hline
&&-&-&-&  \quart{0.0}{100.0}{41.99381417310626}{100}\\ 
&&\cellcolor{steel!10}\textbf{14(12)}&\cellcolor{steel!10}4(0)&\cellcolor{steel!10}0&  \quart{0.0}{45.775119014146}{31.65873239026949}{100}\\ 
&&\cellcolor{steel!10}\textbf{12(10)}&\cellcolor{steel!10}6(6)&\cellcolor{steel!10}0&  \quart{18.99523943416169}{74.61564813827286}{48.4394307143219}{100}\\ &&&&&  \squart{1}{1}{1}{100}\\ 
        \end{tabular} &
        &
       \begin{tabular}{lcP{0.73cm}P{0.73cm}P{0.3cm}cl}
                      \cellcolor[gray]{1}\textbf{} & &\cellcolor[gray]{1}$\mathbfcal{W}$ &\cellcolor[gray]{1}$\mathbfcal{L}$ &\cellcolor[gray]{1}$\mathbfcal{T}$ & \cellcolor[gray]{1}\textbf{Quantiles}  \\
            \hline
&&-&-&-&   \quart{0.0}{70.19184494596652}{33.33333333333336}{100}\\ 
&&\cellcolor{red!10}4(4)&\cellcolor{red!10}\textbf{14(12)}&\cellcolor{red!10}0&  \quart{4.7256693134176935}{77.9352437497407}{43.98118763998505}{100}\\ 
&&\cellcolor{red!10}6(4)&\cellcolor{red!10}\textbf{12(10)}&\cellcolor{red!10}0&  \quart{4.7256693134176935}{77.9352437497407}{43.98118763998505}{100}\\ &&&&&  \squart{1}{1}{1}{100}\\ 
        \end{tabular} &
        &
       \begin{tabular}{lcP{0.73cm}P{0.73cm}P{0.3cm}cl}
                        \cellcolor[gray]{1}\textbf{} & &\cellcolor[gray]{1}$\mathbfcal{W}$ &\cellcolor[gray]{1}$\mathbfcal{L}$  &\cellcolor[gray]{1}$\mathbfcal{T}$& \cellcolor[gray]{1}\textbf{Quantiles}  \\
            \hline
&&-&-&-&  \quart{90}{0.0}{90}{100}\\ 
&&0(0)&0(0)&18&  \quart{90}{0.0}{90}{100}\\ 
&&0(0)&0(0)&18&  \quart{90}{0.0}{90}{100}\\ &&&&&  \squart{1}{1}{1}{100}\\ 
        \end{tabular}&
        &
       \begin{tabular}{lcP{0.73cm}P{0.73cm}P{0.3cm}cl}
                        \cellcolor[gray]{1}\textbf{} & &\cellcolor[gray]{1}$\mathbfcal{W}$ &\cellcolor[gray]{1}$\mathbfcal{L}$  
                   &\cellcolor[gray]{1}$\mathbfcal{T}$& \cellcolor[gray]{1}\textbf{Quantiles}  \\
            \hline
&&-&-&-&  \quart{0.4552246733198046}{99.5447753266802}{50.227612336659895}{100}\\ 
&&\cellcolor{steel!10}\textbf{12(12)}&\cellcolor{steel!10}6(4)&\cellcolor{steel!10}0&  \quart{0.0}{44.23494303894311}{23.97000756628423}{100}\\ &&\cellcolor{steel!10}\textbf{12(10)}&\cellcolor{steel!10}6(6)&\cellcolor{steel!10}0&  \quart{0.27373687649958905}{54.469635735723934}{25.36167917974791}{100}\\&&&&&  \squart{1}{1}{1}{100}\\ 
        \end{tabular}  \\
            (a). SCT: \textsc{wc-c1-3d}&& (b). SCT: \textsc{wc-c3-3d} && (c). SCT: \textsc{wc-c4-3d}&& (d). SCT: \textsc{wc-c5-5d}&& (e). SCT: \textsc{rs-cd-6d}

\\
        \\
      
           \begin{tabular}{llP{0.73cm}P{0.73cm}P{0.3cm}c}
             \cellcolor[gray]{1}\textbf{} & &\cellcolor[gray]{1}$\mathbfcal{W}$ &\cellcolor[gray]{1}$\mathbfcal{L}$  
            &\cellcolor[gray]{1}$\mathbfcal{T}$ & \cellcolor[gray]{1}\textbf{Quantiles}  \\
            \hline
W$_{best}$&&-&-&-&  \quart{26.15490424487061}{73.84509575512938}{54.1898623095828}{100}\\ NSGA-II&&\cellcolor{steel!10}\textbf{18(16)}&\cellcolor{steel!10}0(0)&\cellcolor{steel!10}0&\quart{0.0}{37.10531371327055}{19.19040507868589}{100}\\ 
MOEA/D&&\cellcolor{steel!10}\textbf{16(16)}&\cellcolor{steel!10}2(2)&\cellcolor{steel!10}0&  \quart{8.996094613485251}{38.23841978108813}{26.15490424487061}{100}\\ &&&&&  \squart{1}{1}{1}{100}\\ 
        \end{tabular} & 
        &
       \begin{tabular}{lcP{0.73cm}P{0.73cm}P{0.3cm}cl}
         \cellcolor[gray]{1}\textbf{} & &\cellcolor[gray]{1}$\mathbfcal{W}$ &\cellcolor[gray]{1}$\mathbfcal{L}$  
            &\cellcolor[gray]{1}$\mathbfcal{T}$ & \cellcolor[gray]{1}\textbf{Quantiles}  \\
            \hline
&&-&-&-&  \quart{0.0}{10.404380791912407}{8.141968095225216}{100}\\ 
&&\cellcolor{steel!10}\textbf{14(2)}&\cellcolor{steel!10}2(0)&\cellcolor{steel!10}2&   \quart{0.0}{10.404380791912407}{8.141968095225216}{100}\\ 
&&\cellcolor{red!10}0(0)&\cellcolor{red!10}\textbf{18(16)}&\cellcolor{red!10}0&  \quart{15.32461885479797}{84.67538114520204}{36.88887681060954}{100}\\ &&&&&  \squart{1}{1}{1}{100}\\ 
        \end{tabular} &
        &
       \begin{tabular}{lcP{0.73cm}P{0.73cm}P{0.3cm}cl}
         \cellcolor[gray]{1}\textbf{} & &\cellcolor[gray]{1}$\mathbfcal{W}$ &\cellcolor[gray]{1}$\mathbfcal{L}$  
            &\cellcolor[gray]{1}$\mathbfcal{T}$ & \cellcolor[gray]{1}\textbf{Quantiles}  \\
            \hline
&&-&-&-&  \quart{0.0}{100.0}{50.06774998689252}{100}\\ 
&&\cellcolor{steel!10}\textbf{18(16)}&\cellcolor{steel!10}0(0)&\cellcolor{steel!10}0&  \quart{0.0}{99.10761622234632}{49.59764842656817}{100}\\ 
&&\cellcolor{steel!10}\textbf{14(12)}&\cellcolor{steel!10}4(4)&\cellcolor{steel!10}0&  \quart{0.18138749376844515}{99.73575815937716}{50.06774998689252}{100}\\ &&&&&  \squart{1}{1}{1}{100}\\ 
        \end{tabular} &
        &
       \begin{tabular}{lcP{0.73cm}P{0.73cm}P{0.3cm}cl}
         \cellcolor[gray]{1}\textbf{} & &\cellcolor[gray]{1}$\mathbfcal{W}$ &\cellcolor[gray]{1}$\mathbfcal{L}$  
            &\cellcolor[gray]{1}$\mathbfcal{T}$ & \cellcolor[gray]{1}\textbf{Quantiles}  \\
            \hline
&&-&-&-&  \quart{10.629125761774882}{23.157801473072578}{12.80374709596595}{100}\\  
&&\cellcolor{steel!10}\textbf{16(16)}&\cellcolor{steel!10}0(0)&\cellcolor{steel!10}2&   \quart{0.0}{8.69592767933124}{6.7627295968875965}{100}\\ 
&&\cellcolor{red!10}2(2)&\cellcolor{red!10}\textbf{10(10)}&\cellcolor{red!10}6&  \quart{12.80374709596595}{87.19625290403405}{70.93458236532199}{100}\\ 
&&&&&  \squart{1}{1}{1}{100}\\ 

        \end{tabular}&
        &
       \begin{tabular}{lcP{0.73cm}P{0.73cm}P{0.3cm}cl}
         \cellcolor[gray]{1}\textbf{} & &\cellcolor[gray]{1}$\mathbfcal{W}$ &\cellcolor[gray]{1}$\mathbfcal{L}$  
            &\cellcolor[gray]{1}$\mathbfcal{T}$ & \cellcolor[gray]{1}\textbf{Quantiles}  \\
            \hline
&&-&-&-&  \quart{20.31745370232359}{79.68254629767641}{63.603238006396666}{100}\\ 
&&\cellcolor{steel!10}\textbf{12(12)}&\cellcolor{steel!10}0(0)&\cellcolor{steel!10}6&  \quart{0.0}{65.4009079643437}{49.99999999999997}{100}\\  
&&\cellcolor{steel!10}\textbf{12(12)}&\cellcolor{steel!10}0(0)&\cellcolor{steel!10}6& \quart{0.0}{65.4009079643437}{49.99999999999997}{100}\\  &&&&&  \squart{1}{1}{1}{100}\\ 

        \end{tabular}  \\
            (f). SCT: \textsc{wc-c1-6d}&& (g). SCT: \textsc{llvm} && (h). SCT: \textsc{trimesh}&& (i). WSC: \textsc{5as-1} && (j). WSC: \textsc{5as-2}
            
            \\
            \\
            
             \begin{tabular}{llP{0.73cm}P{0.73cm}P{0.3cm}c}
             \cellcolor[gray]{1}\textbf{} & &\cellcolor[gray]{1}$\mathbfcal{W}$ &\cellcolor[gray]{1}$\mathbfcal{L}$  
            &\cellcolor[gray]{1}$\mathbfcal{T}$ & \cellcolor[gray]{1}\textbf{Quantiles}  \\
            \hline

W$_{best}$&&-&-&-&   \quart{0.0}{100.00000000000001}{84.08008315852828}{100}\\ 
NSGA-II&&\cellcolor{steel!10}\textbf{10(10)}&\cellcolor{steel!10}0(0)&\cellcolor{steel!10}8&  \quart{0.0}{88.72985704424919}{50.000000000000014}{100}\\ 
MOEA/D&&\cellcolor{steel!10}\textbf{10(10)}&\cellcolor{steel!10}0(0)&\cellcolor{steel!10}8&  \quart{0.0}{88.72985704424919}{50.000000000000014}{100}\\  &&&&&  \squart{1}{1}{1}{100}\\ 

        \end{tabular} & 
        &
       \begin{tabular}{lcP{0.73cm}P{0.73cm}P{0.3cm}cl}
         \cellcolor[gray]{1}\textbf{} & &\cellcolor[gray]{1}$\mathbfcal{W}$ &\cellcolor[gray]{1}$\mathbfcal{L}$  
            &\cellcolor[gray]{1}$\mathbfcal{T}$ & \cellcolor[gray]{1}\textbf{Quantiles}  \\
            \hline
&&-&-&-& \quart{15.729642544723808}{34.27035745527619}{18.94778268377171}{100}\\  
&&\cellcolor{steel!10}\textbf{16(16)}&\cellcolor{steel!10}0(0)&\cellcolor{steel!10}2&   \quart{0.0}{12.868775575368625}{10.007908606013443}{100}\\ 
&&\cellcolor{red!10}6(6)&\cellcolor{red!10}12(6)&\cellcolor{red!10}0&  \quart{7.2909894686511985}{92.7090105313488}{50.0}{100}\\  &&&&&  \squart{1}{1}{1}{100}\\ 

        \end{tabular} &
        &
       \begin{tabular}{lcP{0.73cm}P{0.73cm}P{0.3cm}cl}
         \cellcolor[gray]{1}\textbf{} & &\cellcolor[gray]{1}$\mathbfcal{W}$ &\cellcolor[gray]{1}$\mathbfcal{L}$  
            &\cellcolor[gray]{1}$\mathbfcal{T}$ & \cellcolor[gray]{1}\textbf{Quantiles}  \\
            \hline
&&-&-&-&   \quart{3.6115400743751955}{96.3884599256248}{63.98826580754962}{100}\\
&&\cellcolor{steel!10}\textbf{12(10)}&\cellcolor{steel!10}0(0)&\cellcolor{steel!10}6&  \quart{0.0}{67.5269274758966}{38.05197580912431}{100}\\ 
&&\cellcolor{steel!10}\textbf{12(10)}&\cellcolor{steel!10}6(0)&\cellcolor{steel!10}0&   \quart{0.0}{67.5269274758966}{38.05197580912431}{100}\\ &&&&&  \squart{1}{1}{1}{100}\\ 
      \end{tabular} &
        &
       \begin{tabular}{lcP{0.73cm}P{0.73cm}P{0.3cm}cl}
         \cellcolor[gray]{1}\textbf{} & &\cellcolor[gray]{1}$\mathbfcal{W}$ &\cellcolor[gray]{1}$\mathbfcal{L}$  
            &\cellcolor[gray]{1}$\mathbfcal{T}$ & \cellcolor[gray]{1}\textbf{Quantiles}  \\
            \hline
&&-&-&-&  \quart{23.510923145377856}{76.48907685462216}{38.87672304832789}{100}\\ 
&&\cellcolor{steel!10}\textbf{16(16)}&\cellcolor{steel!10}0(0)&\cellcolor{steel!10}2&  \quart{0.0}{40.992973756904206}{25.99213373260572}{100}\\  
&&\cellcolor{steel!10}\textbf{16(16)}&\cellcolor{steel!10}2(0)&\cellcolor{steel!10}0&  \quart{0.0}{40.992973756904206}{25.99213373260572}{100}\\  &&&&&  \squart{1}{1}{1}{100}\\ 

        \end{tabular}&
        &
       \begin{tabular}{lcP{0.73cm}P{0.73cm}P{0.3cm}cl}
         \cellcolor[gray]{1}\textbf{} & &\cellcolor[gray]{1}$\mathbfcal{W}$ &\cellcolor[gray]{1}$\mathbfcal{L}$  
            &\cellcolor[gray]{1}$\mathbfcal{T}$ & \cellcolor[gray]{1}\textbf{Quantiles}  \\
            \hline
&&-&-&-&  \quart{15.729642544723822}{34.27035745527618}{18.94778268377171}{100}\\ 
&&\cellcolor{steel!10}\textbf{16(16)}&\cellcolor{steel!10}2(0)&\cellcolor{steel!10}0&  \quart{0.0}{14.299209060046243}{10.007908606013471}{100}\\
&&\cellcolor{steel!10}\textbf{10(10)}&\cellcolor{steel!10}8(6)&\cellcolor{steel!10}0&   \quart{7.2909894686511985}{92.7090105313488}{50.0}{100}\\&&&&&  \squart{1}{1}{1}{100}\\ 

        \end{tabular}  \\
            (k). WSC: \textsc{5as-3}&& (l). WSC: \textsc{10as-1} && (m). WSC: \textsc{10as-2}&& (n). WSC: \textsc{10as-3} && (o). WSC: \textsc{15as-1}
            
            \\
            \\
            
                         \begin{tabular}{llP{0.73cm}P{0.73cm}P{0.3cm}c}
             \cellcolor[gray]{1}\textbf{} & &\cellcolor[gray]{1}$\mathbfcal{W}$ &\cellcolor[gray]{1}$\mathbfcal{L}$  
            &\cellcolor[gray]{1}$\mathbfcal{T}$ & \cellcolor[gray]{1}\textbf{Quantiles}  \\
            \hline

W$_{best}$&&-&-&-&   \quart{0.0}{100.0}{50.000000000000014}{100}\\ 
NSGA-II&&\cellcolor{red!10}2(0)&\cellcolor{red!10}16(0)&\cellcolor{red!10}0&  \quart{0.0}{100.0}{50.000000000000014}{100}\\ 
MOEA/D&&\cellcolor{red!10}1(0)&\cellcolor{red!10}16(0)&\cellcolor{red!10}1&  \quart{0.0}{100.0}{50.000000000000014}{100}\\  &&&&&  \squart{1}{1}{1}{100}\\

        \end{tabular} & 
        &
       \begin{tabular}{lcP{0.73cm}P{0.73cm}P{0.3cm}cl}
         \cellcolor[gray]{1}\textbf{} & &\cellcolor[gray]{1}$\mathbfcal{W}$ &\cellcolor[gray]{1}$\mathbfcal{L}$  
            &\cellcolor[gray]{1}$\mathbfcal{T}$ & \cellcolor[gray]{1}\textbf{Quantiles}  \\
            \hline
&&-&-&-&  \quart{0.0}{100.0}{62.53384108803873}{100}\\ 
&&\cellcolor{red!10}\textbf{7(6)}&\cellcolor{red!10}11(0)&\cellcolor{red!10}0&  \quart{0.0}{84.39600341616747}{62.53384108803873}{100}\\ 
&&\cellcolor{steel!10}\textbf{9(6)}&\cellcolor{steel!10}9(0)&\cellcolor{steel!10}0&  \quart{0.0}{84.2945627111433}{62.53384108803873}{100}\\  &&&&&  \squart{1}{1}{1}{100}\\

        \end{tabular} &
        &
       \begin{tabular}{lcP{0.73cm}P{0.73cm}P{0.3cm}cl}
         \cellcolor[gray]{1}\textbf{} & &\cellcolor[gray]{1}$\mathbfcal{W}$ &\cellcolor[gray]{1}$\mathbfcal{L}$  
            &\cellcolor[gray]{1}$\mathbfcal{T}$ & \cellcolor[gray]{1}\textbf{Quantiles}  \\
            \hline
&&-&-&-&  \quart{0.0}{100.0}{70.7693722804753}{100}\\ 
&&\cellcolor{steel!10}\textbf{11(10)}&\cellcolor{steel!10}7(6)&\cellcolor{steel!10}0&  \quart{30.638816710274803}{64.69060013027794}{62.521661110875016}{100}\\
&&\cellcolor{steel!10}\textbf{11(9)}&\cellcolor{steel!10}7(7)&\cellcolor{steel!10}0&  \quart{30.965061969711215}{63.68631616551249}{63.41653525211633}{100}\\  &&&&&  \squart{1}{1}{1}{100}\\

      \end{tabular} &
        &
       \begin{tabular}{lcP{0.73cm}P{0.73cm}P{0.3cm}cl}
         \cellcolor[gray]{1}\textbf{} & &\cellcolor[gray]{1}$\mathbfcal{W}$ &\cellcolor[gray]{1}$\mathbfcal{L}$  
            &\cellcolor[gray]{1}$\mathbfcal{T}$ & \cellcolor[gray]{1}\textbf{Quantiles}  \\
            \hline
&&-&-&-&  \quart{30.965061969711215}{63.68631616551249}{63.41653525211633}{100}\\ 
&&\cellcolor{steel!10}\textbf{16(15)}&\cellcolor{steel!10}2(2)&\cellcolor{steel!10}0&  \quart{0.0}{40.42570825691311}{25.559355276075046}{100}\\ 
&&\cellcolor{steel!10}\textbf{15(14)}&\cellcolor{steel!10}3(2)&\cellcolor{steel!10}0&  \quart{4.325366053417177}{38.22470084960441}{29.047190884802138}{100}\\ &&&&&  \squart{1}{1}{1}{100}\\

        \end{tabular}&
        &
       \begin{tabular}{lcP{0.73cm}P{0.73cm}P{0.3cm}cl}
         \cellcolor[gray]{1}\textbf{} & &\cellcolor[gray]{1}$\mathbfcal{W}$ &\cellcolor[gray]{1}$\mathbfcal{L}$  
            &\cellcolor[gray]{1}$\mathbfcal{T}$ & \cellcolor[gray]{1}\textbf{Quantiles}  \\
            \hline
&&-&-&-&  \quart{32.83360009253271}{67.16639990746728}{88.24896914662551}{100}\\ 
&&\cellcolor{steel!10}\textbf{16(16)}&\cellcolor{steel!10}2(2)&\cellcolor{steel!10}0&   \quart{3.1061958217611685}{29.64492872360162}{22.31088712674834}{100}\\ 
&&\cellcolor{steel!10}\textbf{16(16)}&\cellcolor{steel!10}2(2)&\cellcolor{steel!10}0&  \quart{0.0}{30.512462699784894}{19.04381833384292}{100}\\  &&&&&  \squart{1}{1}{1}{100}\\

        \end{tabular}  \\
            (p). WSC: \textsc{15as-2}&& (q). WSC: \textsc{15as-3} && (r). WSC: \textsc{50as}&& (s). NRP: \textsc{nrp-e1} && (t). NRP: \textsc{nrp-e2}
\\
\\

           \begin{tabular}{llP{0.73cm}P{0.73cm}P{0.3cm}c}
             \cellcolor[gray]{1}\textbf{} & &\cellcolor[gray]{1}$\mathbfcal{W}$ &\cellcolor[gray]{1}$\mathbfcal{L}$  
            &\cellcolor[gray]{1}$\mathbfcal{T}$ & \cellcolor[gray]{1}\textbf{Quantiles}  \\
            \hline

W$_{best}$&&-&-&-&  \quart{6.920048710619705}{93.07995128938029}{62.46183343326794}{100}\\
NSGA-II&&\cellcolor{steel!10}\textbf{18(18)}&\cellcolor{steel!10}0(0)&\cellcolor{steel!10}0&  \quart{0.0}{56.50795455831963}{36.340785236776}{100}\\ 
MOEA/D&&\cellcolor{steel!10}\textbf{18(18)}&\cellcolor{steel!10}0(0)&\cellcolor{steel!10}0&  \quart{0.0}{56.07539724667316}{36.340785236776}{100}\\  &&&&&  \squart{1}{1}{1}{100}\\

        \end{tabular} & 
        &
       \begin{tabular}{lcP{0.73cm}P{0.73cm}P{0.3cm}cl}
         \cellcolor[gray]{1}\textbf{} & &\cellcolor[gray]{1}$\mathbfcal{W}$ &\cellcolor[gray]{1}$\mathbfcal{L}$  
            &\cellcolor[gray]{1}$\mathbfcal{T}$ & \cellcolor[gray]{1}\textbf{Quantiles}  \\
            \hline
&&-&-&-&  \quart{46.59729063120105}{53.40270936879895}{79.66055514338136}{100}\\ 
&&\cellcolor{steel!10}\textbf{16(14)}&\cellcolor{steel!10}2(2)&\cellcolor{steel!10}0&   \quart{2.7426945114962353}{43.50023787700335}{32.26364924291622}{100}\\ 
&&\cellcolor{steel!10}\textbf{16(16)}&\cellcolor{steel!10}2(2)&\cellcolor{steel!10}0&  \quart{0.0}{39.88683028414246}{29.03089855157857}{100}\\  &&&&&  \squart{1}{1}{1}{100}\\

        \end{tabular} &
        &
       \begin{tabular}{lcP{0.73cm}P{0.73cm}P{0.3cm}cl}
         \cellcolor[gray]{1}\textbf{} & &\cellcolor[gray]{1}$\mathbfcal{W}$ &\cellcolor[gray]{1}$\mathbfcal{L}$  
            &\cellcolor[gray]{1}$\mathbfcal{T}$ & \cellcolor[gray]{1}\textbf{Quantiles}  \\
            \hline
&&-&-&-&   \quart{13.319297683647273}{86.68070231635272}{85.17210754527957}{100}\\ 
&&\cellcolor{steel!10}\textbf{18(18)}&\cellcolor{steel!10}0(0)&\cellcolor{steel!10}0&  \quart{0.0}{65.92803332429138}{42.06789891414017}{100}\\ 
&&\cellcolor{steel!10}\textbf{18(18)}&\cellcolor{steel!10}0(0)&\cellcolor{steel!10}0&  \quart{0.0}{65.20541398438753}{42.06789891414017}{100}\\  &&&&&  \squart{1}{1}{1}{100}\\

      \end{tabular} &
        &
       \begin{tabular}{lcP{0.73cm}P{0.73cm}P{0.3cm}cl}
         \cellcolor[gray]{1}\textbf{} & &\cellcolor[gray]{1}$\mathbfcal{W}$ &\cellcolor[gray]{1}$\mathbfcal{L}$  
            &\cellcolor[gray]{1}$\mathbfcal{T}$ & \cellcolor[gray]{1}\textbf{Quantiles}  \\
            \hline
&&-&-&-&  \quart{48.62843253803415}{51.37156746196585}{73.19710991960542}{100}\\ 
&&\cellcolor{steel!10}\textbf{18(18)}&\cellcolor{steel!10}0(0)&-
\cellcolor{steel!10}0&    \quart{0.0}{39.30905217174939}{25.53092433947371}{100}\\ 
&&\cellcolor{steel!10}\textbf{16(16)}&\cellcolor{steel!10}2(1)&\cellcolor{steel!10}0&  \quart{4.543997869101188}{34.765054302648196}{27.315987544208514}{100}\\ &&&&&  \squart{1}{1}{1}{100}\\ 

        \end{tabular}&
        &
       \begin{tabular}{lcP{0.73cm}P{0.73cm}P{0.3cm}cl}
         \cellcolor[gray]{1}\textbf{} & &\cellcolor[gray]{1}$\mathbfcal{W}$ &\cellcolor[gray]{1}$\mathbfcal{L}$  
            &\cellcolor[gray]{1}$\mathbfcal{T}$ & \cellcolor[gray]{1}\textbf{Quantiles}  \\
            \hline

&&-&-&-&  \quart{41.43226930185585}{58.56773069814415}{83.91828691591101}{100}\\
&&\cellcolor{steel!10}\textbf{16(16)}&\cellcolor{steel!10}2(2)&\cellcolor{steel!10}0&  \quart{0.0}{42.939529464869004}{29.149254577690396}{100}\\ 
&&\cellcolor{steel!10}\textbf{16(15)}&\cellcolor{steel!10}2(2)&\cellcolor{steel!10}0&  \quart{0.3771372080138106}{46.91808025620484}{32.69201893004855}{100}\\  &&&&&  \squart{1}{1}{1}{100}\\

        \end{tabular}  \\
            (u). NRP: \textsc{nep-e3}&& (v). NRP: \textsc{nrp-e4} && (w). NRP: \textsc{nrp-g1}&& (x). NRP: \textsc{nrp-g2} && (y). NRP: \textsc{nrp-g3}
\\
\\

           \begin{tabular}{llP{0.73cm}P{0.73cm}P{0.3cm}c}
             \cellcolor[gray]{1}\textbf{} & &\cellcolor[gray]{1}$\mathbfcal{W}$ &\cellcolor[gray]{1}$\mathbfcal{L}$  
            &\cellcolor[gray]{1}$\mathbfcal{T}$ & \cellcolor[gray]{1}\textbf{Quantiles}  \\
            \hline
W$_{best}$&&-&-&-&   \quart{10.151792699162305}{89.84820730083769}{82.15996777077831}{100}\\ 
NSGA-II&&\cellcolor{steel!10}\textbf{12(12)}&\cellcolor{steel!10}6(6)&\cellcolor{steel!10}0&  \quart{1.8868977410485386}{64.51439751446374}{40.5425290197043}{100}\\ 
MOEA/D&&\cellcolor{steel!10}\textbf{12(12)}&\cellcolor{steel!10}6(6)&\cellcolor{steel!10}0&  \quart{0.0}{54.37216531864449}{37.7178676762304}{100}\\ 

&&&&&  \squart{1}{1}{1}{100}\\

        \end{tabular} & 
        &
       \begin{tabular}{lcP{0.73cm}P{0.73cm}P{0.3cm}cl}
         \cellcolor[gray]{1}\textbf{} & &\cellcolor[gray]{1}$\mathbfcal{W}$ &\cellcolor[gray]{1}$\mathbfcal{L}$  
            &\cellcolor[gray]{1}$\mathbfcal{T}$ & \cellcolor[gray]{1}\textbf{Quantiles}  \\
            \hline
&&-&-&-&   \quart{24.402661162616496}{75.59733883738349}{75.32027130724813}{100}\\ 
&&\cellcolor{steel!10}\textbf{17(16)}&\cellcolor{steel!10}1(0)&\cellcolor{steel!10}0&   \quart{0.0}{21.277499275146614}{13.11202152317879}{100}\\ 
&&\cellcolor{steel!10}\textbf{16(16)}&\cellcolor{steel!10}2(2)&\cellcolor{steel!10}0&  \quart{2.3531421520813076}{18.489456666381955}{13.533814407458978}{100}\\

&&&&&  \squart{1}{1}{1}{100}\\

        \end{tabular} &
        &
       \begin{tabular}{lcP{0.73cm}P{0.73cm}P{0.3cm}cl}
         \cellcolor[gray]{1}\textbf{} & &\cellcolor[gray]{1}$\mathbfcal{W}$ &\cellcolor[gray]{1}$\mathbfcal{L}$  
            &\cellcolor[gray]{1}$\mathbfcal{T}$ & \cellcolor[gray]{1}\textbf{Quantiles}  \\
            \hline
&&-&-&-&   \quart{24.402661162616496}{75.59733883738349}{75.32027130724813}{100}\\ 
&&\cellcolor{steel!10}\textbf{18(18)}&\cellcolor{steel!10}0(0)&\cellcolor{steel!10}0&  \quart{0.0}{21.277499275146614}{13.11202152317879}{100}\\ 
&&\cellcolor{steel!10}\textbf{16(16)}&\cellcolor{steel!10}2(2)&\cellcolor{steel!10}0&  \quart{2.3531421520813076}{18.489456666381955}{13.533814407458978}{100}\\
&&&&&  \squart{1}{1}{1}{100}\\ 
      \end{tabular} &
        &
       \begin{tabular}{lcP{0.73cm}P{0.73cm}P{0.3cm}cl}
         \cellcolor[gray]{1}\textbf{} & &\cellcolor[gray]{1}$\mathbfcal{W}$ &\cellcolor[gray]{1}$\mathbfcal{L}$  
            &\cellcolor[gray]{1}$\mathbfcal{T}$ & \cellcolor[gray]{1}\textbf{Quantiles}  \\
            \hline
&&-&-&-&  \quart{7.400698725886091}{92.59930127411391}{72.95697588816797}{100}\\ 
&&\cellcolor{steel!10}\textbf{18(18)}&\cellcolor{steel!10}0(0)&\cellcolor{steel!10}0&   \quart{0.0}{46.69818687520137}{27.702542397093794}{100}\\ 
&&\cellcolor{steel!10}\textbf{18(18)}&\cellcolor{steel!10}0(0)&\cellcolor{steel!10}0&  \quart{0.0}{46.58302772213142}{27.702542397093794}{100}\\ 
&&&&&  \squart{1}{1}{1}{100}\\ 
        \end{tabular}&
        &
       \begin{tabular}{lcP{0.73cm}P{0.73cm}P{0.3cm}cl}
         \cellcolor[gray]{1}\textbf{} & &\cellcolor[gray]{1}$\mathbfcal{W}$ &\cellcolor[gray]{1}$\mathbfcal{L}$  
            &\cellcolor[gray]{1}$\mathbfcal{T}$ & \cellcolor[gray]{1}\textbf{Quantiles}  \\
            \hline

&&-&-&-&  \quart{48.83113727494124}{51.16886272505875}{86.64892632351008}{100}\\ 
&&\cellcolor{steel!10}\textbf{16(16)}&\cellcolor{steel!10}2(2)&\cellcolor{steel!10}0&  \quart{0.9015579168067611}{22.04550313087328}{15.41973666811713}{100}\\ 
&&\cellcolor{steel!10}\textbf{16(16)}&\cellcolor{steel!10}2(2)&\cellcolor{steel!10}0&  \quart{0.0}{20.244989916514292}{13.428781044430545}{100}\\  &&&&&  \squart{1}{1}{1}{100}\\

        \end{tabular}

        \\

            (z). NRP: \textsc{nep-g4}&& (aa). NRP: \textsc{nrp-m1} && (bb). NRP: \textsc{nrp-m2}&& (cc). NRP: \textsc{nrp-m3} && (dd). NRP: \textsc{nrp-m4}
     
    \end{tabular}
  \end{adjustbox}
   \end{center}
\end{table*}

\begin{table*}[t!]
\caption{Pairwise comparisons (by $\hat{A}_{12}$ and Wilcoxon rank-sum test) and the weighted scores for Pareto and weighted search on three objectives SBSE problems. A pair summarizes 8 cases (4 weight vectors $\times$ 2 types of budget) on 100 runs each. Formats are the same as Table~\ref{tb:rq2}.}
    \label{tb:rq2-3}
    \footnotesize
    \setlength{\tabcolsep}{1mm}
  \begin{center}
    \begin{adjustbox}{max width = \textwidth}

    \begin{tabular}{cc@{}cc@{}cc@{}cc}
        \begin{tabular}{llP{0.73cm}P{0.73cm}P{0.5cm}c}
            \cellcolor[gray]{1}\textbf{} & &\cellcolor[gray]{1}$\mathbfcal{W}$ &\cellcolor[gray]{1}$\mathbfcal{L}$ 
              &\cellcolor[gray]{1}$\mathbfcal{T}$& \cellcolor[gray]{1}\textbf{Quantiles}  \\
            \hline

W$_{best}$&&-&-&-&   \quart{18.30053822550159}{81.69946177449842}{52.95839735880666}{100}\\
NSGA-II&&\cellcolor{steel!10}\textbf{3(3)}&\cellcolor{steel!10}1(0)&\cellcolor{steel!10}4&   \quart{0.0}{59.5277189336304}{21.896693330916122}{100}\\ 
MOEA/D&&\cellcolor{steel!10}3(1)&\cellcolor{steel!10}1(1)&\cellcolor{steel!10}4&   \quart{11.744010073276266}{88.25598992672373}{33.131256935435616}{100}\\  &&&&&  \squart{1}{1}{1}{100}\\

        \end{tabular} & 
        &
       \begin{tabular}{lcP{0.73cm}P{0.73cm}P{0.5cm}cl}
                        \cellcolor[gray]{1}\textbf{} & &\cellcolor[gray]{1}$\mathbfcal{W}$ &\cellcolor[gray]{1}$\mathbfcal{L}$   &\cellcolor[gray]{1}$\mathbfcal{T}$& \cellcolor[gray]{1}\textbf{Quantiles}  \\
            \hline
&&-&-&-&  \quart{0.0}{9.400488731488188}{2.3247415703029293}{100}\\ 
&&\cellcolor{red!10}1(0)&\cellcolor{red!10}3(0)&\cellcolor{red!10}4&   \quart{0.48444140225249377}{7.016756816597187}{3.1283841944097452}{100}\\
&&\cellcolor{red!10}0(0)&\cellcolor{red!10}4(0)&\cellcolor{red!10}4&    \quart{0.5404454333480704}{99.45955456665192}{5.198784719232289}{100}\\ &&&&&  \squart{1}{1}{1}{100}\\

        \end{tabular} &
        &
       \begin{tabular}{lcP{0.73cm}P{0.73cm}P{0.5cm}cl}
                      \cellcolor[gray]{1}\textbf{} & &\cellcolor[gray]{1}$\mathbfcal{W}$ &\cellcolor[gray]{1}$\mathbfcal{L}$ &\cellcolor[gray]{1}$\mathbfcal{T}$ & \cellcolor[gray]{1}\textbf{Quantiles}  \\
            \hline
&&-&-&-&  \quart{38.0788163813284}{61.9211836186716}{41.9128213120969}{100}\\ 
&&\cellcolor{steel!10}\textbf{6(6)}&\cellcolor{steel!10}0(0)&\cellcolor{steel!10}2&  \quart{0.0}{71.61359471946008}{41.9128213120969}{100}\\ 
&&\cellcolor{steel!10}\textbf{6(6)}&\cellcolor{steel!10}0(0)&\cellcolor{steel!10}2&  \quart{0.0}{71.61359471946008}{41.9128213120969}{100}\\  &&&&&  \squart{1}{1}{1}{100}\\

        \end{tabular} &
        &
       \begin{tabular}{lcP{0.73cm}P{0.73cm}P{0.5cm}cl}
                        \cellcolor[gray]{1}\textbf{} & &\cellcolor[gray]{1}$\mathbfcal{W}$ &\cellcolor[gray]{1}$\mathbfcal{L}$ &\cellcolor[gray]{1}$\mathbfcal{T}$  & \cellcolor[gray]{1}\textbf{Quantiles}  \\
            \hline
&&-&-&-&  \quart{0.0}{100.0}{16.762910423346792}{100}\\
&&\cellcolor{steel!10}\textbf{8(8)}&\cellcolor{steel!10}0(0)&\cellcolor{steel!10}0&  \quart{0.0}{43.03203557467567}{16.762910423346792}{100}\\ 
&&\cellcolor{steel!10}\textbf{8(8)}&\cellcolor{steel!10}0(0)&\cellcolor{steel!10}0&   \quart{0.0}{43.03203557467567}{16.762910423346792}{100}\\  &&&&&  \squart{1}{1}{1}{100}\\

        \end{tabular}
         \\
         (a). SCT: \textsc{vp8}&& (b). SCT: \textsc{hsqldb} && (c). WSC: \textsc{5as-3o}&& (d). WSC: \textsc{10as-3o}

\\
        \\

           \begin{tabular}{llP{0.73cm}P{0.73cm}P{0.5cm}c}
                \cellcolor[gray]{1}\textbf{} & &\cellcolor[gray]{1}$\mathbfcal{W}$ &\cellcolor[gray]{1}$\mathbfcal{L}$ 
              &\cellcolor[gray]{1}$\mathbfcal{T}$& \cellcolor[gray]{1}\textbf{Quantiles}  \\
            \hline

W$_{best}$&&-&-&-&   \quart{0.0}{100.0}{16.762910423346792}{100}\\  
NSGA-II&&\textbf{4(2)}&4(0)&0&  \quart{0.0}{43.03203557467567}{16.762910423346792}{100}\\ 
MOEA/D&&\cellcolor{red!10}\textbf{3(2)}&\cellcolor{red!10}5(0)&\cellcolor{red!10}0&   \quart{0.0}{43.03203557467567}{16.762910423346792}{100}\\

&&&&&  \squart{1}{1}{1}{100}\\

        \end{tabular} & 
        &
       \begin{tabular}{lcP{0.73cm}P{0.73cm}P{0.5cm}cl}
            \cellcolor[gray]{1}\textbf{} & &\cellcolor[gray]{1}$\mathbfcal{W}$ &\cellcolor[gray]{1}$\mathbfcal{L}$ 
              &\cellcolor[gray]{1}$\mathbfcal{T}$& \cellcolor[gray]{1}\textbf{Quantiles}  \\
            \hline
&&-&-&-& \quart{10.759591001972613}{89.24040899802739}{56.95239316316005}{100}\\ 
&&\cellcolor{steel!10}\textbf{6(4)}&\cellcolor{steel!10}2(1)&\cellcolor{steel!10}0&  \quart{2.3011694326678223}{93.74999906849723}{47.626196697349606}{100}\\
&&\cellcolor{steel!10}\textbf{8(7)}&\cellcolor{steel!10}0(0)&\cellcolor{steel!10}0&  \quart{0.0}{91.31455741362014}{41.860375903019076}{100}\\

&&&&&  \squart{1}{1}{1}{100}\\

        \end{tabular} &
        &
       \begin{tabular}{lcP{0.73cm}P{0.73cm}P{0.5cm}cl}
            \cellcolor[gray]{1}\textbf{} & &\cellcolor[gray]{1}$\mathbfcal{W}$ &\cellcolor[gray]{1}$\mathbfcal{L}$ 
              &\cellcolor[gray]{1}$\mathbfcal{T}$& \cellcolor[gray]{1}\textbf{Quantiles}  \\
            \hline
&&-&-&-&   \quart{1.437591955218987}{70.32125651304491}{33.34064556965155}{100}\\ 
&&\cellcolor{red!10}2(2)&\cellcolor{red!10}\textbf{6(4)}&\cellcolor{red!10}0&     \quart{0.0}{81.52153451721576}{43.8269854049812}{100}\\ 
&&\cellcolor{steel!10}\textbf{5(5)}&\cellcolor{steel!10}3(1)&\cellcolor{steel!10}0&   \quart{1.437591955218987}{98.56240804478101}{28.86527491818925}{100}\\

&&&&&  \squart{1}{1}{1}{100}\\ 
      \end{tabular} &
        &
       \begin{tabular}{lcP{0.73cm}P{0.73cm}P{0.5cm}cl}
            \cellcolor[gray]{1}\textbf{} & &\cellcolor[gray]{1}$\mathbfcal{W}$ &\cellcolor[gray]{1}$\mathbfcal{L}$ 
              &\cellcolor[gray]{1}$\mathbfcal{T}$& \cellcolor[gray]{1}\textbf{Quantiles}  \\
            \hline
&&-&-&-&  \quart{33.139835304596694}{66.86016469540331}{69.32162614653551}{100}\\ 
&&\textbf{4(4)}&4(3)& 0& \quart{10.070365779879168}{75.73729143883996}{68.06726880281305}{100}\\ 
&&\cellcolor{steel!10}\textbf{7(7)}&\cellcolor{steel!10}1(0)&\cellcolor{steel!10}0&  \quart{0.0}{68.1371057926668}{36.80341563903494}{100}\\

&&&&&  \squart{1}{1}{1}{100}\\ 
        \end{tabular}

        \\

              (e). WSC: \textsc{15as-3o}&& (f). NRP: \textsc{nrp-e-3o} && (g). NRP: \textsc{nrp-g-3o}&& (h). NRP: \textsc{nrp-m-3o}

    \end{tabular}
  \end{adjustbox}
   \end{center}
\end{table*}

Of course, whether the quality of solution or the resource required is more important is subject to the stakeholders' preferences and requirements; 
we can, however, suggest the following to the SBSE practitioners:

\begin{quotebox}
   \noindent
   \textit{\textbf{Suggestion:} When the quality of the solution is a primary concern for the multi-objective SBSE problem in hand, use Pareto search by default even if there are clear preferences (weights between the objectives).}
\end{quotebox}

\subsubsection{Reason}

\begin{figure}[t!]
	\centering
	\includegraphics[width=\columnwidth]{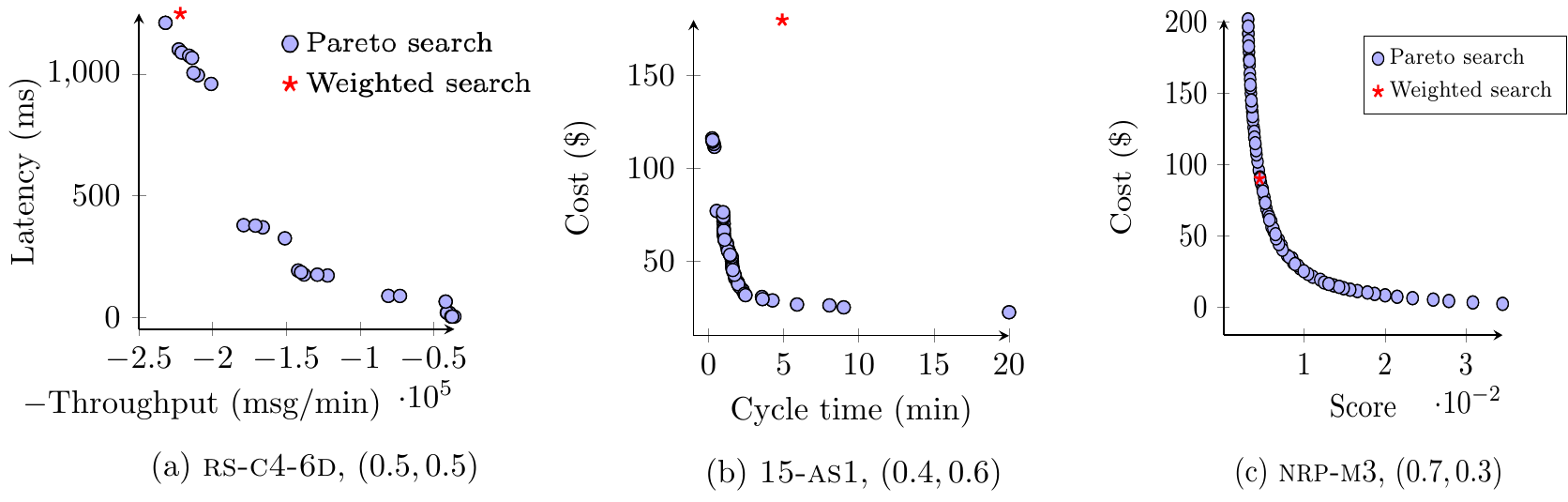}
	\caption{The left (a) and the middle (b) are examples where the result of weighted search 
		is dominated by that of Pareto search; 
		the right (c) is the example wherein a non-dominated solution has been obtained by weighted search, 
		but the weighted fitness is worse than that of Pareto search.}
	\label{fig:dom}
\end{figure}

The above results confirm the theoretical possibility discussed in Section~\ref{sec:theory-reason}.
The diversity of solutions maintained by Pareto search helps to find the global optimum. 
In contrast, 
fine-grained comparability of the scalar fitness in weighted search may be easy to get stuck in local optima. 
especially for SBSE problems wherein the search space is large and/or with many complex local optimal regions. 
Figure~\ref{fig:dom}(a) and (b) give the solutions found by Pareto search and weighted search in a typical run 
on two SCT and WSC cases.
As can be seen,
weighted search not only fails to find the global optimum 
but its solution is actually dominated by some solutions of Pareto search.
This implies that weighted search may stagnate very easily during the search and the solution found is a local optimum.       

Another reason of weighted search performs poorly is that, as discussed in Section~\ref{sec:theory-reason},
the stakeholders' preferences between the objectives may not be well reflected during the search 
since the weights are determined on the basis of the real bounds of the problem, 
whereas the normalization is done on the basis of the bound found during the search. 
This has been evident in Figure~\ref{fig:dom}(c) ---
despite the fact that the solution found by weighted search is on the Pareto front, 
it can differ from the one that the stakeholders are really interested in, 
i.e., the weight vector $(0.7,0.3)$.

In addition,
it is worth mentioning that in a small number of cases (e.g., \textsc{wc-c4-3c} for SCT and \textsc{5as-1} for WSC) 
Pareto search performs significantly worse than weighted search.
The reason for this is due to the ``spread'' search manner of Pareto search.
Pareto search maintains a well-diversified population, 
searching in parallel towards a number of locations 
(i.e., diverse trade-offs over the Pareto front).
Such a spread search manner can be somehow detrimental if the goal of the search is to locate one particular solution on the Pareto front.
For example,
the crossover operation, 
which typically operates on two solutions distant from each other in Pareto search,
is less likely to generate good offspring along with one of their parents' directions \cite{Ishibuchi2003}.
In contrast, 
weighted search has the search focus of the specific direction through 
the crossover between similar parent solutions, 
and is more likely to generate promising solutions along that direction.  
Therefore, 
the weighted search may find better solutions when the search landscape is fairly easy (e.g., without many local optima) and the scale of different objectives as well as their Pareto front ranges is commensurable.



~



\subsection{RQ3: Sensitivity to the Weight Vector}
\label{sec:rq3}

\subsubsection{Method}

Answering \textbf{RQ3} requires us to examine whether the general observations from \textbf{RQ2} would change when looking at each specific weight vector. We do that by following the same settings as discussed in Section~\ref{sec:rq2}, including the same $W_{best}$ for each case.

To provide more meaningful and intuitive illustrations, we report on the percentage gain of one's weighted score over the other:
\begin{equation}
\text{ \% Gain} =
{{1\over n} \times \sum^n_{i=1} {{y_i - x_i} \over {y_i }}} \times 100
\end{equation}
whereby $x_i$ and $y_i$ are the weighted score at the $i$th run for the Pareto search and $W_{best}$, respectively, in which the results of different runs are sorted in ascending order. $n$ is the total number of runs (we have $n=100$ in this work). A negative gain means that the Pareto search is even worse off. 


\subsubsection{Result}

\begin{figure}[t!]
\centering
  \includegraphics[width=\textwidth]{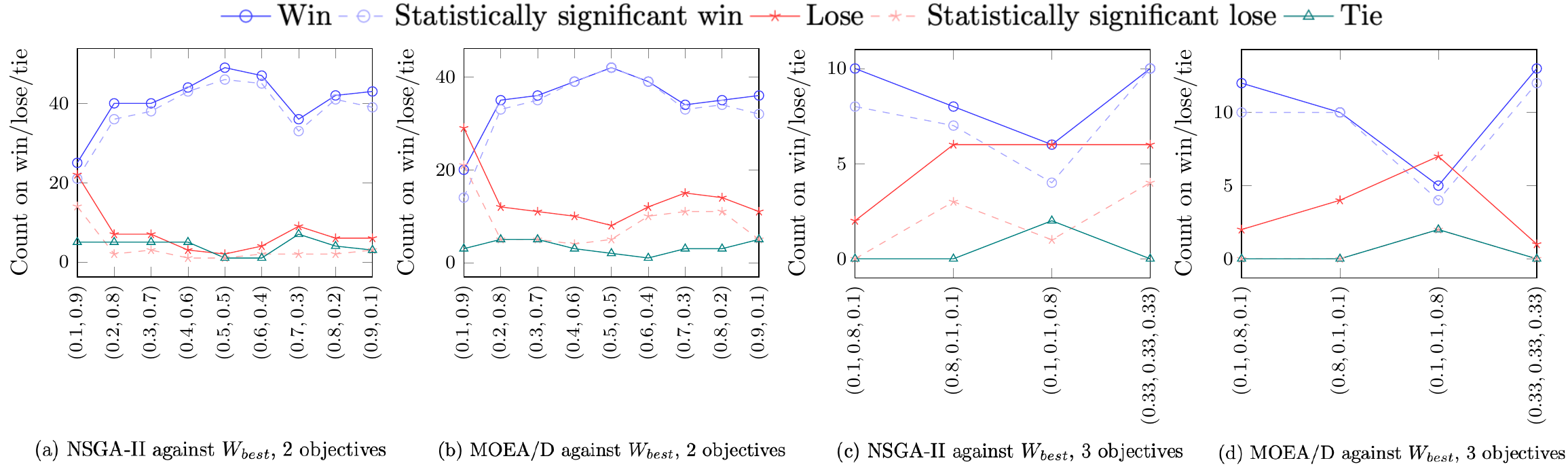}
  \caption{The count on the pairwise comparison results between Pareto search and the best weighted search ($W_{best}$) across the weight vectors. Formats are the same as Figure~\ref{fig:rq2-summary}.}
  \label{fig:rq3-summary}
 \end{figure}

From Figure~\ref{fig:rq3-summary}, we see clearly that the gains achieved by Pareto search (on both NSGA-II and MOEA/D) over its weighted counterpart do fluctuate depending on the weight vector. However, Pareto search remains to win more in general. In particular, the majority of the wins by Pareto search are statistically significant with non-trivial effect size.

\begin{figure}[!t]
  \centering
\includegraphics[width=0.95\textwidth]{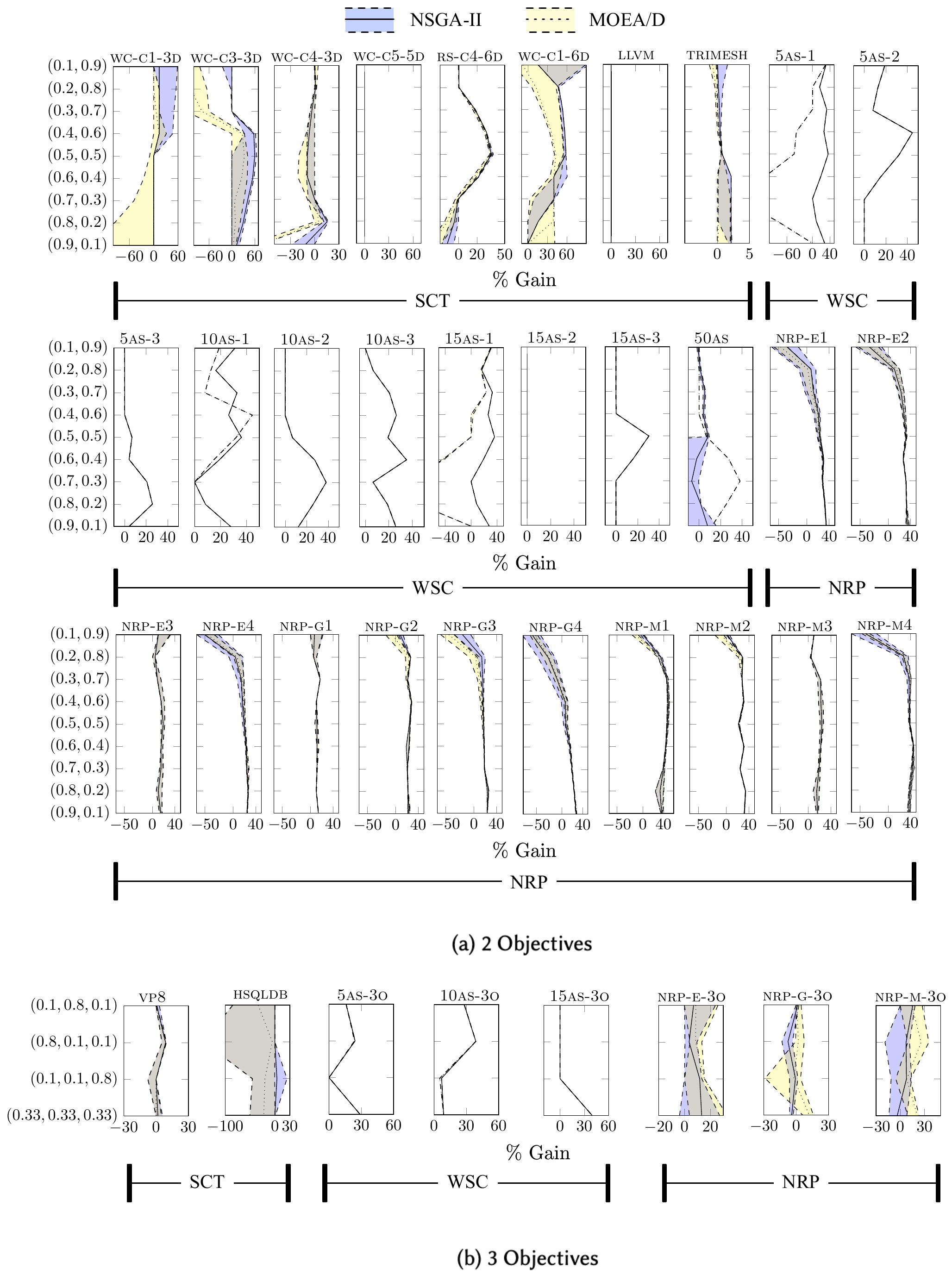}
      \caption{Sensitivity of the 25th, 50th, and 75th percentile gain (\%) of the Pareto search over the best weighted counterpart ($W_{best}$) to different weight vectors. Note that for three objective case, the middle range weight vector lies in the bottom.}
        \label{fig:sen}
  \end{figure}

To better understand the results with respect to the systems/projects and SBSE problems, in Figure~\ref{fig:sen}, we can obtain similar observations with considerably high percentage gain in general. The only case when Pareto search often loses is for NRP under some extreme weight vectors, e.g., $(0.1,0.9)$. We also note that, for both NSGA-II and MOEA/D, the patterns of percentage gains across different systems/projects are more consistent in NRP when compared with SCT and WSC. This makes sense since the systems in the latter two SBSE problems can have a much more radically different nature compared with the project/versions in the former. This is rather clear with SCT where the variations are usually higher. 

The above observations can be seen for both the two and three objective cases.


\begin{quotebox}
   \noindent
   \textit{\textbf{Finding:} The extent to which Pareto search outperforms its weighted counterpart vary depending on the weight vector in the multi-objective SBSE cases. However, the overall result remains the same as we concluded from \textbf{RQ2}.}
\end{quotebox}


\subsubsection{Implication}

We observe from the results that, for each system/project, the best gain achieved by Pareto search is generally centered at a particular range of the weight vectors, mostly around the middle range. For example, in the two objective cases, it can be between $(0.4, 0.6)$ and $(0.6, 0.4)$, for SCT and WSC; or from $(0.3,0.7)$ to $(0.7,0.3)$ for NRP. For three objective case, often the $(0.33,0.33,0.33)$ gives one of the best results. This suggests that different SBSE problems (and their systems/projects) possess different ``comfort zones'' on the given weights for Pareto search to outperform its weighted counterpart. 
However, we see clear patterns on when it is the most difficult for Pareto search to become more beneficial for a system/project, i.e., close to one (or both) extreme weights such as $(0.1,0.9)$, $(0.9,0.1)$, and $(0.1,0.1,0.8)$. 
Of course, this may be asymmetric: on the two objective case, Pareto search may have the lowest gain on $(0.1,0.9)$ while achieving a reasonably well gain on $(0.9,0.1)$ (see NRP), but it does bring it to our attention that one needs to be cautious 
when the given weight vector is close to the edge. As a result, we can make the following advice:

\begin{quotebox}
   \noindent
   \textit{\textbf{Suggestion:} In multi-objective SBSE, 
   	when the given weight vector is close to one extreme  e.g., $(0.1,0.9)$ and $(0.2,0.8)$, 
   	it is useful to further experimentally confirm the benefits of Pareto search in some preliminary runs.}
\end{quotebox}

\subsubsection{Reason}

One reason that Pareto search under extreme weight vectors may not be as effective as under other weight vectors is that 
it can be hard for Pareto search to search for boundary solutions of the Pareto front.
Compared to trade-off solutions (i.e., more central part on the Pareto front), 
on certain multi-objective problems, 
the boundary solutions can be very difficult to be found. 
They may be located in a region that is on the edge of search space, 
far away from the randomly generated initial population.
Finding them needs the focus of the search toward their location.
However,
Pareto search maintains a well-diversified population, 
searching in parallel towards a number of locations (i.e., diverse trade-offs over the Pareto front).
To be more specific,
the crossover operation, 
which typically operates on two solutions distant from each other in Pareto search,
is less likely to produce good offspring along with one of their parents' directions.
In contrast, 
weighted search has the search focus of the specific direction through the crossover between similar parent solutions.

In addition, 
there is another reason, 
related to the search algorithm/optimizer itself, 
that Pareto search may not be that advantageous when the weights are closer to the extremes.
Most multi-objective optimizers do not have a mechanism to preserve boundary solutions, 
such as MOEA/D~\cite{DBLP:journals/tec/ZhangL07} and NSGA-III~\cite{Deb2014}.
That means that even if a boundary solution is generated during the search process,
it may still be eliminated later. 
This is particularly true when the problem's Pareto front is convex 
(such as the WSC as shown in Figure~\ref{fig:dom}b),
where the boundary solutions can be seen ``worse'' than internal solutions in terms of convergence,
thus directly being eliminated by the algorithm's selection criterion
such as $\epsilon$-dominance~\cite{Laumanns2002}, grid ranking~\cite{Yang2013} criteria and shift-based density estimation criterion~\cite{Li2014}. 
It is worth noting that this reason does not apply to the results obtained from NSGA-II since it has an explicit boundary solutions preservation mechanism and 
all nondominated solutions are incomparable in terms of convergence~\cite{Deb2002}. 
However, 
it may apply to other multi-objective optimizers such as MOEA/D. This is why, as shown in Figure~\ref{fig:rq3-summary}a and Figure~\ref{fig:rq3-summary}b, the MOEA/D tends to lose more to the weighted search compared with that of the NSGA-II when the weight vector is close to one extreme.

Lastly, 
it is worth mentioning that the fact that 
Pareto search may not work well in finding the boundary solutions is not alone in the SBSE area. 
Similar observations have been found on generic multi-objective problems in the evolutionary computation area~\cite{Ishibuchi2007,Wang2018}.


  \begin{figure}[!t]
  \centering
\includegraphics[width=\columnwidth]{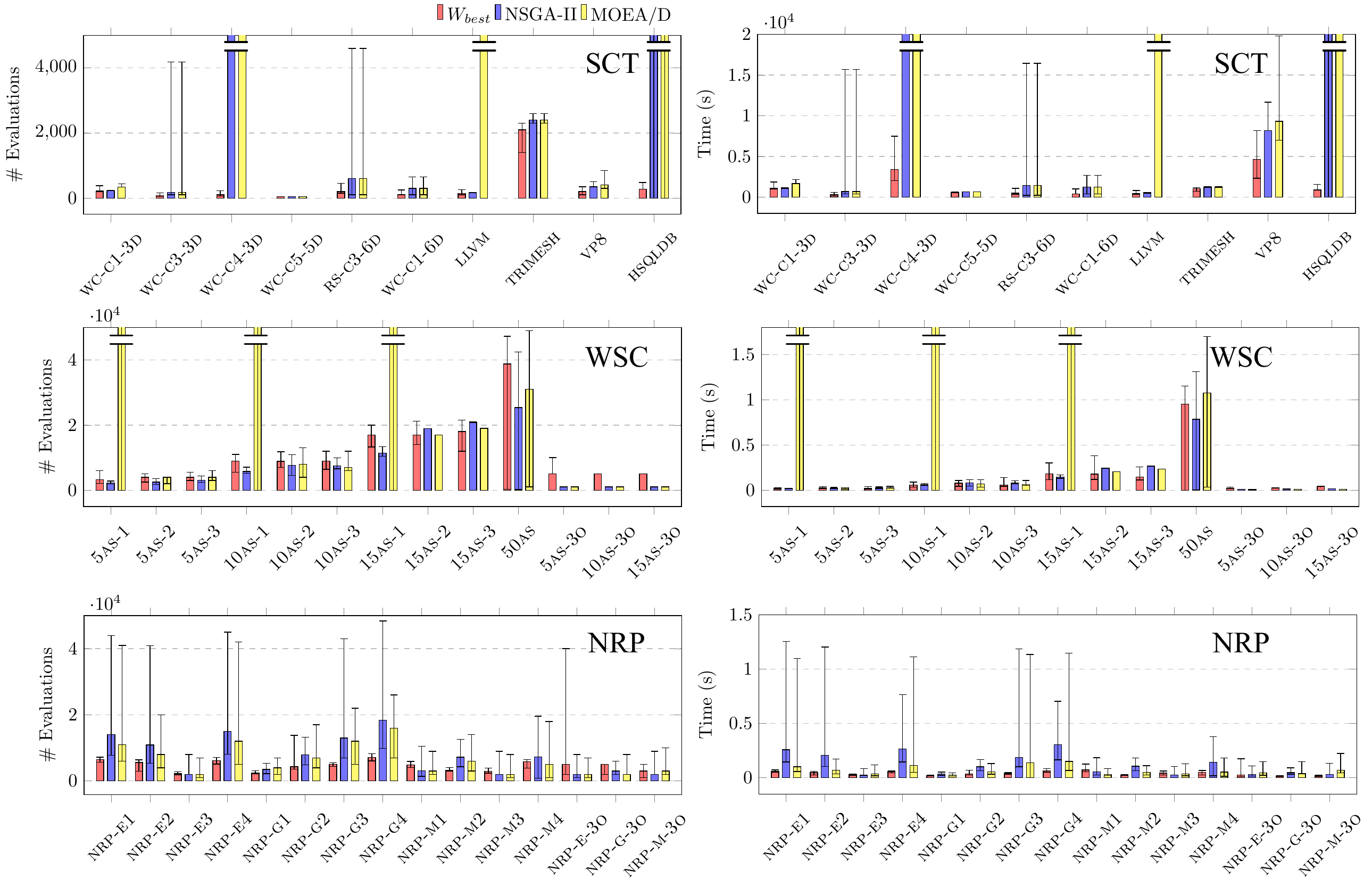}
      \caption{The 25th, 50th, and 75th percentiles (across different weight vectors) of the resources consumed to reach (or outperform) the best median weighted score achieved by $W_{best}$. The broken bar denotes that the Pareto search cannot achieve the same results when the search budget is exhausted.}
        \label{fig:res}
  \end{figure}

\subsection{RQ4: Resource Efficiency}
\label{sec:rq4}

\subsubsection{Method}

To investigate \textbf{RQ4}, we use the same $W_{best}$ for each case from \textbf{RQ2}. Specifically, in each case, we measure the resource efficiency in terms of the evaluations/time used to reach a certain level as:

\begin{enumerate}
    \item Identify a baseline, $b$, taken as the smallest amount of search budget (evaluations and time) that $W_{best}$ consumes to achieve its best median result over 100 runs (says $T$).
    \item For each of the optimizers studied, find the smallest number of evaluations (and amount of time), $m$, at which the median weighted score (over 100 runs) is equivalent to or better than  $T$. 
    \item Report $m$ and $b$ across the different weight vectors for each system/project.
\end{enumerate}


\subsubsection{Result}

As shown from Figure~\ref{fig:res}, we see that the Pareto search (both NSGA-II and MOEA/D) tends to be more resource-efficient on most system workflows in WSC, when using the number of evaluations as the search budget. However, the benefit remains unclear as the variations across different weight vectors remain high. In most of the remaining cases, the weighted search often consumes much less resources at the 25th, 50th, and 75th percentiles. This is particularly true for both SCT and NRP, where the same trends have been observed across the majority of the systems/projects and types of search budget.

The above observations still hold for both two and three objective cases.


\begin{quotebox}
   \noindent
   \textit{\textbf{Finding:} Weighted search is often more resource-efficient for multi-objective SBSE.}
\end{quotebox}

\subsubsection{Implication}

The observation for \textbf{RQ4} is an interesting one: it reveals that weighted search does have its advantages over Pareto search; that is, it allows to converge to its best results with less resources. Yet, it is worth noting that, if we allow the search to continue, Pareto search would often reach a degree of quality that the best weighted counterpart would have never been able to achieve. 
However, the fact that weighed search is more resource-efficient in reaching its best is more preferable to some contexts under a limited search budget, meaning that the \textit{weighted search first} belief is not entirely meaningless. We, therefore, suggest the following:

\begin{quotebox}
   \noindent
   \textit{\textbf{Suggestion:} When the resource efficiency (search budget) is a more important factor for the multi-objective SBSE problem in hand, sticking with the existing belief to use weighted search.}
\end{quotebox}

\subsubsection{Reason}

The reason that weighted search appears to be more efficient than Pareto search is easy to understand. 
Weighted search is conducted for the optimizer seeking the exclusive target (point) in the space based on the given weight vector, 
whereas Pareto search is conducted for the optimizer seeking the entire Pareto front,
thus a waste of lots of recourse on solutions irrelevant to the weight vector. 


In addition, 
a secondary reason for weighted search being less time consuming
is that the fitness comparison between solutions in the population in weighted search requires $O(mn)$ computations 
($m$ is the number of objectives and $n$ is the population size of the optimizer),  
less than that of NSGA-II in which the Pareto-based fitness comparison requires $O(mn^2)$ computations 
and that of MOEA/D in which the neighborhood-based fitness comparison requires $O(kmn)$ computations, 
where $k$ is the neighborhood size. 

\section{A Pragmatic Guidance}
\label{sec:guide}

Drawing on the findings for our RQs, it is clear that a \textit{``weighted search first''} belief under clear preferences can be harmful to SBSE. A more systematic justification is, therefore, required for making such an engineering decision. To that end, we codify pragmatic guidance that outlines the key processes.

\begin{figure*}[!t]
  \centering
\includegraphics[width=\textwidth]{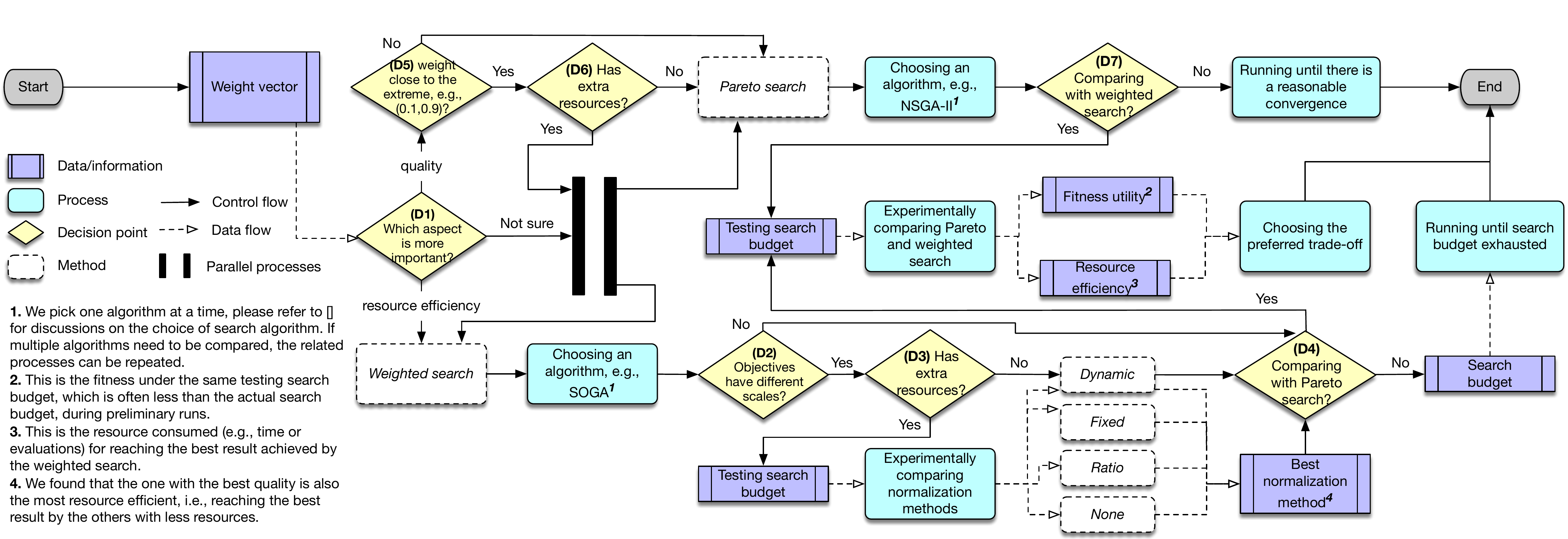}
      \caption{A pragmatic guidance of how to choose between Pareto and weighted search under clear preference of weights in multi-objective SBSE.}
        \label{fig:guide}
  \end{figure*}

As shown in Figure~\ref{fig:guide}, the guidance starts by asking the stakeholders to feed a known weight vector into the first decision point (\textbf{D1}). Here, one can decide on whether the quality is more important than the resource taken, or vice versus. This is important, as different SBSE problems and contexts may impose different preferences.

If the resource efficiency (e.g., time or evaluations) is deemed as more important (choosing \textit{resource efficiency} at \textbf{D1}), e.g., for SBSE problems like SCT where a single evaluation can be rather expensive, we recommend staying with the classic weighted search (based on \textbf{RQ4}). We can then choose an appropriate single-objective algorithm, based on theoretical or empirical understanding (Details of this can be found in~\cite{Harman2012,DBLP:journals/tse/HarmanM10,DBLP:journals/infsof/BagnallRW01}, thus we do not cover this here). Next, in \textbf{D2}, we check whether the objectives need to be normalized for making them commensurable. 
If the objectives are commensurable in nature, we proceed to \textbf{D4}, and since we certainly prefer weighted search which performs reasonably better (than Pareto search) in terms of resource efficiency during the search, the process could end. Note that here, the search needs to be completed under a given search budget as resource consumption is more important. 

The above is an ideal case, however, it is likely that most SBSE problems have objectives with radically different scales. Therefore, as revealed by \textbf{RQ1}, the normalization method can play an integral role in such a case. In \textbf{D3}, we ask if there are extra resources that allow for a thorough comparison between normalization methods. If there are not, we suggest using \texttt{Dynamic}, which has been shown to be generally safe in \textbf{RQ1}. Otherwise, experimental comparison among the methods (at least the four we consider in this work) in preliminary runs is desired (under a testing search budget\footnote{This is only for the preliminary runs, which is often smaller than the actual budget for the ultimate run.}). Note that, at this point, \texttt{Ratio} can be ruled out from consideration to save more resources, since it is generally the worst compared with the other three. Here, there is no need to compare with the Pareto search.

In another situation, if the quality is much more preferred regardless of the resources needed, Pareto search can be a more ideal choice (based on \textbf{RQ2}), i.e., choosing \textit{quality} at \textbf{D1}. This is not surprising, because in certain SBSE problems, such as NRP, satisfying the stakeholders with better (even slightly) fitness utility can often bring significantly more revenues~\cite{DBLP:conf/re/FeatherM02}. One would also need to choose a search algorithm for Pareto search (see~\cite{Sayyad2013b,DBLP:journals/infsof/ColanziAVFG20}), although NSGA-II seems to be a more preferred choice according to various SBSE surveys~\cite{Harman2012,Sayyad2013b,DBLP:journals/infsof/ColanziAVFG20}. Note that in this case, there is no fixed search budget given, instead one should allow the search to achieve reasonable convergence (e.g., the solutions do not change for certain generations), as the quality is more important. A particular step required here is that, in \textbf{D5} and \textbf{D6}, we ask whether the weight is close to one extreme and if extra resources are available, respectively. A \textit{No} to either would proceed to the Pareto search directly. Yet, if there are positive answers to both \textbf{D5} and \textbf{D6}, additional studies are recommended. This is because, as we have shown in \textbf{RQ3}, the worst gains of Pareto search over the best weighted counterpart often occur under the weight vector like $(0.2,0.8)$ and $(0.1,0.9)$ (the other extreme is also applied). Therefore, to further ensure the benefit of Pareto search under such cases, additional experiments in preliminary runs to confirm the quality gains of Pareto search are desirable under a testing search budget (answering \textit{Yes} to \textbf{D4} and \textbf{D7}, but only the fitness utility is important).

Indeed, albeit not always possible, experimentally comparing Pareto and weighted search on a case by case manner in preliminary runs can be helpful on their selection. Therefore, in such case the answer to \textbf{D4} and \textbf{D7} would be \textit{Yes}. In fact, this is also the only way to go in the case that SBSE practitioners have absolutely no clue about whether the quality or resource efficiency is more important, i.e., choosing \textit{not sure} at \textbf{D1}. In such a situation, as shown from the guidance, we suggest one to obtain two outcomes: (1) the quality of results when both Pareto and weighted search under a testing search budget and 
(2) the resource consumed by both in order to reach the best result achieved by weighted search. One can then pick the preferred trade-off with respect to the quality and resource consumed, as achieved by Pareto search and weighted search.

\section{Discussion}
\label{sec:disccusion}

In this section, we discuss a few important factors in our study.

\subsection{Sensitivity of Parameters}
\label{sec:sen-p}

While we have followed the parameter settings used by existing work, it is important to confirm the sensitivity of parameters to the results, particularly on the Pareto optimizers. To that end, for both NSGA-II and MOEA/D, we examine some common settings of the parameters: mutation rate of $\{0.01,0.1\}$, crossover rate of $\{0.8,0.9\}$, and population size of $\{X,{X \over 2}\}$ (where $X$ is the size used in Table~\ref{tb:settings}, depending on the problem), leading to a total of 8 combinatorial settings.

\begin{figure}[t!]
\centering
\includegraphics[width=\columnwidth]{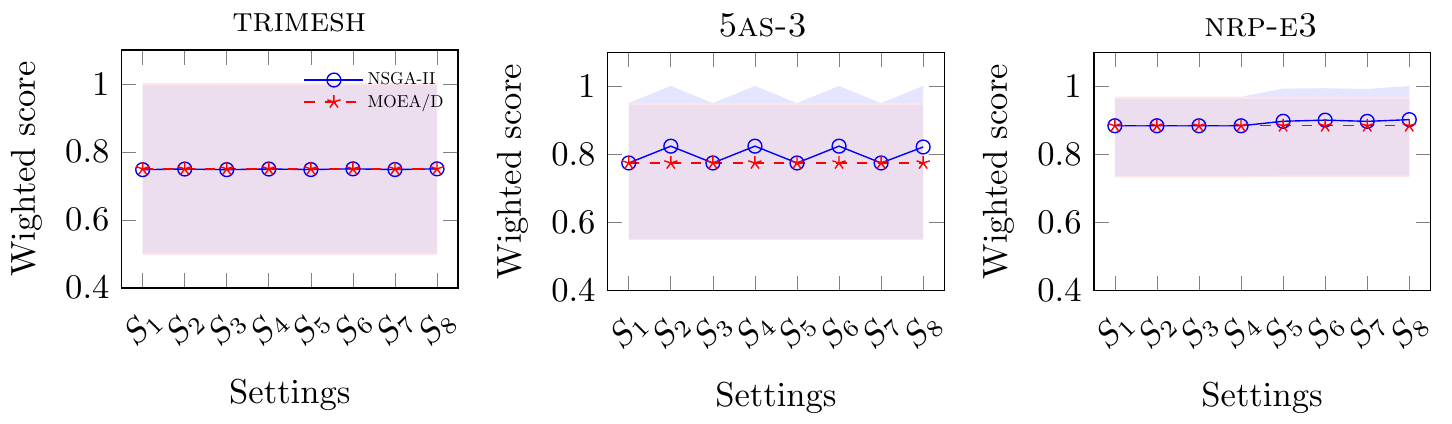}
\caption{The parameter sensitivity of Pareto search on the most obvious cases over 8 different settings of population size, crossover rate, and mutation rate. The line shows the median while the highlighted area indicates the 25th and 75th percentiles of the weighted score over 9 weight vectors and 100 runs each. $S_1$ denotes the setting we applied in this work. }
\label{fig:sen-p}
\end{figure}

We found that for each SBSE problem, the trends are consistent across the systems/projects (regardless of the number of objectives), hence in Figure~\ref{fig:sen-p}, we plot the most obvious case from each problem (i.e., the one with the highest deviation across the settings). From this, we obtained two observations:

\begin{itemize}
    \item The resulted weighted score across different settings do not vary much.
    \item The setting we applied for the \textbf{RQs} (i.e., $S_1$) is one of the best among the others.
\end{itemize}

The above indicate that the conclusions drawn for the \textbf{RQs} are stable, as a reasonable change of the parameter setting leads to very little impact on the result.

\subsection{On $> 3$ Objectives}
\label{sec:obj-num}


We have shown that our conclusions are valid for both two and three objective cases in multi-objective SBSE. 
Yet, certain SBSE problems could consider more than three objectives, 
i.e., what has been known as many-objective search/optimization problems~\cite{DBLP:conf/cec/IshibuchiTN08}.
Many-objective optimization poses big challenges to any algorithms 
aiming to find/approximate the whole Pareto front of the problem.

We anticipate that the benefits of Pareto search would be blurred in such a many-objective setting, 
as what we have shown in Section~\ref{sec:results}, especially for certain optimizers such as NSGA-II.
In fact, 
it has been well studied in the evolutionary computation community 
that the performance of many well-established multi-objective evolutionary algorithms falls rapidly with the increase of the number of objectives,
particularly for Pareto-based search algorithms (e.g., NSGA-II)
which only use the Pareto dominance relation to distinguish between solutions with respect to their convergence~\cite{Wagner2007,Li2013b}.
Recent studies even show that mainstream algorithms like NSGA-II, MOEA/D, and IBEA even completely fail on some four-objective problems~\cite{Li2018a}.

In contrast,
increasing search space has much less effect on weighted search 
since it aggregates the objectives into a scalar value by a weight vector and aims to locate a point of the Pareto front.
However, 
a downside of using weighted search in the many-objective setting is that
the higher the objective dimension the more difficult it is for the stakeholders to specify a weight vector a priori~\cite{DBLP:conf/promise/Harman10}.

%

\subsection{Threats to Validity}
\label{sec:tov}

To ensure \textbf{construct validity}, we use the weighted sum of a given weight vector as the sole metric, which matches precisely with the need to verify our hypothesis. To mitigate threats caused by the stochastic nature of the optimizers, we repeat 100 runs for each case, with validation from Wilcoxon rank-sum test, $\hat{A}_{12}$ effect size, and Scott-Knott test, as commonly recommended for SBSE~\cite{DBLP:journals/infsof/KampenesDHS07,DBLP:conf/icse/ArcuriB11,MittasA13}. To ensure the strongest statistical power, we conduct pairwise comparisons in our study.


Two aspects may form threats to \textbf{internal validity} in our study:

\begin{itemize}
    \item \textbf{Optimizer setting:} In this work, we follow what has been shown to be effective for a SBSE problem in the literature, as our aim is to compare the most common practices. The only part we could not have found for sure is the search budget, which is highly problem-dependent. To tackle such, we have followed carefully designed criteria (Section~\ref{sec:settings}), including both evaluation and time budget, that strike for a balance between reasonable convergence and the time required. We have found that the parameter settings tend to be appropriate and the sensitivity of Pareto search to the parameters are marginal, as discussed in Section~\ref{sec:sen-p}.
    
    \item \textbf{Weight vector:} We used the most common weight vector for a pragmatic reason. That is, in the two objective cases, we choose nine key vectors that are evenly spread across the space; for three objective cases, we used the edge and middle vectors, e.g., $(0.1,0.1,0.8)$ and  $(0.33,0.33,0.33)$. Indeed, this list cannot cover all the possible scenarios, but they are good representatives of the most likely cases.
\end{itemize}


Threats to \textbf{external validity} can come from various sources, including:

\begin{itemize}
    \item \textbf{SBSE problem:} In this work, we select the most representative SBSE problems from several surveys~\cite{Harman2012,Sayyad2013b,Li2020,DBLP:journals/infsof/ColanziAVFG20} based on carefully codified rules (Section~\ref{sec:cases}). Indeed, this list of the studied problem is not exhaustive, and we did not consider some popular problems, such as TCG (due to the nonmonotonic relation between objective and evaluation metric) and automatic refactoring~\cite{DBLP:journals/infsof/MarianiV17} (which is the 6th most popular one from the surveys, but we limited to the top 5 due to resource constraint). We hope our work serves as a first step to open a dialogue on this important topic for the SBSE community, based on which future work can extend this study to cover the SBSE problems that we omitted.
    
    \item \textbf{Optimizer:} In this study, four widely used optimizers for the weighted search are examined, together with four different normalization methods, which are concluded from well-known SBSE surveys. For the Pareto search, we choose NSGA-II as the representative of the Pareto search due mainly to its prevalence and similarity in terms of algorithmic design to the weighted counterpart. We also examine MOEA/D because it uses multiple weights vector to reveal the Pareto front --- a similar property in weighted search. We acknowledge that different optimizers may have diverse ``comfort zones'' for a given SBSE problem, and an extended study may be required for future work.
    
    \item \textbf{Number of objectives:} Our study covers two and three objective cases in SBSE, hence the results may not be generilizable to higher dimension cases of the objectives. However, two or three objectives are the most common studied problem for SBSE, as summarized by Sayyad and Ammar~\cite{Sayyad2013b}. Further, as we discussed in~\ref{sec:obj-num}, there are known studies that confirm some Pareto optimizers can be severely affected by the number of objectives, e.g., NSGA-II.
    
     \item \textbf{SBSE constraints:} We do not consider any constraint for the three multi-objective SBSE problems, which have also been studied in other SBSE work~\cite{Chen2018FEMOSAA}. For example, in SCT, a configuration option cannot be used unless another has been turned on. Indeed, those constraints, when considered, may affect the results on both Pareto and weighted search, as they would inevitably complicate the problem and potentially make the global optimum even more difficult to find. Since this study is the first comprehensive work to compare Pareto and weighted search for SBSE, we started from the simplest assumption where the constraints are omitted. Some of our findings are still exciting, for example, we have shown that, even with such a simpler case, the weighted search has been outperformed by Pareto search in terms of the exact weights that guide it. This can then serve as a foundation for future work to consider a more abnormal landscape of the SBSE problems, i.e., by having complex constraints.
\end{itemize}

We would like to stress that, in this work, we do not aim to exhaustively verify our hypothesis across all situations but to examine whether it is the case under the representative scenarios of SBSE. Nonetheless, we do agree that additional replication studies that extend all (or some) of the above aspects may prove fruitful.



\section{Related Work}
\label{sec:related}

Here we review the prior work for multi-objective SBSE in relation to the purpose of this work.


\subsection{Multi-Objective SBSE with Weighted Search}

Indeed, it is not uncommon to assume clear weights for different objectives under multi-objective SBSE, such as software configuration tuning~\cite{DBLP:conf/icac/RamirezCMB10,DBLP:conf/sigsoft/ShahbazianKBM20,DBLP:conf/ssbse/BowersFC18}, web service composition~\cite{DBLP:conf/gecco/CanforaPEV05,Wagner2012Multi}, next release planning~\cite{DBLP:conf/re/FeatherM02}, software project scheduling~\cite{DBLP:journals/infsof/ChangJDZG08}, and software modularization~\cite{DBLP:journals/isci/HuangL16a}; in fact, this is occasionally referred to as an advantage rather than a limitation. For example, \citeauthor{DBLP:conf/ssbse/BowersFC18}~\cite{DBLP:conf/ssbse/BowersFC18} and \citeauthor{DBLP:conf/sigsoft/ShahbazianKBM20}~\cite{DBLP:conf/sigsoft/ShahbazianKBM20} argue that, in software configuration tuning, being able to specify weights provides more flexibility for the stakeholder to freely set preferences depending on the context.

Existing work applies various normalization methods for the weighted search in multi-objective SBSE when the objectives do not naturally commensurable. For example, \citeauthor{DBLP:conf/sigsoft/ShahbazianKBM20}~\cite{DBLP:conf/sigsoft/ShahbazianKBM20} dynamically update the weights during search such that the different performance objectives are rescaled when tuning software performance. In test case generation, \citeauthor{DBLP:journals/tec/WangAYL18}~\cite{DBLP:journals/tec/WangAYL18} and \citeauthor{DBLP:conf/gecco/PradhanWAY16}~\cite{DBLP:conf/gecco/PradhanWAY16} use $v \over {1+v}$ to normalize an objective's value $v$, yet this method only converts the values into $[0,1]$ without standardizing them, and hence the fitness can still be dominated by the objectives with relatively larger magnitude (e.g., cost over coverage); this, as we have shown in Section~\ref{sec:results}, tends to severely affect the result for the SBSE problems studied. 

\subsection{Multi-Objective SBSE with Pareto Search}

In multi-objective SBSE problems, Pareto search is often regarded as a better strategy when it is impossible to clearly quantify the weights, or it is desirable for the stakeholders to examine the whole Pareto front. Indeed, this is often the case when the number of objectives is high (e.g., three or more) and it has been becoming the standard for certain SBSE problems, such as the software product line engineering~\cite{lian2017,Olaechea2014Comparison,Sayyad2013Optimum,Sayyad2013Scalable} and code refactoring~\cite{Mansoor2015Multi,DBLP:conf/gecco/HarmanT07,DBLP:journals/ese/MkaouerKBCD16}.

\subsection{Comparison on Pareto and Weighted Search}

The first wave of work that compares Pareto and weighted search in multi-objective SBSE appeared more than a decade ago~\cite{Zhang2007The,DBLP:conf/gecco/HarmanT07,DBLP:conf/gecco/LakhotiaHM07} till more recently~\cite{10.1145/3392031}, each of which studies a different SBSE problem, such as next release planning and test case generation. By plotting the result on each objective, the above work demonstrates an obvious but perhaps ``new result" in SBSE by that time: the Pareto search provides better insights on the trade-off surface as it approximates the Pareto front. More recently, \citeauthor{DBLP:journals/tec/WangAYL18}~\cite{DBLP:journals/tec/WangAYL18} and \citeauthor{DBLP:conf/gecco/PradhanWAY16}~\cite{DBLP:conf/gecco/PradhanWAY16} conduct empirical studies that compare weighted search (i.e., \textit{FW} in their work\footnote{This is because \textit{FW} is the only one that assumes clear preferences without approximating the Pareto front.}) with Pareto search using hypervolume (HV) over a few multi-objective SBSE problems, based on which they unsurprisingly concluded that the Pareto search is better as it has higher HV value. 


Those studies, albeit offering interesting findings, are unfair when comparing Pareto and weighted search. This is because they overlook the fact that only a particular solution is of interest to the stakeholders in the presence of clear preferences (rather than the whole Pareto front). 
As a result, interpreting each objective individually from the Pareto front approximation cannot well respect the given preferences. Likewise, 
HV is not suitable since it measures how well the solution set approximates the Pareto front~\cite{Zitzler1998}. 
Another unfairness raised from the fact that the time budget has not been considered, which, as we have shown, could be consumed quite differently even with the same number of evaluations. Our study addresses all of the above issues from prior work. 

\citeauthor{DBLP:journals/ese/MkaouerKBCD16}~\cite{DBLP:journals/ese/MkaouerKBCD16} attempt to achieve a more fair comparison on the software refactoring problem by contrasting the knee solution from the Pareto search to the solution obtained by an equally-weighted search. They concluded that the Pareto search is better, as it leads to a better average of the objectives' values. 
However, 
the knee solution may not be fully in line with the solution obtained by equally weighted search (as it can be most fitted by some other weight vector). 
A direct comparison between them may not be well justified.

A recent study by \citeauthor{DBLP:conf/scam/AlizadehFK19}~\cite{DBLP:conf/scam/AlizadehFK19} seeks to compare both strategies by allowing developers to qualitatively evaluate the solution(s). Since the Pareto search favors many solutions at each run, as expected, the developer concluded that the results of the weighted search are closer to their preferences since there is less ``cognition noise" involved. While this may be the most direct way to assess their usefulness, the evaluation can be, however, biased by human judgment.

The safest option is probably to quantitatively compare them via the weight vector that is used to guide the weighted-search, as what has been done by \citeauthor{Praditwong2011Software}~\cite{Praditwong2011Software} and this work. Yet, unlike our work, \citeauthor{Praditwong2011Software}~\cite{Praditwong2011Software} study the software modularization problem by comparing NSGA-II with hill-climbing\textemdash two fundamentally different optimizers. However, their work differs from ours in the sense that (1) it is questionable that whether the simple hill-climbing can be a good representative of the weighted search; (2) only one SBSE problem is studied; (3) the best normalization method has not been justifiably chosen; and (4) an identical time budget has not been considered.

Overall, through extensive experiments, our work differs from prior studies in that we provide the explicit answer, explanation, and insights to an unexplored question: given clear preferences, identical search budget, can Pareto search converge to the same or better result than weighted search in terms of a given weight vector?

\section{Conclusion}
\label{sec:conclusion}

In this paper, we systematically compare Pareto search and weighted search under clear preferences on a large scale empirical study with 604 cases, including 38 systems/projects from three representative multi-objective SBSE problems, different weight vectors, and two search budget types. Our key finding challenges the existing \textit{weighted search first} belief: 
we show that, although the weighted search is more resource-efficient for reaching certain levels of the result, 
Pareto search is most of the time (up to 77\%) significantly better under a sufficient search budget. 
In particular, 
the more search budget required by Pareto search may be practically trivial, 
e.g., in a magnitude of seconds or less. 
Drawing on the findings, 
we suggest for the first time the following suggestions to the practitioners of multi-objective SBSE:

\begin{itemize}

\item When the quality of the solution is a primary concern for the multi-objective SBSE problem in hand, using Pareto search by default even if there are clear preferences.

\item When the specified weight vector is close to one extreme e.g. $(0.1, 0.9)$ and $(0.2, 0.8)$, 
it is desirable to further experimentally confirm the benefits of Pareto search in preliminary runs.

\item When the resource efficiency (search budget) is a more important factor for the multi-objective SBSE problem in hand, sticking with the existing belief to use weighted search, yet it is non-trivial to decide on how the objectives need to be normalized when necessary.

\item For weighted search, experimentally comparing the normalization methods (at least the four in this work) whenever the conditions permitted; 
the \texttt{Ratio} can be omitted in case the resource is limited. Otherwise, using \texttt{Dynamic} by default.

\end{itemize} 

We codify the above as pragmatic guidance, 
hoping to provide a clear view on the choice between Pareto search and weighted search for the community. 
Several future opportunities can be derived from this work to advance the understanding of the topic, 
such as extending the study to wider range of SBSE problems and scenarios; replacing the weighted sum with other scalarizing functions (e.g., Tchebycheff) in weighted search; and linking the findings to the shape of the SBSE problem's Pareto front. 


\bibliographystyle{ACM-Reference-Format}
\bibliography{references}

\end{document}